\pdfoutput=1
\documentclass[preprint,3p,12pt]{elsarticle}

\journal{Annals of Physics (N.Y.)}
\usepackage{amssymb}
\usepackage{amsmath}
\usepackage{array}
\usepackage{epsfig}

\def\abs#1{\left|#1\right|}
\def\ket#1{|#1\rangle}
\def\bra#1{\langle#1|}

\def\bz{\beta_0}

\def\lm{(\lambda,\mu)}
\def\su3{\mathrm{SU3}}
\def\u5{\mathrm{U5}}
\def\Hone{\hat{H}_1(\rho)}
\def\Htwo{\hat{H}_2(\xi)}

\newcommand{\ba}{\begin{eqnarray}}
\newcommand{\ea}{\end{eqnarray}}
\newcommand{\bmath}{\begin{mathletters}}
\newcommand{\emath}{\end{mathletters}}
\newcommand{\ban}{\begin{eqnarray*}}
\newcommand{\ean}{\end{eqnarray*}}

\newcommand{\tlb}{\tilde{\beta}}
\newcommand{\bsub}{\begin{subequations}}
\newcommand{\esub}{\end{subequations}}

\begin{document}

\begin{frontmatter}

\title{First-order quantum phase transitions: test ground for\\
emergent chaoticity, regularity and persisting symmetries}
\author[huji]{M. Macek\corref{cor1}}
\ead{mmacek@Racah.phys.huji.ac.il}

\author[huji]{A. Leviatan\corref{cor2}}
\ead{ami@phys.huji.ac.il}

\cortext[cor1]{Current address: Center for Theoretical Physics, Sloane Physics 
Laboratory, 
Yale University, New Haven, CT 06520-8120, USA}
\cortext[cor2]{Corresponding author.}

\address[huji]{Racah Institute of Physics, The Hebrew University,
Jerusalem 91904, Israel}

\begin{abstract}
We present a comprehensive analysis of the emerging
order and chaos and enduring symmetries, accompanying
a generic (high-barrier) first-order quantum phase transition (QPT).
The interacting boson model Hamiltonian employed, describes a QPT
between spherical and deformed shapes, associated with
its U(5) and SU(3) dynamical symmetry limits.
A~classical analysis of the intrinsic
dynamics reveals a rich but simply-divided phase space structure
with a H\'enon-Heiles type of chaotic dynamics ascribed to the
spherical minimum and a robustly regular dynamics ascribed to the deformed
minimum. The simple pattern of mixed but well-separated dynamics persists
in the coexistence region and traces the crossing of the two
minima in the Landau potential. A quantum analysis
discloses a number of regular low-energy U(5)-like multiplets
in the spherical region, and regular SU(3)-like rotational bands
extending to high energies and angular momenta, in the deformed region.
These two kinds of regular subsets of states retain their identity
amidst a complicated environment of other states and both occur
in the coexistence region. A symmetry analysis of their wave functions
shows that they
are associated with partial U(5) dynamical symmetry (PDS) and
SU(3) quasi-dynamical symmetry (QDS), respectively.
The pattern of mixed but well-separated dynamics and the PDS or QDS
characterization of the remaining regularity, appear to be robust
throughout the QPT.
Effects of kinetic collective rotational terms, which may disrupt
this simple pattern, are considered.
\end{abstract}

\begin{keyword}

Quantum shape-phase transitions; Regularity and chaos;
Partial and quasi-dynamical symmetries; Interacting boson model (IBM).

\vspace{12pt}
\PACS 21.60.Fw; 05.45.Mt; 05.30.Rt; 21.10.Re
\end{keyword}

\end{frontmatter}

\newpage

\section{Introduction}

Quantum phase transitions (QPTs) are qualitative changes in the properties
of a physical system induced by a variation of parameters $\lambda$ in the
quantum Hamiltonian $\hat{H}(\lambda)$~\cite{Hert76,Gilm78,Gilm79}.
Such ground-state transformations have received considerable attention in
recent years and have found a variety of applications in many areas of
physics and chemistry~\cite{Carr10}.
These structural modifications occur at zero-temperature
in diverse dynamical systems
including spin lattices~\cite{Sachdev99},
ensembles of ultracold atoms~\cite{Grei02,Baumann10} and
atomic nuclei~\cite{Cejn10}.

The particular type of QPT is reflected in the topology of the
underlying mean-field (Landau) potential $V(\lambda)$.
Most studies have focused on second-order
(continuous) QPTs, where $V(\lambda)$ has a single minimum
which evolves continuously into another minimum.
The situation is more complex for
discontinuous (first-order) QPTs, where
$V(\lambda)$ develops multiple minima
that coexist in a range of $\lambda$ values
and cross at the critical point, $\lambda\!=\!\lambda_c$.
The competing interactions in the Hamiltonian that drive these
ground-state phase transitions can affect dramatically the nature of
the dynamics and, in some cases, lead to the emergence of quantum
chaos~\cite{Gutz90,Reic92,Licht92}.
This effect has been observed in quantum optics models of $N$
two-level atoms interacting with a
single-mode radiation field~\cite{Emar03a,Emar03b}, where
the onset of chaos is triggered by continuous QPTs.
In the present article,
we examine similar effects for the less-studied discontinuous QPTs,
and explore the nature of the underlying classical and quantum
dynamics in such circumstances.

The interest in first-order quantum phase transitions
stems from their key role in
phase-coexistence phenomena at zero temperature.
In condensed matter physics, it has been recently recognized that,
for clean samples, the nature of the QPT becomes discontinuous as the
critical-point is approached.
Examples are offered by the metal-insulator Mott
transition~\cite{maria04}, itinerant magnets~\cite{Pf05},
heavy-fermion superconductors~\cite{Pf09},
quantum Hall bilayers~\cite{karm09},
Bose-Einstein condensates~\cite{Phuc13} and Bose-Fermi
mixture~\cite{Bert13}. First-order QPTs are relevant to
shape-coexistence in mesoscopic systems,
such as atomic nuclei~\cite{Cejn10}, and to
optimization problems in quantum computing~\cite{young10}.

Hamiltonians describing first-order QPTs are often non-integrable,
hence their dynamics is mixed. They form a subclass among the family of
generic Hamiltonians with a mixed phase space,
in which regular and chaotic motion coexist.
Mixed phase spaces are often encountered
in billiard systems~\cite{Gutz90,Reic92,Licht92},
which are generated by the free motion of a point particle inside a
closed domain whose geometry governs the
amount of chaoticity. Here, in contrast, we consider
many-body interacting systems undergoing a first-order QPT,
where the onset of chaos is governed by a change of coupling constants
in the Hamiltonian.
The amount of order and disorder in the system is affected by the
relative strengths of different terms in the Hamiltonian which
have incompatible symmetries.
Order, chaos and symmetries are thus intertwined,
and their study can shed light on the structure evolution.
In conjunction with first-order QPTs, this raises a number of key questions.
(i)~How does the interplay of order and chaos reflect the first-order
QPT, in particular, the changing topology of the Landau potential
in the coexistence region.
(ii)~What is the symmetry character (if any)
of the remaining regularity in the system, amidst a complicated environment.
(iii)~What is the effect of kinetic terms, which do not affect
the potential, on the onset of chaos across the QPT.

To address these questions in a transparent manner,
we employ an interacting boson model
(IBM)~\cite{ibm}, which describes
quantum phase transitions between spherical and deformed nuclei.
The model is amenable to both classical and quantum treatments, has
a rich algebraic structure and inherent geometry.
The phases are associated with different nuclear
shapes and correspond to solvable dynamical symmetry limits of the model.
The Hamiltonian accommodates QPTs of first- and second order between
these shapes, by breaking and mixing the relevant limiting
symmetries. These attributes make the IBM an ideal framework
for studying the intricate interplay of order and chaos and
the role of symmetries in such quantum shape-phase transitions.
It is a representative of a wide class of algebraic models used for
describing many-body systems, {\it e.g.}, nuclei~\cite{ibm},
molecules~\cite{vibron} and hadrons~\cite{BIL}.

QPTs have been studied extensively in the IBM
framework~\cite{Cejn10,CejJol09,iac11} and
are manifested empirically in nuclei~\cite{Casten09,CasMcC07}.
The situation is summarized in the indicated review papers where
a complete list of references is given.
Particular attention has been paid to symmetry aspects
(critical point
symmetries~\cite{cps1,cps2}, quasi-dynamical~\cite{qds,Rowe04,RosRow05}
and partial dynamical symmetries~\cite{lev07}),
finite-size effects~\cite{levgin03,zamfir,lev05,lev06,arias07} and
scaling behavior~\cite{RowTur04,dusuel05,williams10}.
Further extensions of the QPT concept to excited states~\cite{caprio}
and to Bose-Fermi systems~\cite{PLI11}, have also been considered.

Chaotic properties of the IBM have been throughly investigated
both classically and quantum
mechanically~\cite{ANW90,AW91,AW91b,ANW92,WA93,Heinze06,
Macek06,Macek07}. All such treatments involved a simplified Hamiltonian
giving rise to
integrable second order QPTs
and to non-integrable first order QPTs with
an extremely low barrier and narrow coexistence region.
A new element in the present treatment, compared to previous works,
is the employment of
IBM Hamiltonians without such restrictions~\cite{lev06}
and their resolution into intrinsic and collective
parts~\cite{kirlev85,lev87}.
This enables a comprehensive analysis of the vibrational and rotational
dynamics across a generic (high-barrier) first-order QPT, both
inside and outside the coexistence region.
Brief accounts of some aspects of this
analysis were reported in~\cite{MacLev11,LevMac12}.

Section~2 reviews the algebraic, geometric and symmetry content
of the IBM. An intrinsic Hamiltonian for a first-order QPT between
spherical and deformed shapes, with an adjustable barrier height, is
introduced in Section~3, and its symmetry properties are discussed.
The  classical limit of the QPT Hamiltonian is derived in Section 4.
The topology of the classical potential is studied in great detail,
identifying the control and order parameters in various structural
regions of the QPT. A comprehensive classical analysis is performed
in Section~5, focusing on regular and chaotic features of the intrinsic
vibrational dynamics across the QPT. Special attention is paid to the
dynamics in the vicinity of minima in the Landau potential
and to resonance effects.
An elaborate quantum analysis is conducted in Section~6 with emphasis on
quantum manifestations of classical chaos and remaining regular features
in the spectrum. A symmetry analysis is performed in Section~7, examining
the symmetry content of eigenstates and the evolution of purity and
coherence throughout the QPT. The impact of different collective
rotational terms on the classical and quantum dynamics is considered
in Section~8. The implications of modifying the barrier height, are
examined in Section~9. The final Section is devoted to a summary and
conclusions. Specific details on the IBM potential surface and on linear
correlation coefficients are collected in Appendix A and B, respectively.

\section{The interacting boson model: algebras, geometry, and symmetries}
\label{sec:IBM}

The interacting boson model (IBM)~\cite{ibm} describes low-lying quadrupole
collective states in nuclei in terms of $N$ interacting monopole
$(s)$ and quadrupole $(d)$ bosons representing valence nucleon pairs.
The bilinear combinations
${\cal G}_{ij}\equiv b^{\dag}_{i}b_j =
\{s^{\dag}s,\,s^{\dag}d_{m},\, d^{\dag}_{m}s,\,
d^{\dag}_{m}d_{m '}\}$ span a U(6) algebra, which
serves as the spectrum generating algebra.
The IBM Hamiltonian is expanded in terms of these generators,
$\hat{H} = \sum_{ij}\epsilon_{ij}\,{\cal G}_{ij}
+ \sum_{ijk\ell}u_{ijk\ell}\,{\cal G}_{ij}{\cal G}_{k\ell}+\ldots$,
and consists of Hermitian, rotational-invariant interactions
which conserve the total number of $s$- and $d$- bosons,
$\hat N = \hat{n}_s + \hat{n}_d =
s^{\dagger}s + \sum_{m}d^{\dagger}_{m}d_{m}$.
A dynamical symmetry (DS) occurs if the Hamiltonian
can be written in terms of the Casimir operators
of a chain of nested sub-algebras of U(6).
The Hamiltonian is then completely solvable in the basis associated with
each chain.
The three dynamical symmetries of the IBM~\cite{ibmu5,ibmsu3,ibmo6}
and corresponding bases are
\bsub
\label{eq:chains}
\ba
&&{\rm U(6) \supset U(5)  \supset O(5) \supset O(3)}
\qquad
\vert N,n_d,\tau,n_{\Delta},L\rangle
\;\;\;\;\,
\label{u5ds}
\;{\rm spherical\;vibrator}
\\
&&{\rm U(6) \supset SU(3) \supset O(3)}
\qquad\qquad\;\;\,
\vert N,(\lambda,\mu),K,L\rangle \quad\;\;
\label{su3ds}
{\rm axially} {\rm-deformed \; rotor}
\\
&&{\rm U(6) \supset O(6)  \supset O(5) \supset O(3)}
\qquad
\vert N,\sigma,\tau,n_{\Delta},L\rangle \qquad
\,\gamma{\rm-unstable\; deformed\; rotor}\qquad\quad
\label{o6ds}
\ea
\esub
The associated analytic solutions
resemble known limits of the geometric model of nuclei~\cite{bohr75},
as indicated above. The basis members are
classified by the irreducible representations (irreps) of the
corresponding algebras. Specifically,
the quantum numbers $N,n_d,(\lambda,\mu),\sigma,\tau$ and $L$
label the relevant irreps of ${\rm U(6),\, U(5),\, SU(3),\, O(6),\, O(5)}$
and O(3), respectively.
$n_{\Delta}$ and $K$ are multiplicity labels needed for complete
classification of selected states in the reductions
${\rm O(5)\supset O(3)}$ and ${\rm SU(3)\supset O(3)}$, respectively.
Each basis is complete and can be used for a numerical diagonalization
of the Hamiltonian in the general case. Relevant information on the
generators, Casimir operators and eigenvalues for the above algebras
is collected in Table~\ref{TabIBMcas}.
Also listed are additional algebras,
$\overline{{\rm O(6)}}$ and $\overline{{\rm SU(3)}}$, obtained by
a phase-change of the $s$-boson.
\begin{table}[t]
\begin{center}
\caption{
\label{TabIBMcas}
\protect\small
Generators, linear and quadratic Casimir operators $\hat{C}_{G}$
and their eigenvalues $\langle\hat{C}_G\rangle$,
for algebras~$G$ in the IBM.
Here
$\hat{n}_{s} = s^{\dag}s$,
$\hat{n}_d = \sqrt{5}\,U^{(0)}$, $\hat{N}= \hat{n}_s + \hat{n}_d$,
$L^{(1)}_{m} = \sqrt{10}\,U^{(1)}_{m}$,
$Q^{(2)}_{m} = \Pi^{(2)}_{m} -{\textstyle\frac{\sqrt{7}}{2}}\,U^{(2)}_{m}$,
$\bar{Q}^{(2)}_{m} = \Pi^{(2)}_{m}
+{\textstyle\frac{\sqrt{7}}{2}}\,U^{(2)}_{m}$,
$\Pi^{(2)}_{m} = d^{\dag}_{m}s + s^{\dag}\tilde{d}_{m}$,
$\bar{\Pi}^{(2)}_{m} = i(d^{\dag}_{m}s - s^{\dag}\tilde{d}_{m})$,
$U^{(\ell)}_{m} = (d^{\dag}\,\tilde{d})^{(\ell)}_{m}$, where
$U^{(\ell)}_{m}$ stands for a spherical tensor operator of rank $\ell$
projection $m$, and $\tilde{d}_{m} = (-1)^{m}d_{-m}$.}
\vspace{1mm}
\centering
\begin{tabular}{llll}
\hline
& & &\\[-3mm]
Algebra & Generators & Casimir operator $\hat{C}_{G}$ &
Eigenvalues $\langle\hat{C}_G\rangle$\\[4pt]
& & & \\[-3mm]
\hline
& & & \\[-2mm]
{\rm O(3)} & $U^{(1)}$ & $L^{(1)}\cdot L^{(1)}$ & L(L+1) \\[2pt]
{\rm O(5)} & $U^{(1)},U^{(3)}$ &
$2\sum_{L=1,3}\,U^{(L)}\cdot U^{(L)}$ & $\tau(\tau+3)$ \\[2pt]
{\rm O(6)} & $U^{(1)},U^{(3)},\Pi^{(2)}$ &
$\hat{C}_{\rm O(5)}
+ \Pi^{(2)}\cdot\Pi^{(2)}$ & $\sigma(\sigma+4)$ \\[2pt]
{\rm SU(3)} & $U^{(1)},Q^{(2)}$ &
$2Q^{(2)}\cdot Q^{(2)} + {\textstyle\frac{3}{4}}\hat{C}_{\rm O(3)}$ &
$\lambda^2 +(\lambda+\mu)(\mu+3)$\\[2pt]
{\rm U(5)} & $U^{(\ell)}$ $\ell=0,...\,,4$ &
$\hat{n}_d,\,\hat{n}_d(\hat{n}_d+4)$ & $n_d,\, n_d(n_d+4)$ \\[2pt]
{\rm U(6)} &
$\hat{n}_s,\,\Pi^{(2)},\, \bar{\Pi}^{(2)},\,
U^{(\ell)}$ $\ell=0,...\,,4$ &
$\hat{N},\,\hat{N}(\hat{N}+5)$ & $N,\, N(N+5)$ \\[2pt]
$\overline{{\rm O(6)}}$ & $U^{(1)},U^{(3)},\bar{\Pi}^{(2)}$ &
$\hat{C}_{\rm O(5)}
+ \bar{\Pi}^{(2)}\cdot\bar{\Pi}^{(2)}$ & $\bar{\sigma}(\bar{\sigma}+4)$ \\[2pt]
$\overline{{\rm SU(3)}}$ & $U^{(1)},\bar{Q}^{(2)}$ &
$2\bar{Q}^{(2)}\cdot \bar{Q}^{(2)}
+ {\textstyle\frac{3}{4}}\hat{C}_{\rm O(3)}$ &
$\bar{\lambda}^2 +(\bar{\lambda}+\bar{\mu})(\bar{\mu}+3)$\\[2pt]
& & &\\[-3mm]
\hline
\end{tabular}
\label{Tab1}
\end{center}
\end{table}

A geometric visualization of the model is obtained by
a potential surface
\ba
V(\beta,\gamma) &=&
\bra{\beta,\gamma;N}\hat{H}\ket{\beta,\gamma;N} ~,
\label{enesurf}
\ea
defined by the expectation value of the Hamiltonian in the
following intrinsic condensate state~\cite{gino80,diep80}
\bsub
\label{condgen}
\ba
\vert\beta,\gamma ; N \rangle &=&
(N!)^{-1/2}[\,\Gamma^{\dagger}_{c}(\beta,\gamma)\,]^N\vert 0\rangle ~,\\
\Gamma^{\dagger}_{c}(\beta,\gamma) &=&
\left [\beta\cos\gamma d^{\dagger}_{0} + \beta\sin{\gamma}
( d^{\dagger}_{2} + d^{\dagger}_{-2})/\sqrt{2}
+ \sqrt{2-\beta^2}s^{\dagger}\right ]/\sqrt{2} ~.
\ea
\esub
Here $(\beta,\gamma)$ are quadrupole shape parameters
analogous to the variables of the collective model of
nuclei~\cite{bohr75}. Their values
$(\beta_{\mathrm{eq}},\gamma_{\mathrm{eq}})$ at the global minimum
of $V(\beta,\gamma)$
define the equilibrium shape for a given Hamiltonian.
For one- and two-body interactions,
the shape can be spherical $(\beta_{\mathrm{eq}}=0)$ or
deformed $(\beta_{\mathrm{eq}}>0)$ with
$\gamma_{\mathrm{eq}}=0$ (prolate),
$\gamma_{\mathrm{eq}}=\pi/3$ (oblate) or $\gamma$-independent.
The parameterization adapted in Eq.~(\ref{condgen})
is particularly suitable for a classical analysis of the model.
An alternative parameterization for the shape parameters
and further properties of the potential surface are discussed
in Appendix~A.

The dynamical symmetries of Eq.~(\ref{eq:chains})
correspond to solvable limits of the model.
The often required symmetry breaking is achieved by including in
the Hamiltonian terms associated with different sub-algebra chains
of U(6). In general, under such circumstances, solvability is lost, there
are no remaining non-trivial conserved quantum numbers
and all eigenstates are expected to be mixed.
However, for particular symmetry breaking, some intermediate symmetry
structure can survive. The latter include partial dynamical
symmetry (PDS)~\cite{pds} and quasi-dynamical symmetry (QDS)~\cite{qds}.
In a PDS, the conditions of an
exact dynamical symmetry (solvability of the complete spectrum and
existence of exact quantum numbers for all eigenstates) are relaxed and
apply to only part of the eigenstates and/or of the quantum numbers.
In a QDS, particular states continue to exhibit
selected characteristic properties ({\it e.g.}, energy
and B(E2) ratios) of the closest dynamical symmetry, in the face
of strong-symmetry breaking interactions.
This ``apparent'' symmetry is due to the coherent nature of the mixing.
Interestingly, both PDS~\cite{lev07} and
QDS~\cite{qds,Rowe04,RosRow05} have been shown to occur
in quantum phase transitions.

In discussing the dynamics of the IBM Hamiltonian,
it is convenient to resolve it into intrinsic and collective
parts~\cite{kirlev85,lev87},
\ba
\hat{H} = \hat{H}_{\mathrm{int}} + \hat{H}_{\mathrm{col}} ~.
\label{eq:H}
\ea
The intrinsic part ($\hat{H}_{\mathrm{int}}$)
determines the potential surface $V(\beta,\gamma)$, Eq.~(\ref{enesurf}),
and is defined to yield zero when acting on the equilibrium condensate
\ba
\hat{H}_{\mathrm{int}}
\vert\beta=\beta_{\mathrm{eq}},
\gamma=\gamma_{\mathrm{eq}} ; N \rangle &=& 0 ~.
\label{Hintcond0}
\ea
For $\beta_{\rm eq}=0$, the condensate is spherical, and consists
of a single state with angular momentum $L=0$ built from $N$ s-bosons.
For $(\beta_{\rm eq}>0,\gamma_{\rm eq}=0)$ the condensate is deformed,
and has angular projection $K=0$ along the symmetry
$z$-axis. States of good $L$ projected from it
span the $K=0$ ground band, and other eigenstates of
$\hat{H}_{\mathrm{int}}$ are arranged in excited $K$-bands.
The collective part ($\hat{H}_{\mathrm{col}}$) has a flat potential surface
and involves collective rotations linked
with the groups in the chain
$\overline{O(6)}\supset O(5)\supset O(3)$. These orthogonal
groups correspond to ``generalized'' rotations associated with the
$\beta$-, $\gamma$- and Euler angles degrees of freedom, respectively.
Apart from constant terms of no significance to the excitation spectrum,
the collective Hamiltonian is composed of the
two-body parts of the respective Casimir operators
\ba
\label{eq:Hcol}
\hat{H}_\mathrm{col} &=&
\bar{c}_3 \left [\, \hat{C}_{O(3)} - 6\hat{n}_d \,\right ] +
\bar{c}_5 \left [\, \hat{C}_{O(5)} - 4\hat{n}_d \,\right ] +
\bar{c}_6 \left [\, \hat{C}_{\overline{O(6)}} - 5\hat{N}\,\right ] ~.
\ea
Here $\hat{C}_{G}$
are defined in Table~\ref{TabIBMcas} and a per-boson scaling
is invoked $\bar{c}_i = c_i/N(N-1)$
to ensure that the bounds of the energy spectrum do not change for
large $N$.
In general, the intrinsic and collective Hamiltonians do not commute
and $\hat{H}_{\rm col}$ splits and mixes the bands generated by
$\hat{H}_{\rm int}$.

\section{First order quantum phase transitions in the IBM}
\label{sec:QPT1}

The dynamical symmetries of the IBM,
Eq.~(\ref{eq:chains}), correspond to phases of the system, and provide
analytic benchmarks for the dynamics of stable nuclear shapes.
Quantum phase transitions (QPTs) between different shapes are
studied~\cite{diep80} by considering Hamiltonians $\hat{H}(\lambda)$
that mix interaction terms from different DS chains, {\it e.g.},
$\hat{H}(\lambda) = \lambda\hat{H}_1+ (1-\lambda)\hat{H}_2$.
The coupling coefficient $(\lambda$) responsible
for the mixing, serves as the control parameter which upon variation
induces qualitative changes in the properties of the system.
The kind of QPT is dictated by the potential surface
$V(\lambda)\equiv V(\lambda;\beta,\gamma)$, Eq.~(\ref{enesurf}),
which serves as a mean-field Landau potential with the equilibrium
deformations ($\beta_{\mathrm{eq}},\gamma_{\mathrm{eq}}$) as order parameters.
The order of the phase transition and the critical point,
$\lambda=\lambda_c$, are determined by the order of the derivative
with respect to $\lambda$ of
$V(\lambda;\beta_{\mathrm{eq}},\gamma_{\mathrm{eq}})$, where discontinuities
first occur.

The IBM phase diagram~\cite{feng81} consists of spherical and deformed
phases separated by a line of first-order transition ending in a point of
second-order transitions in-between the spherical [U(5)] and
deformed $\gamma$-unstable [O(6)] phases. The spherical [U(5)] to
axially-deformed [SU(3)] transition is of first order and the O(6)-SU(3)
transition exhibits a cross-over. In what follows, we examine the nature of
the classical and quantum dynamics across a generic first-order QPT,
with a high-barrier separating the two phases.

\subsection{Intrinsic Hamiltonian in a first order QPT}
\label{sec:HintQPT}

Focusing on first-order QPTs between
stable spherical ($\beta_{\mathrm{eq}}=0$) and prolate-deformed
($\beta_{\mathrm{eq}}>0$, $\gamma_{\mathrm{eq}}=0$) shapes,
the intrinsic Hamiltonian reads
\bsub
\label{eq:Hint}
\ba
\hat{H}_1(\rho)/\bar{h}_2 &=&
2(1\!-\! \rho^2\beta_0^{2})\hat{n}_d(\hat{n}_d \!-\! 1)
+\beta_0^2 R^{\dag}_2(\rho) \cdot\tilde{R}_2(\rho) ~,\qquad
\label{eq:H1} \\
\hat{H}_2(\xi)/ \bar{h}_2 &=&
\xi P^{\dag}_0(\beta_0) P_0(\beta_0) +
P^{\dag}_2(\beta_0) \cdot \tilde{P}_2(\beta_0) ~,
\label{eq:H2}
\ea
\esub
where $\hat{n}_d$
is the $d$-boson number operator and the boson pair operators are
defined as
\bsub
\ba
R^{\dag}_{2\mu}(\rho) &=& \sqrt{2}s^\dag d^\dag_\mu +
\rho\sqrt{7}(d^\dag d^\dag)^{(2)}_\mu ~,
\label{R0}\\
P^{\dag}_0(\beta_0) &=&
d^\dag \cdot d^\dag - \beta_0^2 (s^\dag)^2 ~,
\label{P0}\\
P^{\dag}_{2\mu}(\beta_0) &=&
\sqrt{2}\beta_0 s^\dag d^\dag_\mu +
\sqrt{7}(d^\dag d^\dag)^{(2)}_\mu ~.
\label{P2}
\ea
\label{eq:Ps}
\esub
In Eq.~(\ref{eq:Hint}),
$\tilde{R}_{2\mu} \!=\! (-1)^{\mu}R_{2,-\mu}$,
$\tilde{P}_{2\mu} \!=\! (-1)^{\mu}P_{2,-\mu}$, standard notation of angular
momentum coupling is used and the dot implies a scalar product.
As in Eq.~(\ref{eq:Hcol}),
scaling by $\overline{h}_2\equiv h_2/ N(N-1)$
is used throughout ($h_2>0$), to facilitate the comparison with the
classical limit.
The control parameters that drive the QPT are $\rho$ and $\xi$,
with $0\leq\rho\leq\beta_{0}^{-1}$ and $\xi\geq 0$, while $\beta_0$
is a constant.

The intrinsic Hamiltonian in the spherical phase, $\hat{H}_1(\rho)$,
describes the dynamics of a spherical shape and
satisfies Eq.~(\ref{Hintcond0}), with $\beta_{\rm eq}=0$.
For large $N$, its normal modes
involve five-dimensional quadrupole vibrations
about the spherical global minimum of its potential surface,
with frequency
\ba
\epsilon = 2\bar{h}_2N\beta_{0}^2 ~.
\label{nmodesph}
\ea
The intrinsic Hamiltonian in the deformed phase, $\hat{H}_2(\xi)$,
describes the dynamics of an axially-deformed shape and
satisfies Eq.~(\ref{Hintcond0}), with
$(\beta_{\rm eq} = \sqrt{2}\bz(1+\bz^2)^{-1/2},\gamma_{\rm eq}=0)$.
For large $N$, its normal modes
involve one-dimensional $\beta$ vibration and two-dimensional
$\gamma$ vibrations about the prolate-deformed
global minimum of its potential surface, with frequencies
\bsub
\ba
\epsilon_\beta &=& 2\bar{h}_2 N\bz^2 (2\xi + 1) ~, \\
\epsilon_\gamma &=& 18 \bar{h}_2 N \bz^2 (1+\bz^2)^{-1} ~.
\ea
\label{nmodedef}
\esub
The two intrinsic Hamiltonians coincide at the critical
point, $\rho_c\!=\!\bz^{-1}$ and $\xi_c \!=\!0$,
\ba
\hat{H}^{\rm int}_{\rm cri} \equiv
\hat{H}_{1}(\rho_c) = \hat{H}_{2}(\xi_c) ~,
\label{Hcri}
\ea
where $\hat{H}^{\rm int}_{\rm cri}$ is the critical-point
intrinsic Hamiltonian considered in~\cite{lev06}.

The collective Hamiltonian, Eq.~(\ref{eq:Hcol}), does not affect the
shape of the potential surface but can contribute a shift to the
normal-mode frequencies, Eqs.~(\ref{nmodesph})-(\ref{nmodedef}),
by the amount
\bsub
\ba
\epsilon^{c} &=& 2N\bar{c}_6 ~,\\
\epsilon^{c}_{\beta} &=& 2N\bar{c}_6 ~, \\
\epsilon^{c}_{\gamma} &=&
2N \left [\,\bar{c}_6 + \bar{c}_5\bz^2(1+\bz^2)^{-1}\,\right ]~.
\ea
\label{nmodescol}
\esub
In general, given an Hamiltonian $\hat{H}$,
the intrinsic and collective parts,
Eq.~(\ref{eq:H}), are fixed by the condition of Eq.~(\ref{Hintcond0})
and by requiring $\hat{H}_{\rm int}$ and $\hat{H}$ to have the same shape
for the potential surface.
For example, an Hamiltonian~\cite{ECQF}
frequently used in the study of QPTs~is
\ba
\hat{H} &=& \epsilon\,\hat{n}_d - \kappa\, Q(\chi)\cdot Q(\chi) ~,
\label{HndQQ}
\ea
where $Q(\chi) = d^{\dag}s+s^{\dag}\tilde{d}
+ \chi(d^{\dag}\tilde{d})^{(2)}$ is the quadrupole operator
and $\epsilon\geq 0$, $\kappa\geq 0$,
$ -\tfrac{\sqrt{7}}{2} \leq \chi < 0$. The critical-point Hamiltonian
is obtained for a specific relation among $\epsilon,\,\kappa$ and $\chi$,
\ba
\hat{H}_{\rm cri}:&&\quad
\epsilon =\kappa \left [
(\tfrac{2}{7}\chi^2 + 4)N + \tfrac{5}{7}\chi^2 - 8 \right ]~.
\label{HcrindQQ}
\ea
The parameters of the intrinsic and collective Hamiltonians are then
found to be
\bsub
\ba
\hat{H}^{\rm int}_{\rm cri}:&&\quad
\bar{h}_2 = 2\kappa \;\; , \;\;
\rho_c =\bz^{-1}\;\; , \;\;\xi_c=0 \;\; , \;\;
\bz = -\tfrac{1}{\sqrt{14}}\chi ~.\\
\hat{H}^{\rm col}_{\rm cri}:&&\quad
\bar{c}_3 = \kappa(2-\bz^2)\;\; ,\;\;
\bar{c}_5 = \kappa(4\bz^2-5)\;\; ,\;\;
\bar{c}_6 = \kappa ~.
\ea
\esub
In the present study, we adapt a different strategy.
We fix the value of the parameter $\bz$ in the intrinsic Hamiltonian,
Eq.~(\ref{eq:Hint}), so as to ensure a high barrier at the critical point.
We then vary, {\it independently},
the control parameters ($\rho, \xi$) in the intrinsic Hamiltonian
and the parameters $\bar{c}_3,\,\bar{c}_5,\,\bar{c}_6$,
of the collective Hamiltonian, Eq.~(\ref{eq:Hcol}).
This will allow us to examine, separately, the influence
on the dynamics of those terms affecting the Landau potential
and of individual rotational kinetic terms,
in a generic (high-barrier) first order QPT.

\subsection{Symmetry properties and integrability}
\label{sec:Sym}

The symmetry properties of the intrinsic Hamiltonian (\ref{eq:Hint})
depend on the choice of control parameters $(\rho,\xi)$ and of $\bz$.
In general, the dynamical symmetries are completely broken in the
Hamiltonian and hence the underlying dynamics is non-integrable.
However, for particular values of
these parameters, exact dynamical symmetries (DS) or partial dynamical
symmetries (PDS) can occur and their presence affects the integrability
of the system.

The appropriate intrinsic
Hamiltonian in the spherical phase
is $\hat{H}_{1}(\rho)$, Eq.~(\ref{eq:H1}),
with $0\leq\rho\leq\bz^{-1}$.
For $\rho=0$, $\hat{H}_{1}(\rho=0)$ reduces to
\ba
\hat{H}_1(\rho=0)/\bar{h}_2 &=&
2\hat{n}_d(\hat{n}_d - 1)
+2\beta_{0}^2(\hat{N}-\hat{n}_d)\hat{n}_d ~,
\label{H1u5}
\ea
and hence has U(5) DS.
The spectrum is completely solvable
\ba
\vert N,n_d,\tau,n_{\Delta},L\rangle \;\;\;\;
&&E_{\rm DS} =
\left [\, 2n_d(n_d-1) + 2\beta_{0}^2(N-n_d)n_d\, \right ]\bar{h}_2 ~.
\label{U5-DSspec}
\ea
The eigenstates are those of the U(5) chain,
Eq.~(\ref{u5ds}), with $n_d=0,1,2,\ldots, N$ and
$\tau=n_d,\,n_d-2,\dots 1\,{\rm or}\; 0$.
The values of $L$ are obtained by partitioning
$\tau=3n_{\Delta}~+~p$, with $n_{\Delta},\,p\geq 0$
integers, and $L=2p,2p-2,2p-3,\ldots, p$.
The spectrum resembles that of an anharmonic spherical vibrator,
describing quadrupole excitations of a spherical shape.
The lowest U(5) multiplets involve states with quantum numbers~\cite{ibmu5}
\ba
\begin{array}{lll}
n_d=3: & L=6,4,3,0\;\; (\tau=3) & L=2\;\; (\tau=1) \\
n_d=2: & L=4,2\;\; (\tau=2) & L=0\;\; (\tau=0) \\
n_d=1: & L=2\;\; (\tau=1) & \\
n_d=0: & L=0\;\; (\tau=0) &
\end{array}
\label{u5mult}
\ea

The situation changes drastically
when $\rho > 0$, for which the Hamiltonian becomes
\ba
\hat{H}_1(\rho)/\bar{h}_2 &=&
2\hat{n}_d(\hat{n}_d\!-\!1)
+2\beta_{0}^2(\hat{N}-\hat{n}_d)\hat{n}_d
+ \rho^2\beta_{0}^2(2\hat{C}_{O(5)} - \hat{C}_{O(3)} -2\hat{n}_d)
\nonumber\\
&&
+\rho\beta_{0}^2\sqrt{14}
[(d^{\dag}d^{\dag})^{(2)}\cdot \tilde{d}s + H.c.] ~,
\label{Hrho}
\ea
where $H.c.$ means Hermitian conjugate.
In this case, the last term in Eq.~(\ref{Hrho}) breaks the U(5) DS,
and induces U(5) and O(5) mixing subject to
$\Delta n_d = \pm 1$ and
$\Delta\tau=\pm 1, \pm 3$. The explicit breaking of O(5) symmetry
leads to non-integrability and, as will be shown in subsequent discussions,
is the main cause for the onset of chaos in the spherical region.
Although $\hat{H}_1(\rho)$, Eq.~(\ref{Hrho}), is not diagonal in the
U(5) chain, it retains the following selected solvable U(5) basis
states~\cite{pds}
\bsub
\ba
\label{ePDSu5L0}
\vert [N], n_d=\tau=L=0\rangle \;\;\;\;
&&E_{\rm PDS} = 0 ~,\\
\vert [N], n_d=\tau=L=3\rangle \;\;\;\;
&&E_{\rm PDS}
= 6\left [ 2 +\beta_{0}^2(N-3) + 3\rho^2\beta_{0}^2\right ]\bar{h}_2 ~,
\label{ePDSu5L3}
\ea
\label{ePDSu5}
\esub
while other eigenstates are mixed. As such, it exhibits U(5) partial dynamical
symmetry [U(5)-PDS].

In the deformed phase, the appropriate intrinsic Hamiltonian
is $\hat{H}_{2}(\xi)$, Eq.~(\ref{eq:H2}), with $\xi\geq 0$.
The latter has a zero-energy ground band composed of states with
$L=0,2,4,\ldots,2N$, projected from the intrinsic state,
Eq.~(\ref{condgen}), with
$(\beta_{\rm eq} = \sqrt{2}\bz(1+\bz^2)^{-1/2},\gamma_{\rm eq}=0)$.
For $\beta_0=\sqrt{2}$ and $\xi=1$, the Hamiltonian reduces to
\ba
\hat{H}_2(\xi=1;\beta_0=\sqrt{2}) /\bar{h}_2 &=&
-\hat{C}_{SU(3)} + 2\hat{N}(2\hat{N}+3) ~,
\ea
and hence has SU(3) DS. The spectrum is completely solvable
\ba
\vert N,(\lambda,\mu),K,L\rangle \;\;\;\;
&&E_{\rm DS} /\bar{h}_2 = \left [\,-c(\lambda,\mu)
+ 2N(2N+3)\right ]\bar{h}_2 ~,
\label{SU3-DSspec}
\ea
where $c(\lambda,\mu)$ are the eigenvalues of the SU(3) Casimir operator
listed in Table~1. The eigenstates are those of the SU(3) chain,
Eq.~(\ref{su3ds}), with $(\lambda,\mu)=(2N-4k-6m,2k)$,
with $k,m$ non-negative integers, such that, $\lambda,\mu\geq 0$.
The values of $L$ contained in these SU(3) irreps
are $L=K,K+1,K+2,\ldots,K+{\rm max}\{\lambda,\mu\}$, where
$K=0,\, 2,\ldots, {\rm min}\{\lambda,\mu\}$;
with the exception of $K=0$ for which
$L=0,\, 2,\ldots, {\rm max}\{\lambda,\mu\}$.
The spectrum resembles that of an axially-deformed rotor with
degenerate K-bands arranged in SU(3) multiplets,
$K$ being the angular momentum projection on the symmetry axis.
The lowest SU(3) irreps are $(2N,0)$
which describes the ground band $g(K=0)$, $(2N-4,2)$ which contains
the $\beta(K=0)$ and $\gamma(K=2)$
bands, and $(2N-8,4)$, $(2N-6,0)$, which contain the
$\beta^2(K=0)$, $\beta\gamma(K=2)$, $\gamma^2(K=0,4)$ bands.
The corresponding band-members are~\cite{ibmsu3}
\ba
\begin{array}{lll}
(\lambda,\mu)=(2N-6,0): & L=0,2,4,\ldots  & (K=0)
\qquad\gamma^2\, ,\, \beta^2\\
(\lambda,\mu)=(2N-8,4): & L=4,5,6,\ldots  & (K=4)
\qquad \gamma^2\\
                        & L=2,3,4,\ldots  & (K=2)\qquad \beta\gamma\\
                        & L=0,2,4,\ldots  & (K=0)\qquad \beta^2\,,\,\gamma^2\\
(\lambda,\mu)=(2N-4,2): & L=2,3,4,\ldots  & (K=2)\qquad \gamma\\
                        & L=0,2,4,\ldots  & (K=0)\qquad \beta \\
(\lambda,\mu)=(2N,0): & L=0,2,4,\ldots    &  (K=0)\qquad g
\end{array}
\ea

For $\bz=\sqrt{2}$ and $\xi\neq 1$,
the Hamiltonian becomes
\ba
\hat{H}_2(\xi;\beta_0=\sqrt{2}) /\bar{h}_2 &=&
-\hat{C}_{SU(3)} + 2\hat{N}(2\hat{N}+3) +
(\xi-1) P^{\dag}_0P_0 ~,
\label{H2sqrt2}
\ea
where $P^{\dag}_0\!\equiv\! P^{\dag}_0(\bz\!=\!\sqrt{2})$ of Eq.~(\ref{P0}).
The added term breaks the SU(3) DS of the Hamiltonian
and most eigenstates are mixed with respect to SU(3). However,
the following states~\cite{pds}
\bsub
\ba
&&\vert N,(2N,0)K=0,L\rangle \;\;\;\; L=0,2,4,\ldots, 2N
\nonumber\\
&&E_{\rm PDS} = 0
\label{solsu3g}
\\
&&\vert N,(2N-4k,2k)K=2k,L\rangle
\;\;\;\;
L=K,K+1,\ldots, (2N-2k) \qquad\; k>0 ~,
\qquad\qquad
\nonumber\\
&&
E_{\rm PDS} =  6k \left (2N - 2k+1 \right )\bar{h}_2
\label{solsu3gam}
\ea
\label{solsu3}
\esub
remain solvable with good SU(3) symmetry. As such,
$\hat{H}_2(\xi;\beta_0=\sqrt{2})$ exhibits SU(3) partial dynamical
symmetry [SU(3)-PDS]. The selected states of Eq.~(\ref{solsu3})
span the ground band $g(K=0)$ and $\gamma^{k}(K=2k)$ bands.
Such Hamiltonians with SU(3)-PDS have been used for the spectroscopy
of deformed rotational nuclei~\cite{lev96,levsin99}.
In general, the analytic properties of the solvable states in a PDS,
provide unique signatures for their identification in the quantum
spectrum.

The collective Hamiltonian of Eq.~(\ref{eq:Hcol}) preserves the O(5)
symmetry for any choice of couplings $\bar{c}_i$, hence its
dynamics is integrable with $\tau$ and $L$ as good quantum numbers.
The $\hat{C}_{\rm O(3)}$ and $\hat{C}_{\rm O(5)}$ terms
lead to an $L(L+1)$ and $\tau(\tau+3)$ type of splitting.
In general, integrability is lost when the collective Hamiltonian is
added to the intrinsic Hamiltonian~({\ref{eq:Hint}),
since the latter breaks the O(5) symmetry, and only $L$ remains a good
quantum number for the full Hamiltonian.
A~notable exception is when $\rho=0$, since now all terms
in $\hat{H}_1(\rho=0) + \hat{H}_{\rm col}$, respect the O(5) symmetry.

\section{Classical limit}
\label{sec:Class}

The classical limit of the IBM is obtained through the use of Glauber
coherent states~\cite{Hatch82}.
This amounts to
replacing $(s^{\dagger},\,d^{\dagger}_{\mu})$ by six
c-numbers $(\alpha_{s}^{*},\,\alpha_{\mu}^{*})$ rescaled
by $\sqrt{N}$ and taking $N\rightarrow\infty$, with $1/N$ playing the
role of $\hbar$. Number conservation ensures that
phase space is 10-dimensional and can be phrased in terms of
two shape (deformation) variables, three orientation (Euler) angles
and their conjugate momenta. The shape variables can be identified with the
$\beta, \gamma$ variables introduced through Eq.~(\ref{condgen}).
Setting all momenta to zero, yields the classical potential which
is identical to $V(\beta,\gamma)$ of Eq.~(\ref{enesurf}).
In the classical analysis presented below we consider, for simplicity, the
dynamics of $L=0$ vibrations, for which only two
degrees of freedom are active.
The rotational dynamics with $L>0$ is examined in
the subsequent quantum analysis.

\subsection{Classical limit of the QPT Hamiltonian}
\label{sec:ClassQPT}

For the intrinsic Hamiltonian of Eq.~(\ref{eq:Hint}), constrained to $L=0$,
the above procedure yields the following classical Hamiltonian
\bsub
\label{eq:Hintcl}
\ba
\label{eq:H1cl}
\mathcal{H}_1 (\rho)/h_2 &=&
{\cal H}_{d,0}^2 + \bz^2(1 - {\cal H}_{d,0}){\cal H}_{d,0}
+ \rho^2 \bz^2 p_\gamma^2
\nonumber\\
&& + \rho\bz^2\sqrt{\tfrac{1 - {\cal H}_{d,0}}{2}}
\left [\,(p_\gamma^2/\beta - \beta p^2_{\beta} - \beta^3) \cos{3\gamma}
+ 2p_{\beta} p_\gamma \sin{3\gamma}\,\right] ~,\\
\label{eq:H2cl}
\mathcal{H}_2(\xi)/h_2 &=& {\cal H}_{d,0}^2
+ \bz^2(1\!\!-\!\!{\cal H}_{d,0}){\cal H}_{d,0} + p_\gamma^2
\nonumber\\
&&+ \bz\sqrt{\tfrac{1 - {\cal H}_{d,0}}{2}}
\left [\,(p_\gamma^2/\beta
- \beta p^2_{\beta} - \beta^3) \cos{3\gamma}
+ 2p_{\beta} p_\gamma \sin{3\gamma}\,\right ]
\nonumber \\
&& + \xi\left [\,\beta^2 p_\beta^2 + \tfrac{1}{4}(\beta^2 - T)^2
- \bz^2(1 - {\cal H}_{d,0})(\beta^2 - T)
+ \bz^4(1\!\!-\!\!{\cal H}_{d,0})^2\,\right] ~.
\qquad
\ea
\esub
Here the coordinates $\beta\in[0,\sqrt{2}]$, $\gamma\in[0,2\pi)$
and their canonically
conjugate momenta $p_\beta\in[0,\sqrt{2}]$ and $p_\gamma\in[0,1]$
span a compact classical phase space.
The term
\ba
{\cal H}_{d,0}\equiv (T+\beta^2)/2 \;\; ,\;\;
T = p_{\beta}^2+p_\gamma^2/\beta^2 ~,
\ea
denotes the classical limit of $\hat{n}_d$
(restricted to $L=0$) and forms an isotropic harmonic oscillator
Hamiltonian in the $\beta$ and $\gamma$ variables.
Notice that the classical Hamiltonian of Eq.~(\ref{eq:Hintcl}) contains
complicated momentum-dependent terms originating from the two-body
interactions in the Hamiltonian~(\ref{eq:Hint}), not just the usual
quadratic kinetic energy $T$.
Setting $p_{\beta} = p_\gamma=0$ in Eq.~(\ref{eq:Hintcl}) leads to the
following classical potential
\bsub
\label{eq:Vcl}
\ba
\label{eq:V1}
V_1(\rho)/h_2 &=&
\bz^2 \beta^2 - \rho\bz^2 \sqrt{2\!-\!\beta^2} \beta^3\cos3\gamma
+ \frac{1}{2}(1\!-\!\bz^2)\beta^4 ~,\\
\label{eq:V2}
V_2(\xi)/h_2 &=&
\bz^2[1 - \xi(1\!+\!\bz^2)] \beta^2
-\bz \sqrt{2\!-\!\beta^2} \beta^3\cos3\gamma
\nonumber\\
&&
+ \frac{1}{4}\left [2(1\!-\!\bz^2)
+ \xi(1\!+\!\bz^2)^2\right]\beta^4 + \xi\bz^4 ~.
\ea
\esub
The same expressions can be obtained from Eq.~(\ref{enesurf}) using the
static intrinsic coherent state~(\ref{condgen}).
Notice that the potential of Eq.~(\ref{eq:Vcl})
is independent of $N$ due to the
per-boson scaling used in Eq.~(\ref{eq:Hint}).

The variables $\beta$ and $\gamma$
can be interpreted as polar coordinates
in an abstract plane parameterized by Cartesian coordinates ($x,y)$.
The transformation between these two sets of coordinates and
conjugate momenta is
\bsub
\ba
&& x = \beta\cos\gamma \;\;\; , \;\;\;  y = \beta\sin\gamma ~,\\
&& p_x = p_{\beta}\cos\gamma  - (p_{\gamma}/\beta)\sin\gamma
\;\;\; , \;\;\;
p_y = (p_{\gamma}/\beta)\cos\gamma + p_{\beta}\sin\gamma ~.
\ea
\esub
Using the relations
\bsub
\ba
&& p_{\gamma} = xp_y - yp_x \;\; , \;\;
\beta p_{\beta} = xp_x + yp_y \;\; , \;\;
\beta^2 = x^2 + y^2 \;\; , \;\;
\beta^3\cos 3\gamma = x^3 - 3xy^2 ~,
\qquad\quad\\
&& (p_\gamma^2/\beta - \beta p^2_{\beta} ) \cos{3\gamma}
+ 2p_{\beta} p_\gamma \sin{3\gamma} =
x(p_{y}^2-p_{x}^2) + 2yp_xp_y ~,\\
&&{\cal H}_{d,0} \equiv (T + x^2 + y^2 )/2 \;\; ,\;\;
T = p_{x}^2 + p_{y}^2 ~,\qquad\qquad
\ea
\esub

\begin{figure*}[!t]
\begin{center}
\epsfig{file=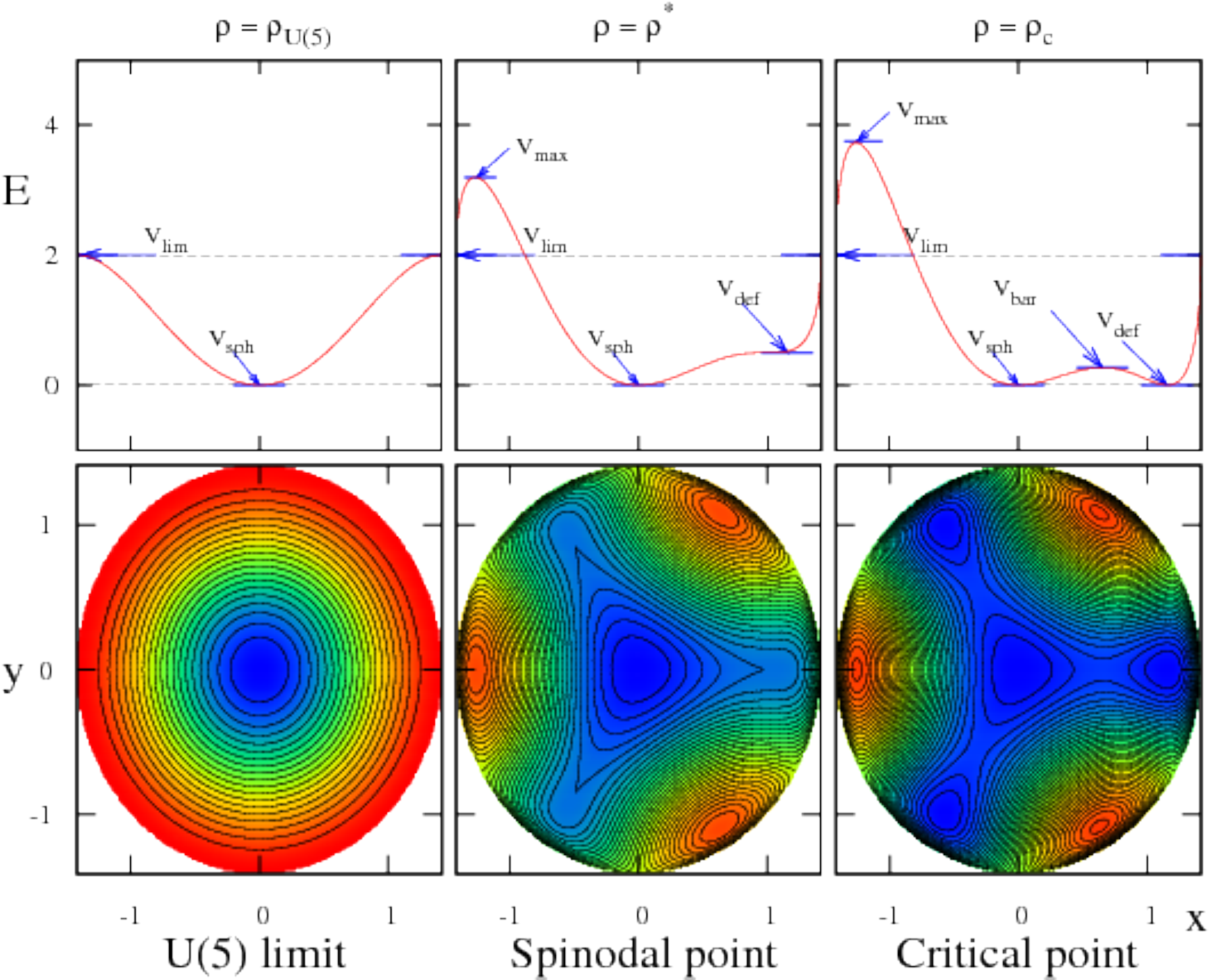,width=0.99\linewidth}
\end{center}
\caption{
Contour plots of the potential $V_{1}(\rho; x,y)$,
Eq.~(\ref{Vcl1xy}), [lower portion], and the section $V_{1}(\rho;x,y=0)$
[upper portion], for $\bz=\sqrt{2}$, $h_2=1$,
relevant to the spherical side of the QPT ($0\leq\rho\leq\rho_c$).
Here $\rho_{\rm U(5)}=0$ corresponds to the U(5) limit,
$\rho^{*}$ the spinodal point, Eq.~(\ref{spinodal}),
and $\rho_c$ the critical point, Eq.~(\ref{critical}).
The stationary values
($V_\mathrm{sph}$,$V_\mathrm{def}$,$V_\mathrm{bar}$,
$V_\mathrm{sad}$, $V_\mathrm{max}$) and boundary values ($V_\mathrm{lim}$)
of $V_{1}(\rho; x,y)$ are marked by arrows.
The energy scale of the color coding in the contour plots
can be inferred from the respective panels above them.}
\label{fig1}
\end{figure*}

one can express the classical Hamiltonians of Eq.~(\ref{eq:Hintcl})
in terms of ($x,\,y,\,p_x,\,p_y$).
Setting $p_x=p_y=0$ in the resulting expressions, we obtain the
classical potential of Eq.~(\ref{eq:Vcl}) in Cartesian form
\bsub
\label{Vclxy}
\ba
\label{Vcl1xy}
V_1(\rho)/h_2 &=& \bz^2 (x^2+y^2)
- \rho\bz^2 \sqrt{2-(x^2+y^2)}
\left [x^3 -3xy^2\right ]
+ \frac{1}{2}(1\!-\!\bz^2)(x^2+y^2)^2 ~,
\qquad\quad\\
V_2(\xi)/h_2 &=&  \bz^2\left [1 - \xi(1\!+\!\bz^2)\right ] (x^2+y^2)
-\bz \sqrt{2\!-(x^2+y^2)} \left[x^3 -3xy^2\right ]
\nonumber\\
&&+ \frac{1}{4}\left[ 2(1\!-\!\bz^2) + \xi(1\!+\!\bz^2)^2 \right ]
(x^2+y^2)^2 + \xi\bz^4 ~.
\label{Vcl2xy}
\ea
\esub

\begin{figure*}[!t]
\begin{center}
\epsfig{file=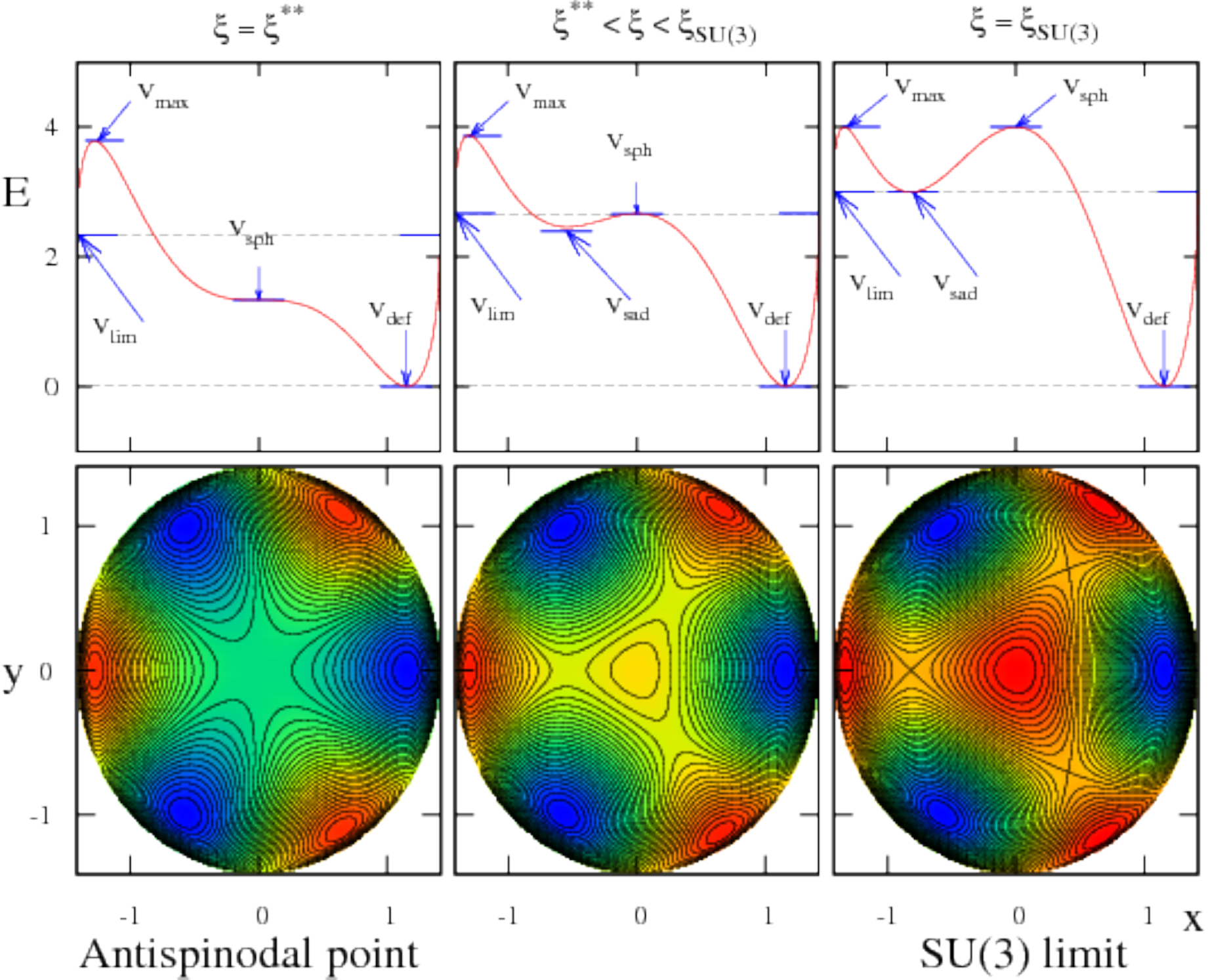,width=0.99\linewidth}
\end{center}
\caption{
Same as in Fig.~1, but for the potential $V_{2}(\xi; x,y)$,
Eq.~(\ref{Vcl2xy}), [lower portion], and the section $V_{2}(\xi;x,y=0)$
[upper portion], relevant to the deformed side of the QPT
($\xi^{**}\leq\xi\leq 1$).
Here $\xi^{**}$ denotes the anti-spinodal point, Eq.~(\ref{antispinodal}),
and $\xi_{\rm SU(3)}=1$ corresponds to the SU(3) limit.}
\label{fig2}
\end{figure*}

Note that the potentials
$V(\beta,\gamma)=V(x,y)$ depend on the combinations
$ \beta^2 \!=\! x^2 + y^2$, $\beta^4= (x^2+y^2)^2$ and
$\beta^3\cos 3\gamma \!=\! x^3 - 3xy^2$.

The classical limit of the collective Hamiltonian, Eq.~(\ref{eq:Hcol}),
constrained to $L=0$, is obtained in a similar manner and is given by
\ba
\label{eq:HColCl}
\mathcal{H}_\mathrm{col} &=&
c_5\, p_\gamma^2 + c_6 \left( -T^2 + 2T - \beta^2 p_\beta^2\right)
\nonumber\\
&=&
c_5\, (xp_y - yp_x)^2
+ c_6 \left [\, -T^2 + 2T - (xp_x + yp_y)^2 \,  \right] ~,
\ea
where $T = p_{\beta}^2+p_\gamma^2/\beta^2 = p_{x}^2 + p_{y}^2$.
The O(3)-rotational $c_3$-term is absent from Eq.~(\ref{eq:HColCl}),
since the classical Hamiltonian  is constrained to angular momentum zero.
The purely kinetic character of the collective terms is evident
from the fact that $\mathcal{H}_\mathrm{col}$
vanishes for $p_\beta=p_\gamma=0$, thus not contributing to
the potential $V(\beta,\gamma)$.

\subsection{Topology of the classical potentials}
\label{sec:Regions123}

The values of the control parameters $\rho$ and $\xi$ determine the
landscape and extremal points of the potentials
$V_{1}(\rho;\beta,\gamma)$ and $V_{2}(\xi;\beta,\gamma)$,
Eq.~(\ref{eq:Vcl}). Important values of these parameters at which
a pronounced change in structure is observed,
are the spinodal point where a second (deformed)
minimum occurs, an anti-spinodal point where the first (spherical) minimum
disappears and a critical point in-between, where the two minima are
degenerate. For the potentials under discussion, the
critical point $(\rho_c,\xi_c)$ given by
\ba
\rho_c =\bz^{-1} \;\;\; {\rm and}\;\;\; \xi_c=0 ~,
\label{critical}
\ea
separates the spherical and deformed phases.
The spinodal point ($\rho^{*}$)
\ba
\rho^* &=& \frac{1}{\sqrt{6}}
\left [\, -(r^2-4r+1)
+(r+1)\sqrt{(r+1)(r-1/3)}\, \right]^{1/2}
\;\; , \;\;
r\equiv\bz^{-2} ~,
\label{spinodal}
\ea
and the anti-spinodal point ($\xi^{**}$)
\ba
\xi^{**} &=& (1+\bz^2)^{-1} ~,
\label{antispinodal}
\ea
embrace the critical point and mark the boundary of the phase coexistence
region. The derivation of these expressions is explained in Appendix~A.

In general, the only $\gamma$-dependence in the potentials~(\ref{eq:Vcl})
is due to the $\sqrt{2-\beta^2}\beta^3\cos 3\gamma$ term.
This induces a three-fold symmetry about the origin $\beta=0$,
as is evident in the contour plots of the potentials shown in Figs~1-2.
As a consequence, the deformed extremal points are obtained for
$\gamma=0,\,\tfrac{2\pi}{3},\,\tfrac{4\pi}{3}$ (prolate shapes),
or $\gamma = \tfrac{\pi}{3},\, \textstyle{\pi},\, \tfrac{5\pi}{3}$
(oblate shapes).
It is therefore possible to restrict the
analysis to $\gamma=0$ and allow for
both positive and negative values of $\beta$, corresponding to prolate
and oblate deformations, respectively.
Henceforth, we occasionally use the shorthand
notation, $V_{1}(\rho;\beta)\equiv V_{1}(\rho;\beta,\gamma=0)$ and
$V_{2}(\xi;\beta)\equiv V_{2}(\xi;\beta,\gamma=0)$.
These $\gamma=0$ sections are shown in the upper portion of Figs.~1-2.

\vspace{8pt}
\noindent
{\it The spherical phase} $(0\leq\rho\leq\rho_c=\bz^{-1})$.
\vspace{6pt}

The relevant potential in the spherical phase is
$V_{1}(\rho;\beta,\gamma)$, Eq.~(\ref{eq:V1}), with
$0\leq\rho\leq\rho_c$.
In this case, $\beta=0$ is a global minimum of the potential
at an energy $V_{\rm sph}$,
representing the equilibrium spherical shape,
\ba
\beta_{\rm eq} = 0:\qquad V_{\rm sph}
= V_{1}(\rho;\beta_{\rm eq}=0,\gamma)=0 ~.
\label{beqsph}
\ea
The limiting value at the domain boundary is
\ba
V_{\rm lim} &=& V_{1}(\rho;\beta=\sqrt{2},\gamma) = 2h_{2} ~.
\label{Vlim1}
\ea
For $\rho=0$, the potential is independent of $\gamma$,
\ba
V_1(\rho=0)/h_2 &=&
\bz^2 \beta^2 + \tfrac{1}{2}(1\!-\!\bz^2)\beta^4 ~,
\label{V1rho0}
\ea
and has $\beta_{\rm eq}=0$ as a single minimum.

For $\rho>0$, the deformed extremal points
($\beta\neq 0$) are given by
\ba
\beta = \sqrt{2}\tlb(1+\tlb^2)^{-1/2} ~,
\label{btlb}
\ea
where $\tlb$ are real solutions of the cubic equation
\ba
\rho\tlb^3+3\eta\tlb^2 -3\rho\tlb + 1 = 0 ~,\qquad\quad
\eta = (2-\bz^2)/3\bz^2 ~.
\label{exteqV1}
\ea
For $\rho<\rho^{*}$, Eq.~(\ref{exteqV1}) has one real
root, $\tlb_1<0$. The corresponding deformation, $\beta_1<0$, obtained
from Eq.~(\ref{btlb}), produces a maximum in $V_1(\rho;\beta)$ at an energy
$V_{\rm max}= F_{1}(\rho,\tlb_1)$, where
\ba
F_{1}(\rho;\tlb)/h_2 =
\bz^{2}(1+\tlb^2)^{-1}\tlb^{2}(1-\rho\tlb) ~.
\label{F1}
\ea

At the spinodal point, $\rho=\rho^{*}$, Eq.~(\ref{exteqV1}) has one
negative root ($\tlb_1<0$) and a doubly-degenerate positive root
($\tlb_2>0$), given by
\bsub
\ba
\tlb_1 &=& -2\sqrt{1+(\eta/\rho^{*})^2} - \eta/\rho^{*} ~,\\
\tlb_2 &=&  \sqrt{1+(\eta/\rho^{*})^2} - \eta/\rho^{*} ~.
\ea
\esub
The corresponding deformations $\beta_1<0$ and $\beta_2>0$,
obtained from Eq.~(\ref{btlb}), correspond to a maximum
of the potential at an energy $V_{\rm max} = F_{1}(\rho^{*};\tlb_1)$,
and to an inflection point at an energy
$V_{\rm def} = F_{1}(\rho^{*};\tlb_2)$, respectively.

For $\rho> \rho^{*}$,  Eq.~(\ref{exteqV1}) has three distinct real roots,
$\tlb_1 < \tlb_2 < \tlb_3$, satisfying $\tlb_1+\tlb_2+\tlb_3 = -3\eta/\rho$,
$\tlb_1\tlb_2 + \tlb_2\tlb_3 + \tlb_3\tlb_1 = -3$ and
$\tlb_1\tlb_2\tlb_3 = \rho^{-1}$.
The extremal points ($\beta_1<0$, $\beta_2>0$ and $\beta_3>0$)
obtained from Eq.~(\ref{btlb}),
correspond to a maximum, a saddle and a minimum point of $V_{1}(\rho,\beta)$,
at energies $V_{\rm max} = F_{1}(\rho;\tlb_1),\,
V_{\rm bar} = F_{1}(\rho;\tlb_2)$, and $V_{\rm def} = F_{1}(\rho;\tlb_3)$,
respectively. The saddle point ($\beta_2$) forms a barrier
between the newly-developed local deformed minimum ($\beta_3$)
and the global minimum at $\beta_{\rm eq}=0$.
As seen in the contour plot of Fig.~1, the potential
near the saddle point decreases towards
the spherical and prolate-deformed minima,
and increases towards the two-out-of-three equivalent
oblate-deformed maxima. Thus a barrier in the $\beta$-direction
at the saddle point, separates the spherical and deformed phases.

\vspace{8pt}
\noindent
{\it The deformed phase} $(\xi\geq\xi_c=0)$.
\vspace{6pt}

The relevant potential in the deformed phase is
$V_{2}(\xi;\beta,\gamma)$, Eq.~(\ref{eq:V2}), with  $\xi\geq\xi_c$.
In this case, $[\beta_{\rm eq}>0,\gamma_{\rm eq}=0]$ is a global
minimum of the potential
at an energy $V_{\rm def}$, representing the equilibrium deformed shape
\ba
[ \beta_{\rm eq} = \sqrt{2}\bz(1+\bz^2)^{-1/2},\gamma_{\rm eq}=0 ]:
\qquad
V_{\rm def} = V_{2}(\xi;\beta_{\rm eq}>0,\gamma_{\rm eq}=0) = 0 ~.
\label{beqdef}
\ea
The limiting value of the domain boundary is
\ba
V_{\rm lim} &=& V_{2}(\xi;\beta=\sqrt{2},\gamma) = (2+\xi)h_2 ~.
\label{Vlim2}
\ea
$\beta=0$ is an extremal point and occurs at an energy $V_{\rm sph}$
\ba
\beta=0:
\qquad
V_{\rm sph} &=& V_{2}(\xi;\beta=0,\gamma) = h_{2}\xi\bz^4~.
\label{bzero}
\ea
It is a local minimum for $\xi<\xi^{**}$
and a maximum for $\xi>\xi^{**}$.
The deformed extremal points ($\beta\neq 0$) are given by
$\beta=\sqrt{2}\tlb(1+\tlb^2)^{-1/2}$, where
$\tlb$ satisfies the following cubic equation,
\ba
\bz\tlb^3 + [\,2-\bz^2 + \xi(1+\bz^2)\,]\tlb^2 - 3\bz\tlb
+ \bz^2 [\,1 - \xi(1+\bz^2)\,] = 0 ~.
\label{exteqV2}
\ea
One solution of Eq.~(\ref{exteqV2}) is $\tlb=\bz$, which yields the
global deformed minimum, $\beta_{\rm eq} = \sqrt{2}\bz(1+\bz^2)^{-1/2}$,
Eq.~(\ref{beqdef}).
The remaining solutions ($\tlb=\tlb_{\pm}$)
satisfy the quadratic equation,
\ba
\bz\tlb^2 + [\,2+\xi(1+\bz^2)\,]\tlb + \bz [\,\xi(1+\bz^2)-1\,] = 0 ~,
\label{quadEq}
\ea
and determine two additional deformed extremal points,
$\beta_{\pm} = \sqrt{2}\tlb_{\pm}(1+\tlb_{\pm}^2)^{-1/2}$.
\begin{figure}[!t]
\begin{center}
\epsfig{file=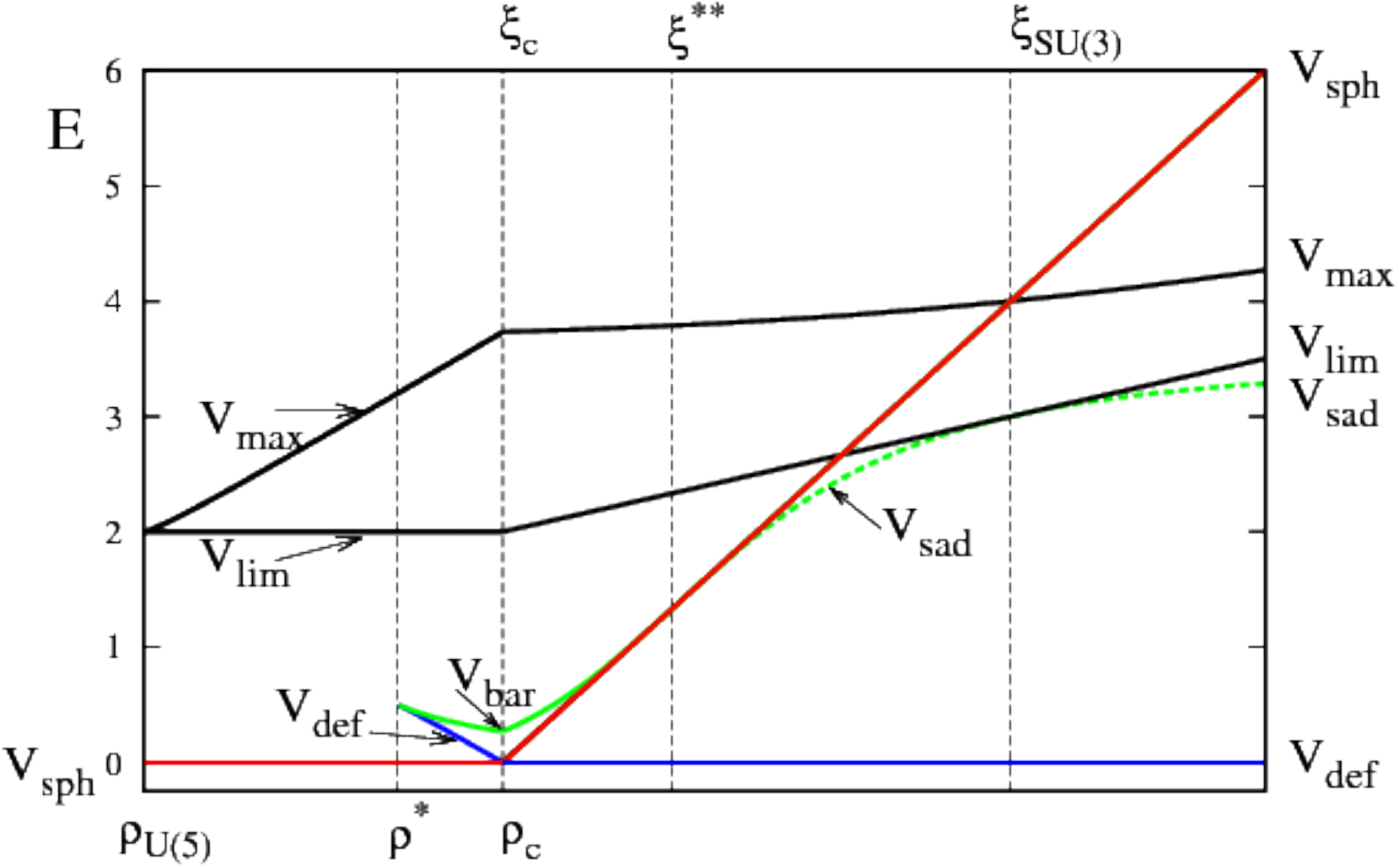,width=0.99\linewidth}
\end{center}
\caption{
Evolution of the stationary and boundary values of the potential
surfaces $V_1(\rho)$ and $V_2(\xi)$, Eq.~(\ref{eq:Vcl}),
with $\bz\!=\!\sqrt{2}$, $h_2\!=\!1$.
$\rho_{\rm U(5)}\!=\!0$ and $\xi_{\rm SU(3)}\!=\!1$ correspond to the U(5)
and SU(3) limits, respectively.
$V_{\rm lim}= V(\beta=\sqrt{2},\gamma)$ are the domain boundaries.
$V_{\rm sph}$ is the energy of the
spherical configuration which is a global minimum for $0\leq\rho<\rho_c$
and a local minimum for $\xi_c\leq \xi <\xi^{**}$.
For $\rho>0$, a deformed maximum occurs at an energy $V_\mathrm{max}$.
Beyond the spinodal point, $\rho>\rho^*$, a local
deformed minimum at an energy $V_\mathrm{def}$ develops,
along with a saddle point which creates a barrier of height $V_\mathrm{bar}$
separating the two minima. The latter cross and become degenerate
($V_\mathrm{sph}=V_\mathrm{def}$) at the critical point $(\rho_c,\xi_c$).
At the anti-spinodal point, $\xi^{**}$,
the spherical configuration changes from a minimum to a maximum,
the deformed configuration remains a single minimum (energy $V_{\rm def}$)
and the saddle point (energy $V_{\rm sad}$),
now separates pairs of equivalent deformed minima (see Fig.~2).
Note that $V_{\rm bar}=V_{\rm def}$ at $\rho^{*}$ and
$V_{\rm bar}=V_{\rm sph}$ at $\xi^{**}$.
The relations $V_\mathrm{sph}=V_\mathrm{max}$ and
$V_\mathrm{sad}=V_\mathrm{lim}$ at $\xi_{\rm SU(3)}=1$ are a specific
property of the SU(3) surface.}
\label{fig3}
\end{figure}

At the critical point ($\xi_c=0$, $\rho_c=\bz^{-1}$),
the potentials in the spherical and deformed phases coincide
$V_2(\xi_c;\beta,\gamma)=V_1(\rho_c;\beta,\gamma)
\equiv V_\mathrm{cri}(\beta,\gamma)$, and read
\ba
V_\mathrm{cri}(\beta,\gamma)/h_2 &=&
\bz^2 \beta^2 - \bz \sqrt{2\!-\!\beta^2} \beta^3\cos3\gamma
+ \frac{1}{2}(1\!-\!\bz^2)\beta^4 ~.
\label{eq:Vcri}
\ea
The topology of $V_\mathrm{cri}(\beta,\gamma)$ is shown on the right
panel in Fig.~1.
In this case, the global deformed
minimum $(\beta_{\rm eq}>0)$ of Eq.~(\ref{beqdef}), is degenerate with
the spherical minimum ($\beta=0$), and both occur at zero energy,
\ba
V_{\rm cri}(\beta=0,\gamma) =
V_{\rm cri}(\beta_{\rm eq}>0,\gamma_{\rm eq}=0)=0~.
\label{Vcrimin}
\ea
The two solutions of Eq.~(\ref{quadEq}), for $\xi_c=0$, are
$\tlb_{\pm} = [\,-1\pm\sqrt{1+\bz^2}\,]/\bz$ and the resulting
additional deformed extremal points,
$\beta_{\pm} = \sqrt{2}\tlb_{\pm}(1+\tlb_{\pm}^2)^{-1/2}$,
are found to be
\ba
\beta_{\pm} &=& \pm \left [\,1 \mp (1+\bz^2)^{-1/2}\,\right ]^{1/2} ~.
\ea
Here $\beta_{-}<0$ corresponds to a maximum of $V_{\rm cri}(\beta,\gamma=0)$
and occurs at an energy $V_{\rm max}$
\ba
V_{\rm max} &=& V_{\rm cri}(\beta_{-},\gamma=0)
= \frac{1}{2}h_2\left (\, 1+\sqrt{1+\bz^2}\, \right)^2 ~.
\ea
$\beta_{+}>0$ corresponds to a saddle point, which creates a barrier
of height $V_{\rm bar}$,
\ba
V_{\rm bar} &=&  V_{\rm cri}(\beta_{+},\gamma=0)
= \frac{1}{2}h_2\left (\,1-\sqrt{1+\bz^2}\,\right )^2 ~,
\label{Vbarcri}
\ea
separating the spherical and deformed minima. The height and width
of the barrier are governed by $\bz$.

For $\xi>\xi_c$, the spherical $\beta=0$ minimum turns local
with an energy $V_{\rm sph}>0$, Eq.~(\ref{bzero}), above that of the
deformed minimum $[\beta_{\rm eq}>0,\gamma_{\rm eq}=0]$, Eq.~(\ref{beqdef}).
The additional deformed extremal points,
$\beta_{\pm} = \sqrt{2}\tlb_{\pm}(1+\tlb_{\pm}^2)^{-1/2}$,
are determined by the solutions of Eq.~(\ref{quadEq})
\bsub
\ba
\tlb_{\pm} &=& \frac{-[2+\xi(1+\bz^2)]\pm \sqrt{\Delta}}{2\bz} ~,\\
\Delta &=& (1+\bz^2)\left [ \,4 + 4\xi(1-\bz^2) +\xi^2(1+\bz^2)
\,\right ] ~.
\ea
\esub
The two solutions $\tlb_{\pm}$ satisfy,
$\tlb_{-}+\tlb_{+} =  -[2+\xi(1-\bz^2)]/\bz$ and $\tlb_{-}\tlb_{+} =
\xi(1+\bz^2)-1$.
The extremal point $\beta_{-}$ corresponds to a maximum of the potential
at an energy $V_{\rm max}$, and $\beta_{+}$ is a saddle point at an
energy $V_{\rm sad}$, given by
\bsub
\ba
&&\beta_{-} = \sqrt{2}\tlb_{-}(1+\tlb_{-}^2)^{-1/2}:\qquad
V_{\rm max} = V_{2}(\beta_{-},\gamma=0) = F_{2}(\tlb_{-}) ~,
\label{Vmax}\\
&&\beta_{+} = \sqrt{2}\tlb_{+}(1+\tlb_{+}^2)^{-1/2}:\qquad
V_{\rm sad}\; = V_{2}(\beta_{+},\gamma=0) = F_{2}(\tlb_{+}) ~,
\label{Vsad}
\ea
\esub
where
\ba
F_{2}(\tlb)/h_2 =  \frac{1}{2}\bz^2\tlb^2
+\frac{1}{2}\xi\bz(1+\bz^2)
(1+\tlb^2)^{-1}\tlb^2(\tlb_{\pm}-\bz) +\xi\bz^4 ~.
\ea

For $0\leq\xi<\xi^{**}$, the local spherical minimum, Eq.~(\ref{bzero}),
coexists with the deformed global minimum, Eq.~(\ref{beqdef}),
and $\beta_{-}\beta_{+}<0$.
At the anti-spinodal point, $\xi=\xi^{**}$, the spherical
minimum disappears and $\beta=0$ becomes an inflection point.
For $\xi >\xi^{**}$, $\beta=0$ becomes a maximum,
$[\beta_{\rm eq}>0,\gamma_{\rm eq}=0]$ remains a single minimum of
the potential, and $\beta_{-}\beta_{+} >0$.
In this case, as seen in the contour plot in Fig.~2, the potential
near the saddle point ($\beta_{+}$) increases both towards the spherical
maximum ($\beta=0$) and the oblate-deformed maximum ($\beta_{-}$),
and decreases towards the two-out-of-three equivalent
prolate-deformed global minima ($\beta_{\rm eq}>0$).
The saddle point has now a different character from that encountered in
the coexistence region, accommodating a barrier in the $\gamma$-direction
between pairs of equivalent prolate-deformed minima.
\begin{table}
\begin{center}
\caption{
\label{Tab2}
\protect\small
Significant values of the control parameters ($\rho,\xi)$
and order parameters, ($\beta_{\rm eq},\gamma_{\rm eq}$)
for a first order QPT between a spherical [U(5)]
phase and an axially-deformed [SU(3)] phase. The relevant
intrinsic Hamiltonians are $\hat{H}_1(\rho)$, Eq.~(\ref{eq:H1}) and
$\hat{H}_2(\xi)$, Eq.~(\ref{eq:H2}) with $\bz=\sqrt{2}$.
At the critical point, the barrier height is $V_{\rm bar} = 0.268h_2$,
and the domain boundary is $V_{\rm lim}=2h_2$.}
\vspace{1mm}
\centering
\begin{tabular}{l|ll}
\hline
& & \\[-3mm]
Special points & Control parameters & Order parameters\\[4pt]
& & \\[-3mm]
\hline
& & \\[-2mm]
Spherical phase & $0\leq\rho\leq\frac{1}{\sqrt{2}}$
& $\beta_{\rm eq} =0$  \\[3pt]
Deformed phase & $0\leq\xi\leq 1$ &
$[\beta_{\rm eq}=\frac{2}{\sqrt{3}},\gamma_{\rm eq} =0]$  \\[3pt]
& & \\[-2mm]
\hline
& & \\[-2mm]
U(5) DS & $\rho_{\rm U(5)}=0$ & \\[3pt]
Spinodal point & $\rho^{*} = \frac{1}{2}$ & \\[3pt]
Critical point & $\rho_c = \frac{1}{\sqrt{2}}$  , $\xi_c=0$ & \\[3pt]
Anti-spinodal point & $ \xi^{**} = \frac{1}{3}$ & \\[3pt]
SU(3) DS & $\xi_{\rm SU(3)}=1$ & \\
 & & \\
\hline
\end{tabular}
\end{center}
\end{table}

The evolution of the various
stationary and asymptotic values of the Landau potentials
($V_{\rm sph}$, $V_{\rm max}$, $V_{\rm def}$, $V_{\rm bar}$,
$V_{\rm sad}$, $V_{\rm lim}$) as a function of the control
parameters $\rho$ and $\xi$, is depicted in Fig.~3.
Most of these quantities, depend also on the parameter $\bz$ of the
Hamiltonian (\ref{eq:Hint}).
In particular, $\bz$ determines the equilibrium deformation
in the deformed phase $\beta_{\rm eq}>0$, Eq.~(\ref{beqdef}),
the height of the barrier at the critical point $V_{\rm bar}$,
Eq.~(\ref{Vbarcri}), and the width of the coexistence region through
the values of the spinodal point $\rho^{*}$,
Eq.~(\ref{spinodal}), and anti-spinodal point $\xi^{**}$,
Eq.~(\ref{antispinodal}). In the present work,
we choose $\bz=\sqrt{2}$, for which the intrinsic Hamiltonian interpolates
between the U(5) and SU(3) dynamical symmetries and various expressions
simplify, since $\eta=0$ in Eq.~(\ref{exteqV1}).
For convenience, Table~2 lists the values of the relevant
control and order parameters when $\bz=\sqrt{2}$.

\subsection{Structural regions of the QPT and order parameters}

\begin{figure}[!t]
\begin{center}
\epsfig{file=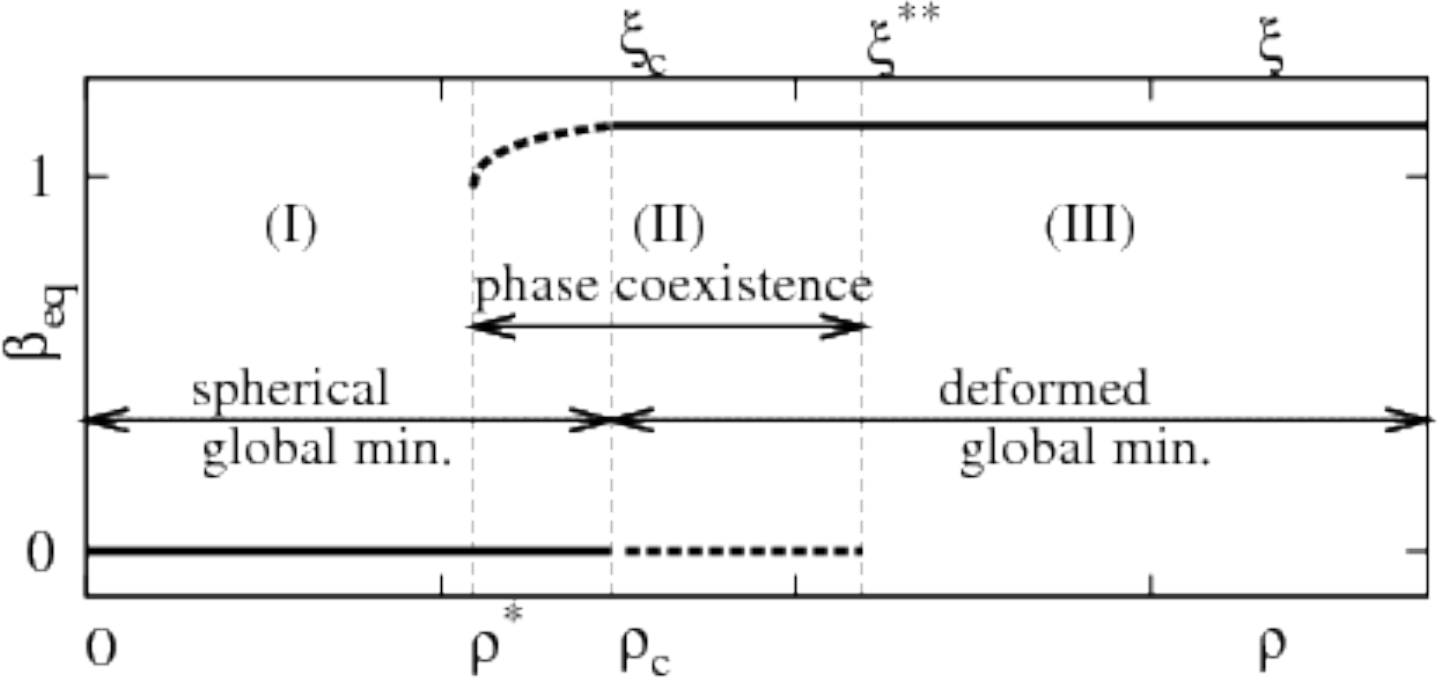,width=0.99\linewidth}
\end{center}
\caption{
Behavior of the order parameter, $\beta_{\mathrm{eq}}$,
as a function of the control parameters ($\rho,\xi$) of the
intrinsic Hamiltonian~(\ref{eq:Hint}), with $\bz=\sqrt{2}$.
Here
$\rho^{*},\, (\rho_c,\,\xi_c),\, \xi^{**}$, are the spinodal, critical
and anti-spinodal points, respectively, with values given in Table~2.
The deformation at the
global (local) minimum of the Landau potential (\ref{eq:Vcl}) is marked
by solid (dashed) lines.
$\beta_{\mathrm{eq}}\!=\!0$ ($\beta_{\mathrm{eq}}\!=\tfrac{2}{\sqrt{3}}$)
on the spherical (deformed) side of the QPT.
Region I (III) involves a single spherical (deformed) shape,
while region~II involves shape-coexistence.}
\label{fig4}
\end{figure}
The preceding classical analysis of the potential surfaces has
identified three regions with distinct structure.
\begin{enumerate}[I.]
\item{
The region of a stable spherical phase, $\rho \in [0,\rho^*]$, where
the potential has a single spherical minimum.}
\item{
The region of phase coexistence, $\rho \in (\rho^*,\rho_c]$ and
$\xi \in [\xi_c,\xi^{**})$, where
the potential has both spherical and deformed minima which
cross and become degenerate at the critical point.}
\item{
The region of a stable deformed phase, $\xi\geq \xi^{**}$, where
the potential has a single deformed minimum.}
\end{enumerate}

The potential surface in each region serves as the Landau potential of the
QPT, with the equilibrium deformations as order parameters. The latter
evolve as a function of the control parameters ($\rho,\xi$)
and exhibit a discontinuity typical of a first order transition.
As depicted in Fig~4, the  order parameter ${\beta_{\mathrm eq}}$ is a
double-valued function in the coexistence region (in-between $\rho^{*}$
and $\xi^{**}$) and a step-function outside it. In what follows we examine
the nature of the classical dynamics in each region.

\section{Regularity and chaos: classical analysis}
\label{sec:Chaos}

Hamiltonians with dynamical symmetry are always completely
integrable~\cite{zhang88}.
The Casimir invariants of the algebras in the chain provide a complete
set of constants of the motion in involution. The classical motion is
purely regular. A dynamical symmetry-breaking is usually connected
to non-integrability and may give rise to chaotic
motion~\cite{zhang88,zhang89,zhang90}.
This is the situation encountered in a QPT, which occurs as a result of a
competition between terms in the Hamiltonian with incompatible symmetries.

Regular and chaotic properties of the IBM have been studied extensively,
employing various measures of classical and quantum
chaos~\cite{ANW90,AW91,AW91b,ANW92,WA93,Heinze06,Macek06,Macek07}.
All such treatments involved the simplified Hamiltonian
of Eq.~(\ref{HndQQ}), giving rise to an extremely low barrier and narrow
coexistence region. For that reason, the majority of studies focused on
the regions I and III of stable phases, while far less effort was devoted
to the dynamics inside the region~II of phase-coexistence.
Considerable attention has been paid to integrable paths
(the U(5)-O(6) transition for $\chi=0$ in
Eq.~(\ref{HndQQ})~\cite{Heinze06,Macek06})
and to specific sets of parameters leading to an enhanced regularity
(``arc of regularity''~\cite{AW91b,Macek07}) within these regions.
Similar type of analysis was performed in the framework of
the geometric collective model of
nuclei~\cite{Bere04,CejStr04,CejStr06,Stran09a,Stran09b}.

In the present work, we consider the evolution of order and chaos across
a generic first order quantum phase transition, with particular
emphasis on the role of a high barrier separating the two phases.
For that purpose, we
employ the intrinsic Hamiltonian of Eq.~(\ref{eq:Hint})
with $\bz\!=\!\sqrt{2}$. In this case, the height of the barrier at the
critical point, Eq.~(\ref{Vbarcri}), is $V_{\rm bar}/h_2 = 0.268$,
substantially higher than
barrier heights encountered in previous works. In comparison, for the
Hamiltonian of Eq.~(\ref{HcrindQQ}) with $\chi=-\tfrac{\sqrt{7}}{2}$,
the corresponding quantities are
$\bz=\tfrac{1}{2\sqrt{2}}$ and $V_{\rm bar}/h_2=0.0018$.
A~high barrier will allow us to uncover
a rich pattern of regularity and chaos in region~II of shape-coexistence.

The classical dynamics of $L\!=\!0$ vibrations,
associated with the Hamiltonian~(\ref{eq:Hint}),
can be depicted conveniently
via Poincar\'e surfaces of sections~\cite{Gutz90,Reic92,Licht92}.
The latter are chosen in the plane $y=0$ which passes through all
the various types of stationary points (minimum, maximum, saddle)
in the Landau potential~(\ref{eq:Vcl}). The values of $x$ and
$p_x$ are plotted
each time a trajectory intersects the plane.
The method of Poincar\'e sections provides a snapshot of the
dynamics at a given energy.
Regular trajectories are bound to toroidal manifolds within the phase
space and their intersections with the plane of section lie on
one-dimensional curves (ovals). In contrast, chaotic trajectories
diverge exponentially and randomly
cover kinematically accessible areas of the section.
Although restricted to $L=0$, the method is particularly valuable
to the present study, due to its ability to identify
different forms of dynamics occurring at the same energy
in separate regions of phase space. Standard global classical measures of
chaos, such as, the fraction of chaotic volume and the average largest
Lyapunov exponent, are insensitive such local variations.
We first discuss distinctive features of the
dynamics in each region and relate them to the morphology of the Landau
potential. This will provide the necessary background for understanding
the complete evolution of the dynamics across the~QPT.

\subsection{Characteristic features of the dynamics in the
vicinity of minima}

Considerable insight into the nature of the classical dynamics at low energy
can be gained by examining the topology of the Landau
potential in the vicinity of its minima.
A sample of representative Poincar\'e sections for the classical
Hamiltonian constrained to $L=0$, Eq.~(\ref{eq:Hintcl}),
are depicted in Figs.~5-6, along with selected trajectories.
\begin{figure*}[!t]
\begin{center}
\epsfig{file=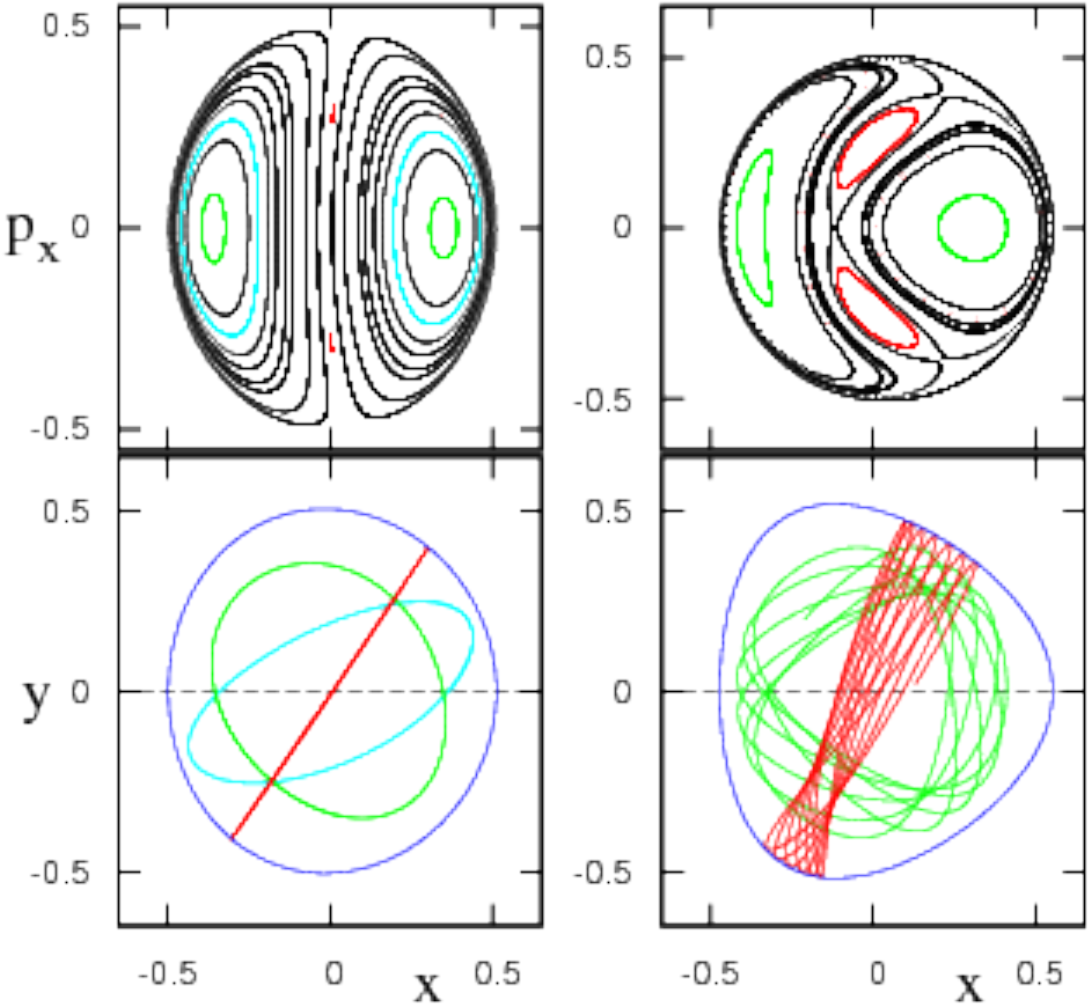,width=0.495\linewidth}
\epsfig{file=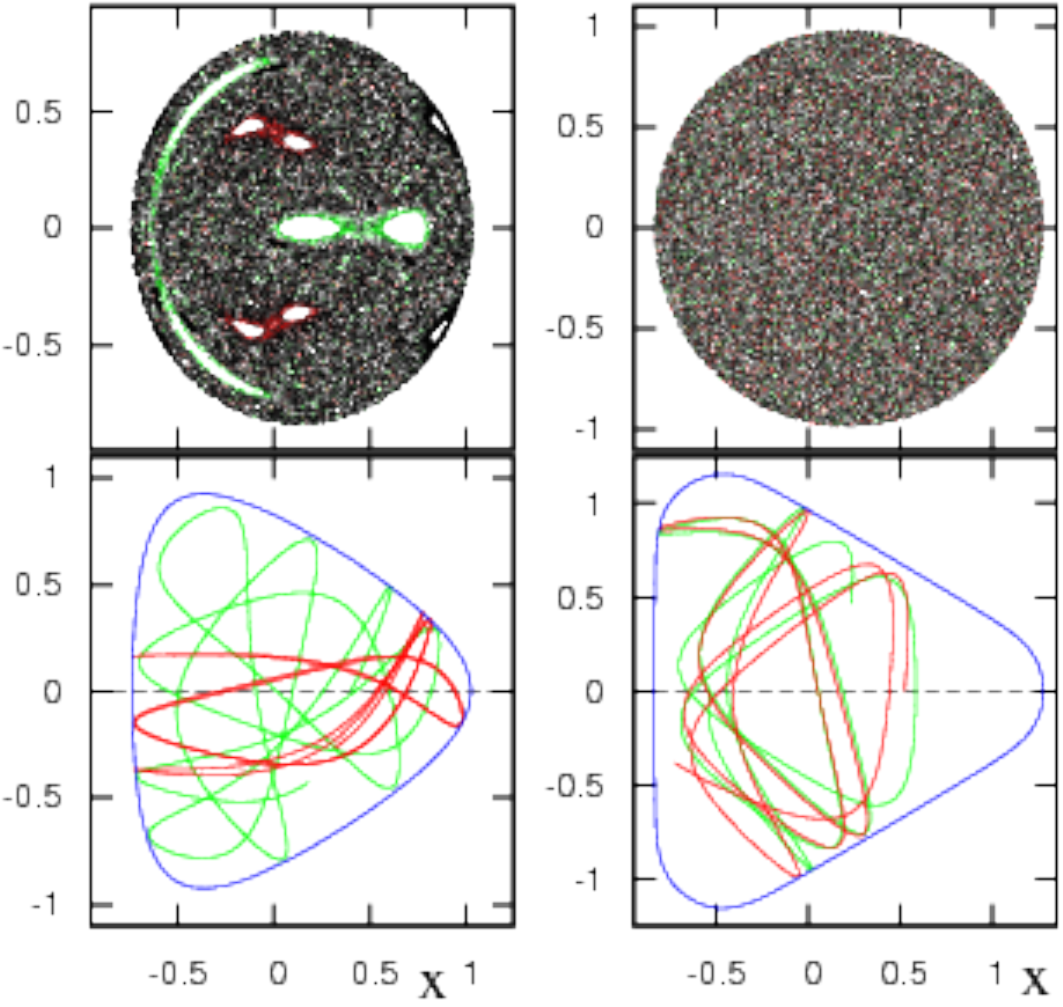,width=0.482\linewidth}
\end{center}
\caption{
Classical dynamics of $\mathcal{H}_1 (\rho)$,
Eq.~(\ref{eq:H1cl}) with $\bz=\sqrt{2}$, $h_2=1$,
for several values of $\rho$ and energies $E$, in the vicinity
of a spherical minimum.
Selected trajectories (bottom row) and
corresponding Poincar\'e sections (top row), portray
the motion in the presence of a single spherical minimum.
(a)~$\rho=0.03$, $E=0.5$. (b)~$\rho=0.2$, $E=0.5$. (c)~$\rho=0.2$, $E=1.114$.
(d)~$\rho=0.2$, $E=1.5$.}
\label{fig5}
\end{figure*}
\begin{figure}
\epsfig{file=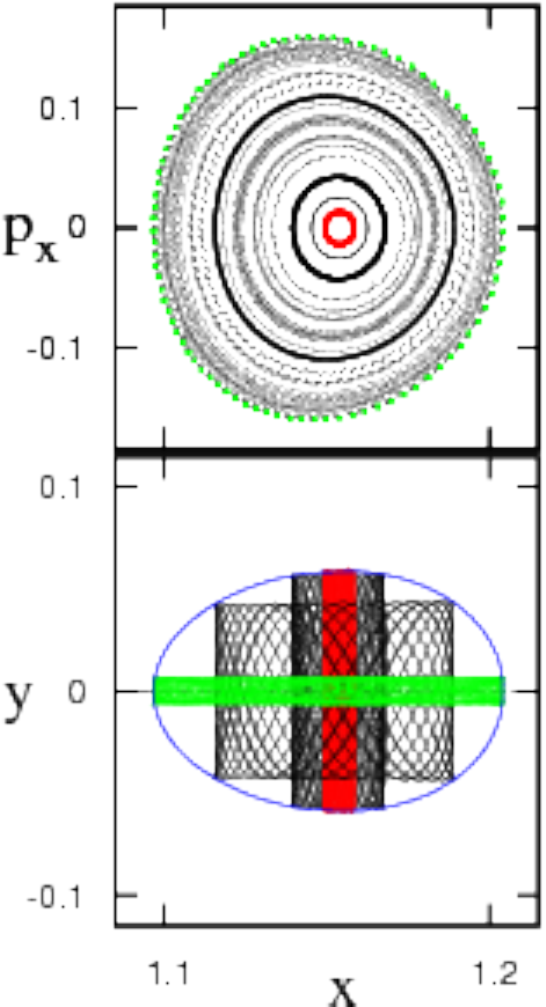,height=0.335\textheight}
\epsfig{file=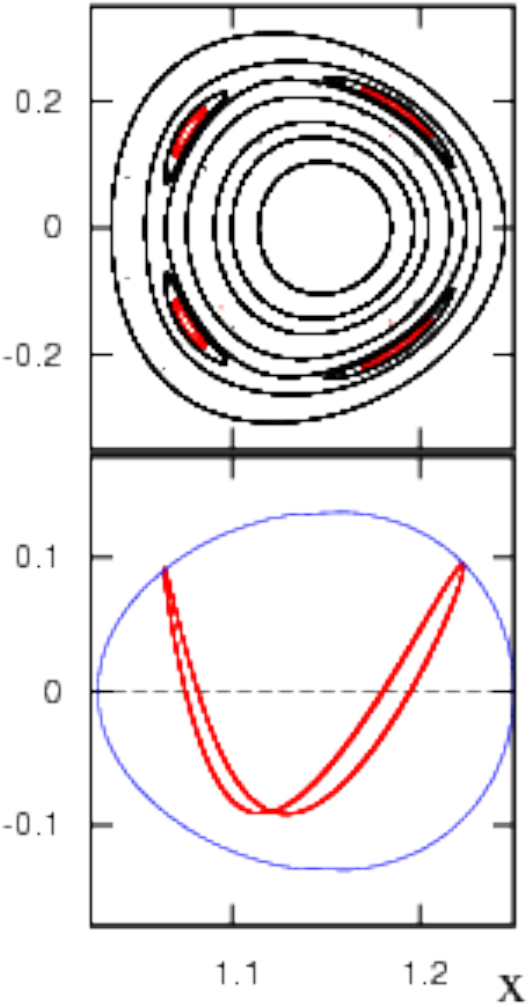,height=0.335\textheight}
\epsfig{file=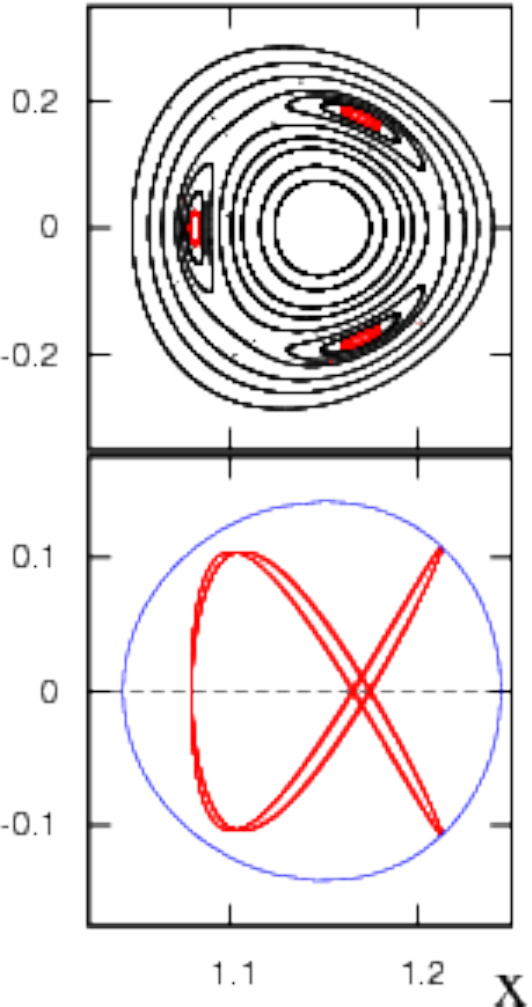,height=0.335\textheight}
\epsfig{file=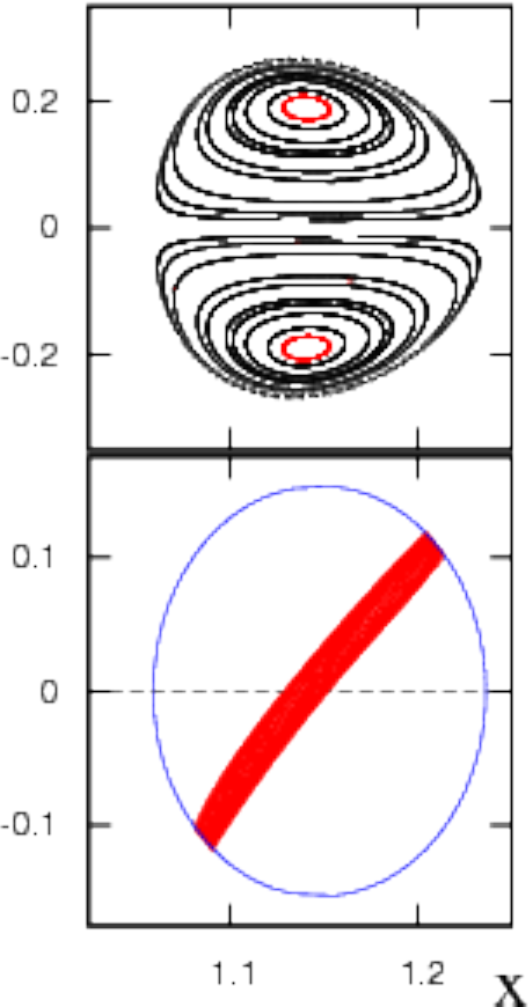,height=0.335\textheight}
\caption{
Classical dynamics of $\mathcal{H}_2 (\xi)$,
Eq.~(\ref{eq:H2cl}) with $\bz=\sqrt{2}$, $h_2=1$, for several values
of $\xi$ and energies $E$, in the vicinity of a deformed minimum.
Selected trajectories (bottom row) and
corresponding Poincar\'e sections (top row), portray
the motion in the presence of a single deformed minimum.
(a)~$\xi=0.11$, $E=0.02$. (b)~$\xi=0.258$, $E=0.11$.
(c)~$\xi=0.505$, $E=0.12$.
(d)~$\xi=1$, $E=0.15$.
Panel~(a) shows a typical pattern encountered for most values of
$\xi\geq\xi^{**}$.
Panels~(b)-(c)-(d) correspond to
$R=\epsilon_{\beta}/\epsilon_{\gamma} =
1/2,\,2/3,\,1$ in Eq.~(\ref{R}), and
show the effect of local degeneracies of normal modes.
Panel~(d) corresponds to the SU(3) DS limit.}
\label{fig6}
\end{figure}

The spherical configuration ($\beta=0$) is a global minimum
of the potential $V_{1}(\rho;\beta,\gamma)$, Eq.~(\ref{eq:V1}),
on the spherical side of the QPT ($0\leq\rho\leq\rho_c$).
For $\rho=0$, the system has U(5) DS and hence is integrable.
The potential $V_{1}(\rho=0)$, Eq.~(\ref{V1rho0}),
is $\gamma$-independent and exhibits $\beta^2$ and $\beta^4$ dependence.
As shown in Fig.~5(a), the sections, for small $\rho$,
show the phase space portrait typical of a weakly perturbed anharmonic
(quartic) oscillator with two major regular islands and quasi-periodic
trajectories. The effect of increasing $\rho$ on the dynamics
in the vicinity of the spherical ($s$) minimum ($\beta\approx 0$),
can be inferred from a small $\beta$-expansion of the potential,
\ba
\label{eq:VExpB0}
V_{1,s}(\rho) &\approx&
\bz^2 \beta^2
- \rho\sqrt{2}\bz^2\beta^3\cos3\gamma ~.
\label{eq:V1b0}
\ea
To order $\beta^3$,
$V_{1,s}(\rho)$ has the same functional form as the the well-known
H\'enon-Heiles (HH) potential~\cite{Heno64},
which in polar coordinates $(r,\phi$) reads
\ba\label{eq:HH}
V_\mathrm{HH} = r^2 - \alpha\, r^3\cos 3\phi\,,
\ea
with $\alpha > 0$.
The latter potential serves as a paradigm of a system that exhibits
a transition from regular to chaotic dynamics as the energy
increases~\cite{Gutz90,Reic92,Licht92}.
As shown for $\rho\!=\!0.2$, at low energy [Fig.~5(b)],
the dynamics remains regular, and two additional islands show up.
The four major islands surround stable fixed points and
unstable (hyperbolic) fixed points occur in-between.
At higher energy [Fig.~5(c)], one observes a marked onset
of chaos and an ergodic domain.
This typical HH-type of behavior persists
in the vicinity of the spherical minimum
throughout the coexistence region, including the critical point
($\rho_c,\xi_c$) and beyond
where the spherical minimum is only local ($0\leq\xi\leq\xi^{**}$).
This can be inferred from a similar small $\beta$-expansion
of the relevant
potential $V_{2}(\xi;\beta,\gamma)$, Eq.~(\ref{eq:V2}),
\ba
V_{2,s}(\xi)  &\approx&
\xi\bz^4 +
\bz^2[1 - \xi(1\!+\!\bz^2)]\beta^2
- \sqrt{2}\bz\beta^3\cos3\gamma ~.
\label{eq:V2b0}
\ea
It should be noted that although the
expansions in Eqs.~(\ref{eq:V1b0}) and~(\ref{eq:V2b0}) are similar
in form to the H\'enon-Heiles potential (\ref{eq:HH}),
the full potentials, Eq.~(\ref{eq:Vcl}), have a finite domain
and include a $\beta^4$ term, thus ensuring that the motion is bounded
at all energies.

The deformed configuration ($\beta_{\rm eq}>0,\gamma_{\rm eq}=0$),
Eq.~(\ref{beqdef}), is a global minimum of the potential
$V_{2}(\xi;\beta,\gamma)$,
Eq.~(\ref{eq:V2}), on the deformed side of the QPT ($\xi\geq\xi_c$).
The classical dynamics in its vicinity ($x\approx 1$)
has a very different character, being robustly regular.
At low energy, the motion reflects the $\beta$ and $\gamma$ normal
mode oscillations about the deformed minimum.
As shown in  Fig.~6(a),
the family of regular trajectories has a particular simple structure.
It forms a single set of concentric loops
around a single stable (elliptic) fixed point.
They portray $\gamma$-vibrations at the center
of the surface ($p_x\approx 0$) and $\beta$-vibrations at the perimeter
(large $\vert p_x\vert$).
This regular pattern of the dynamics is found for most values of
$\xi\geq 0$ both inside and outside the phase coexistence region.
The dynamics remains regular but its pattern changes in the presence of
resonances. The latter appear when the ratio
of normal mode frequencies, Eq.~(\ref{nmodedef}),
is a rational number
\ba
R \equiv \frac{\epsilon_{\beta}}{\epsilon_{\gamma}} =
\frac{1}{9}(1+\bz^2)(2\xi+1) ~.
\label{R}
\ea
Panels (b)-(c)-(d) of Fig.~6 
show examples of such scenario for $\bz=\sqrt{2}$ and
$\xi$-values corresponding to
$R=1/2,\,2/3,\,1$. The corresponding
surfaces exhibit four, three and two islands, respectively.
The phase space portrait for ($\xi=1,R=1$), shown in Fig.~6(d), 
corresponds to the integrable SU(3) DS limit.
These additional chains of regular islands will be considered
in more detail in Section~5.3.
\begin{figure}
\centering
\vspace{-2cm}
\epsfig{file=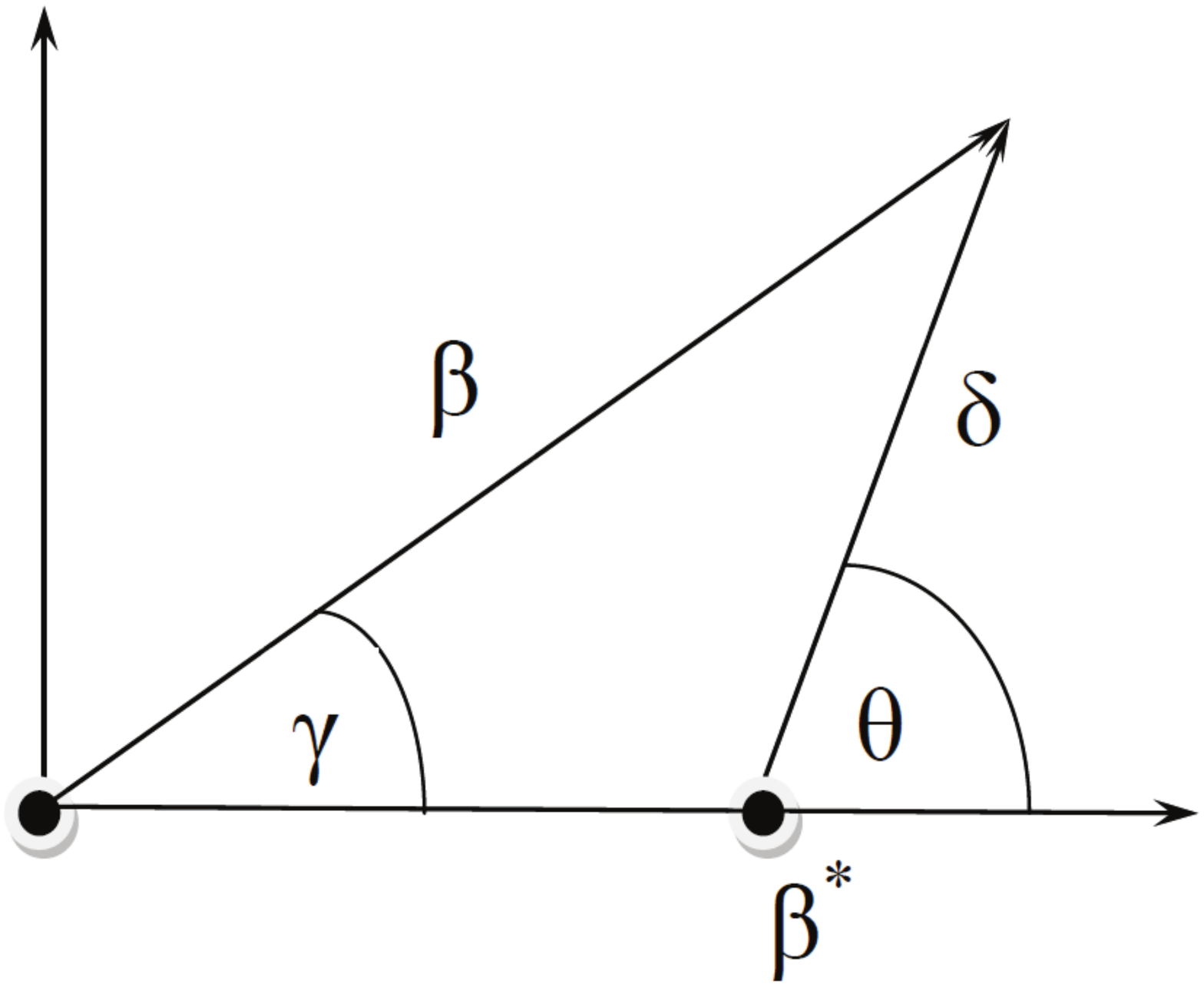,width=0.5\linewidth}
\caption{
Local coordinates $(\delta,\theta)$, Eq.~(\ref{localcor}),
about a deformed minimum $(\beta^*,\gamma^*=0)$.}
\label{fig7}
\end{figure}

Similar trends are observed in the region ($\rho^{*}< \rho \leq\rho_c$),
where the deformed minimum is only local.
A regular dynamics is thus an inherent feature of a deformed
minimum and, at low energy, reflects the behavior of the Landau potential
in its vicinity.
The structure of the latter is revealed in an expansion of the potential
in local coordinates.
Consider a deformed minimum (global or local)
of the Landau potential
characterized by the deformation
($\beta^{*}\!>\!0,\gamma^{*}=0$). The local coordinates ($\delta,\theta$)
about it,
shown in Fig~7, are defined by the relations
\bsub
\ba
\beta\cos\gamma &=& \beta^{*} + \delta\cos\theta ~,\\
\beta\sin\gamma &=& \delta\sin\theta ~.
\ea
\label{localcor}
\esub
\begin{figure}[!t]
\vspace{-2cm}
\begin{center}
\epsfig{file=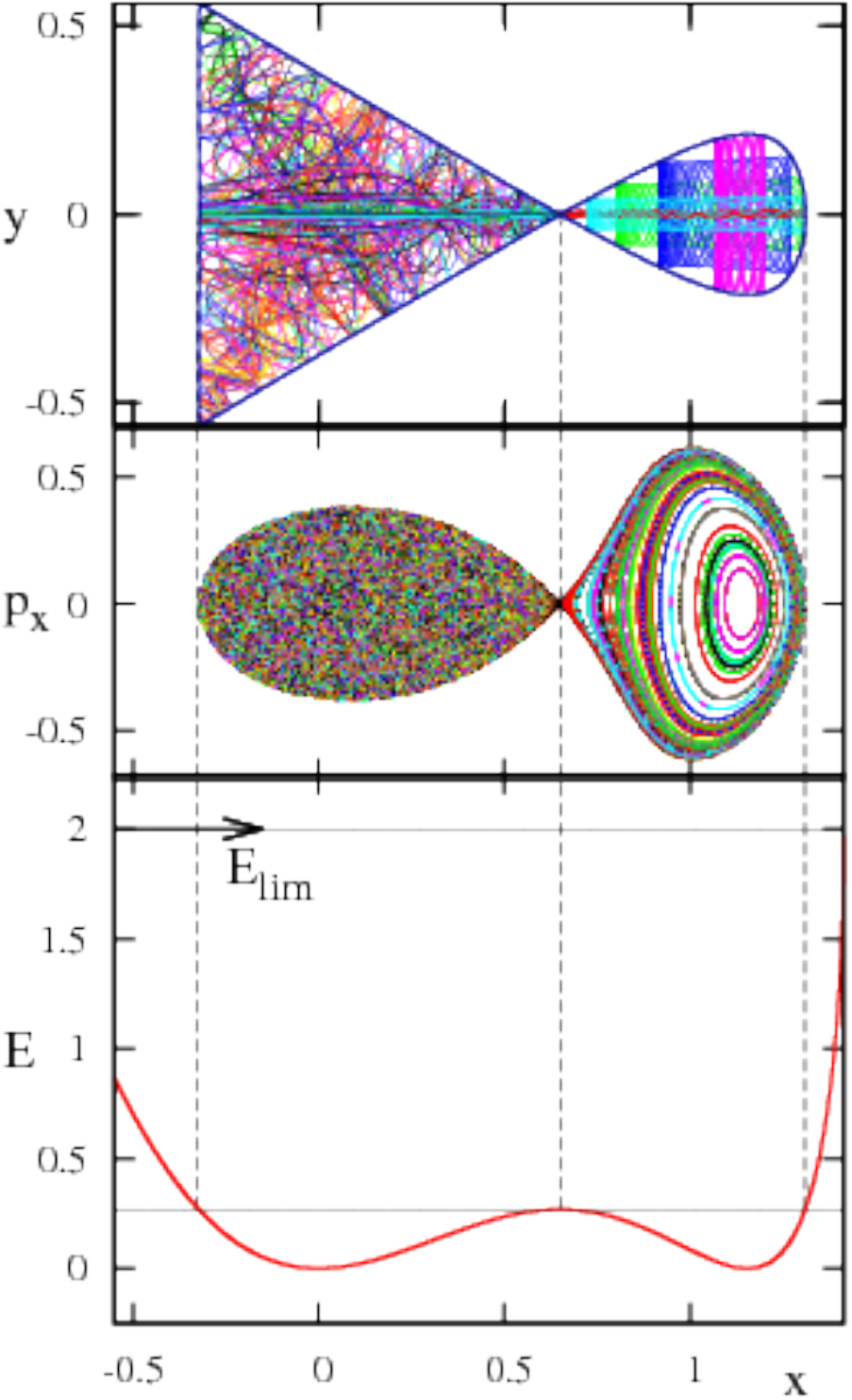,height=10.3cm}
\end{center}
\caption{
Classical dynamics of the critical point Hamiltonian,
${\cal H}_{\rm cri} = {\cal H}_{1}(\rho_c) = {\cal H}_{2}(\xi_c)$,
Eq.~(\ref{eq:Hintcl}), at the energy of the barrier.
The lower, middle, and upper portions depict
$V_\mathrm{cri}(\beta,\gamma=0)$, Eq.~(\ref{eq:Vcri}),
the Poincar\'e section and selected trajectories,
respectively. The pattern of mixed but well-separated chaotic and regular
dynamics, associated with the spherical and deformed minima,
appears throughout the coexistence region.}
\label{fig8}
\end{figure}
A small $\delta$-expansion of the potential
about this minimum (to order $\delta^3$ ), reads
\ba
V(\beta,\gamma) \approx K_1 +
\delta^2 \left ( K_2 + K_3 \cos^2\theta \right )
+ \delta^3 ( K_4\cos\theta + K_5\cos^3\theta) ~.
\label{Vdelta}
\ea
Here $V(\beta,\gamma)$ stands for $V_{1}(\rho;\beta,\gamma)$ in the
spherical phase and $V_{2}(\xi;\beta,\gamma)$ in the deformed phase.
In general, the coefficients $K_i$ depend on $\beta^{*}$ and the
control parameters, {\it e.g.}, $K_1 = V(\beta^{*},\gamma=0)$.
In the deformed phase, where the deformed minimum is global,
$\beta^{*}=\beta_{\rm eq} = \sqrt{2}\bz(1+\bz^2)^{-1/2}$,
Eq.~(\ref{beqdef}), and the $K_i$ coefficients are given by
\ba
&&K_1 = 0 \;\;\; , \;\;\; K_2 = \tfrac{9\bz^2}{1+\bz^2}\;\;\; , \;\;\;
K_3 = \tfrac{\bz^2}{1+\bz^2}
\left [ (\bz^2-2)(\bz^2+4) + 2\xi(1+\bz^2)^2\right ] ~,\qquad
\nonumber\\
&&K_4 = \tfrac{\sqrt{2}\bz}{2\sqrt{1+\bz^2}}
\left [ (\bz^2-2)(\bz^2-5) + 2\xi(1+\bz^2)^2\right ]\;\;\; ,
\nonumber\\
&&K_5 = \tfrac{\sqrt{2}\bz}{2\sqrt{1+\bz^2}}
\left [ (\bz^2-2)(\bz^2+2)^2 + \bz^2(\bz^2+16)\right ] ~.
\ea
For $\bz=\sqrt{2}$, these expressions simplify to
\ba
V_{2}(\xi) \approx 6\delta^2\left [1 + 2\xi\cos^{2}\theta\right ]
+ 6\sqrt{3}\delta^3\left [2\cos^{3}\theta +\xi\cos\theta\right ] ~.
\label{V2delta}
\ea
The expansions in Eqs.~(\ref{Vdelta}) and (\ref{V2delta}) contain
terms with $\cos\theta$, $\cos 2\theta$ and $\cos 3\theta$ dependence.
The presence of lower harmonics destroys, locally,
the three-fold symmetry
encountered near the spherical minimum,
Eqs.~(\ref{eq:V1b0})-(\ref{eq:V2b0}),
due to the $\cos 3\gamma$ term.
This asymmetry is clearly seen in the contour plots of Figs.~1-2.

Both spherical and deformed minima of the Landau potentials
$V_{1}(\rho;\beta,\gamma)$ and $V_{2}(\xi;\beta,\gamma)$,
are present in the coexistence region, $\rho^{*}< \rho\leq\rho_c$
and $\xi_c\leq\xi<\xi^{**}$.
In this case, each minimum preserves its own characteristic dynamics
resulting in a marked separation between a H\'enon-Heiles type of chaotic
motion in the vicinity of the spherical minimum and a regular motion
in the vicinity of the deformed minimum.
Such mixed form of dynamics occurring at the same energy in different
regions of phase space, is demonstrated in Fig.~8.
The latter depicts the potential landscape at the critical point
($\rho_c,\xi_c$), along with
the Poincar\'e section and selected trajectories at the barrier energy.
In this case, the spherical (s) and deformed (d) minima
are degenerate, and for $\bz=\sqrt{2}$, the expansions of
the corresponding Landau potential
in their vicinity exhibit a different morphology
\bsub
\ba
V_{\rm s, cri} &\approx&
2\beta^2  -2\beta^3\cos 3\gamma ~,\\
V_{\rm d, cri} &\approx&
6\delta^2 + 3\sqrt{3}\delta^3
\left [\cos3\theta + 3\cos\theta\right ] ~.
\ea
\esub
The critical-point potential near the spherical minimum
($V_{\rm s,cri}$) has a 3-fold symmetry and its contours are either
concave or convex towards the origin (see Fig.~1).
The former contours lead to divergence of trajectories, a characteristic
property of chaotic motion. In contrast, the critical-point
potential near the deformed minimum ($V_{\rm d,cri}$) has an egg-shape,
without a local 3-fold symmetry. The potential contours are convex and
tend to focus the trajectories towards the minimum, resulting in a
confined regular motion.

\subsection{Evolution of the classical dynamics across the QPT}

\begin{figure*}
\begin{center}
\epsfig{file=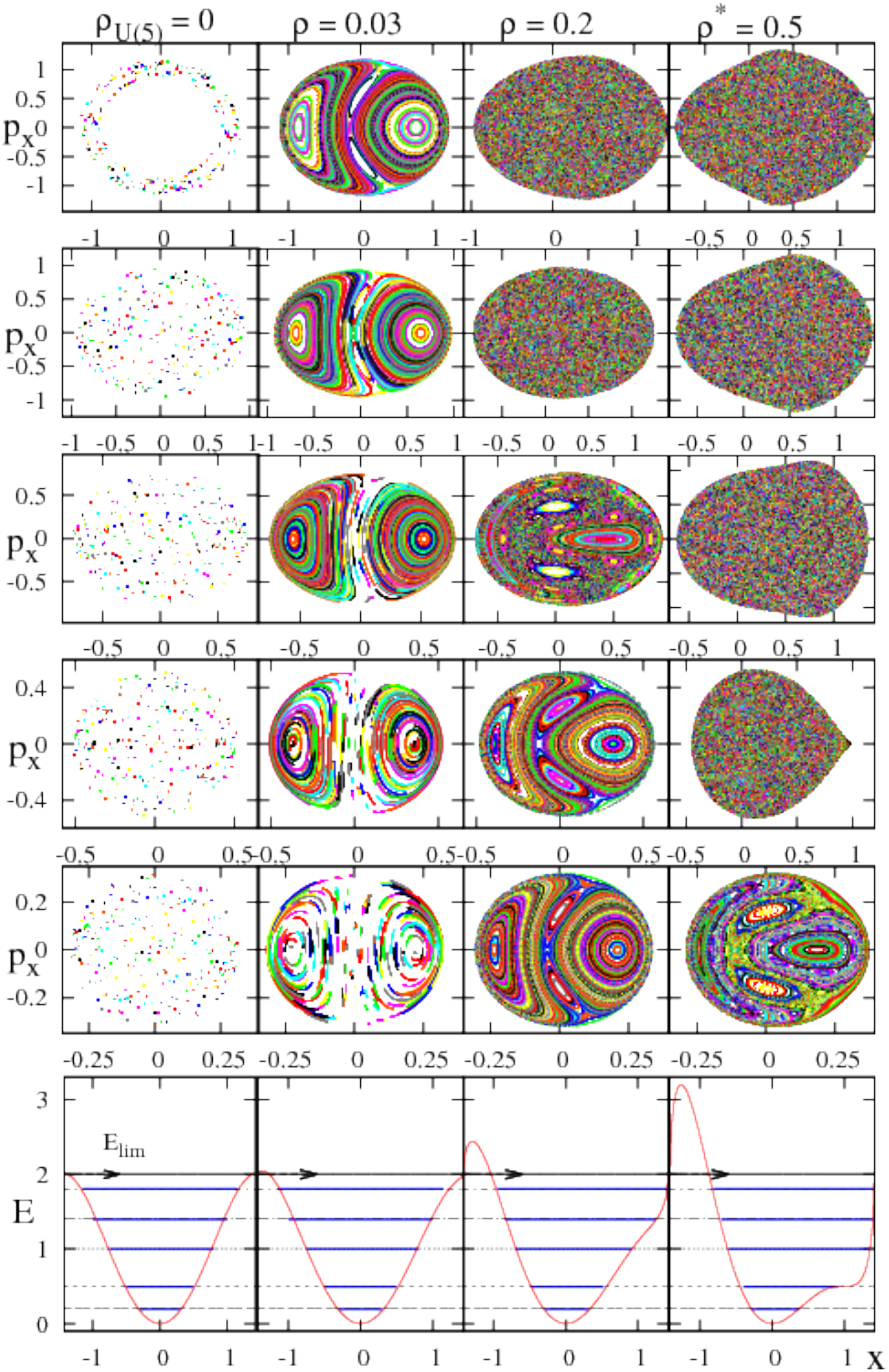,width=0.7\linewidth}
\end{center}
\caption{
Poincar\'e sections in the stable spherical phase (region~I).
Upper five rows depict the classical dynamics of
${\cal H}_{1}(\rho)$~(\ref{eq:H1cl})
with $h_2=1$ and $\bz=\sqrt{2}$, for several values of $\rho\leq\rho^{**}$.
The bottom row displays the corresponding classical potentials
$V_{1}(\rho;x,y=0)$~(\ref{Vcl1xy}).
The five energies, below $E_{\rm lim}=2h_2$,
at which the sections were calculated consecutively, are indicated
by horizontal lines.
The left column ($\rho_{\rm U(5)}=0$) corresponds to the integrable
U(5) DS limit. The right column ($\rho^{*}$) corresponds to the
spinodal point.}
\label{fig9}
\end{figure*}
\begin{figure*}
\begin{center}
\epsfig{file=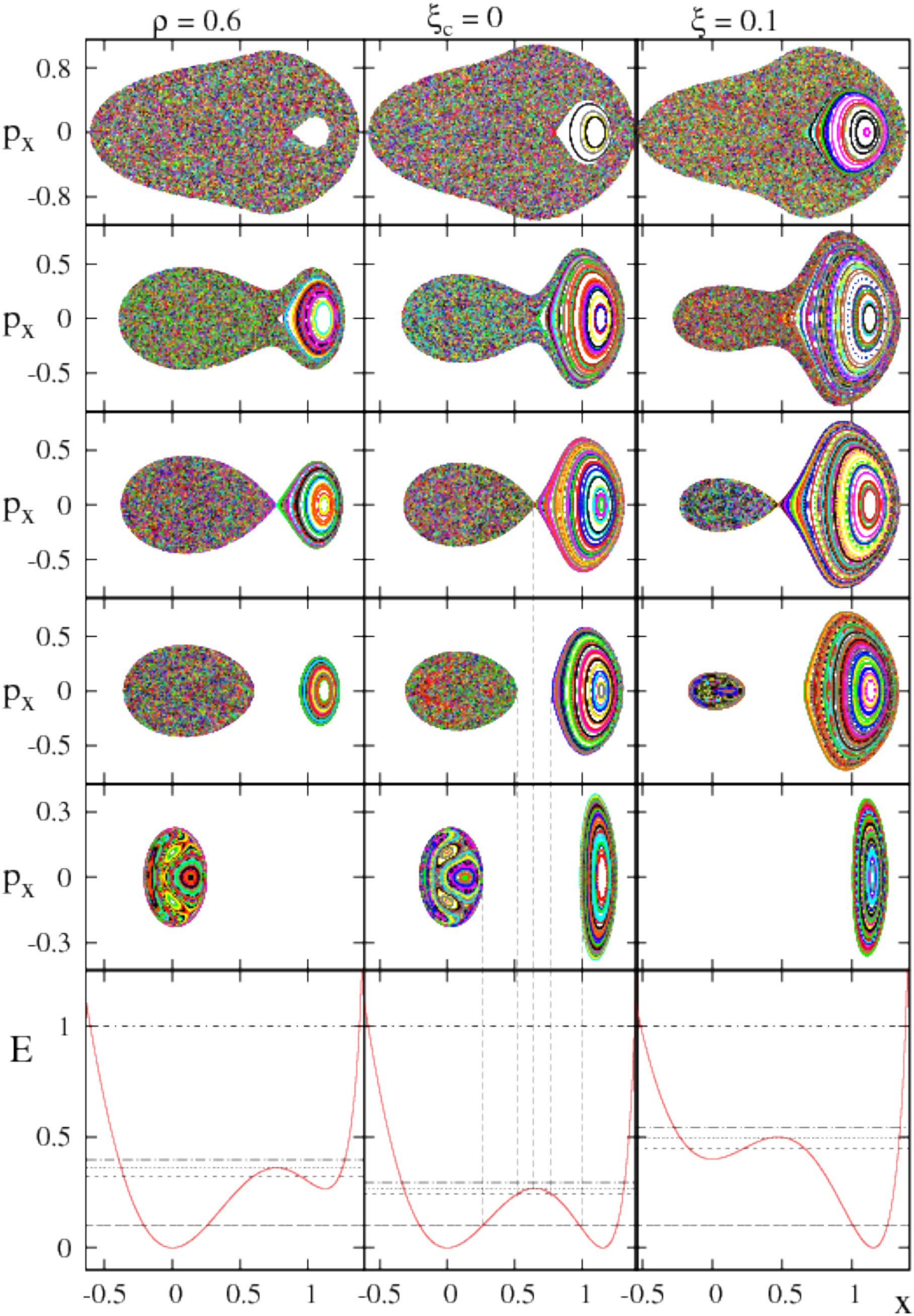,width=0.7\linewidth}
\end{center}
\caption{
Poincar\'e sections in the region of phase-coexistence (region~II).
The panels are as in Fig.~9, but for
${\cal H}_1(\rho)$~(\ref{eq:H1cl}) with $\rho^{**}<\rho \leq\rho_c$,
and ${\cal H}_2(\xi)$~(\ref{eq:H2cl}) with $\xi_c\leq \xi < \xi^{**}$.
The classical potentials are $V_{1}(\rho;x,y=0)$~(\ref{Vcl1xy})
and $V_{2}(\xi;x,y=0)$~(\ref{Vcl2xy}), respectively.
The middle column corresponds to the critical point $(\rho_c,\xi_c)$,
and the vertical lines mark the turning points for energies
below or equal to the barrier height.}
\label{fig10}
\end{figure*}
\begin{figure*}
\begin{center}
\epsfig{file=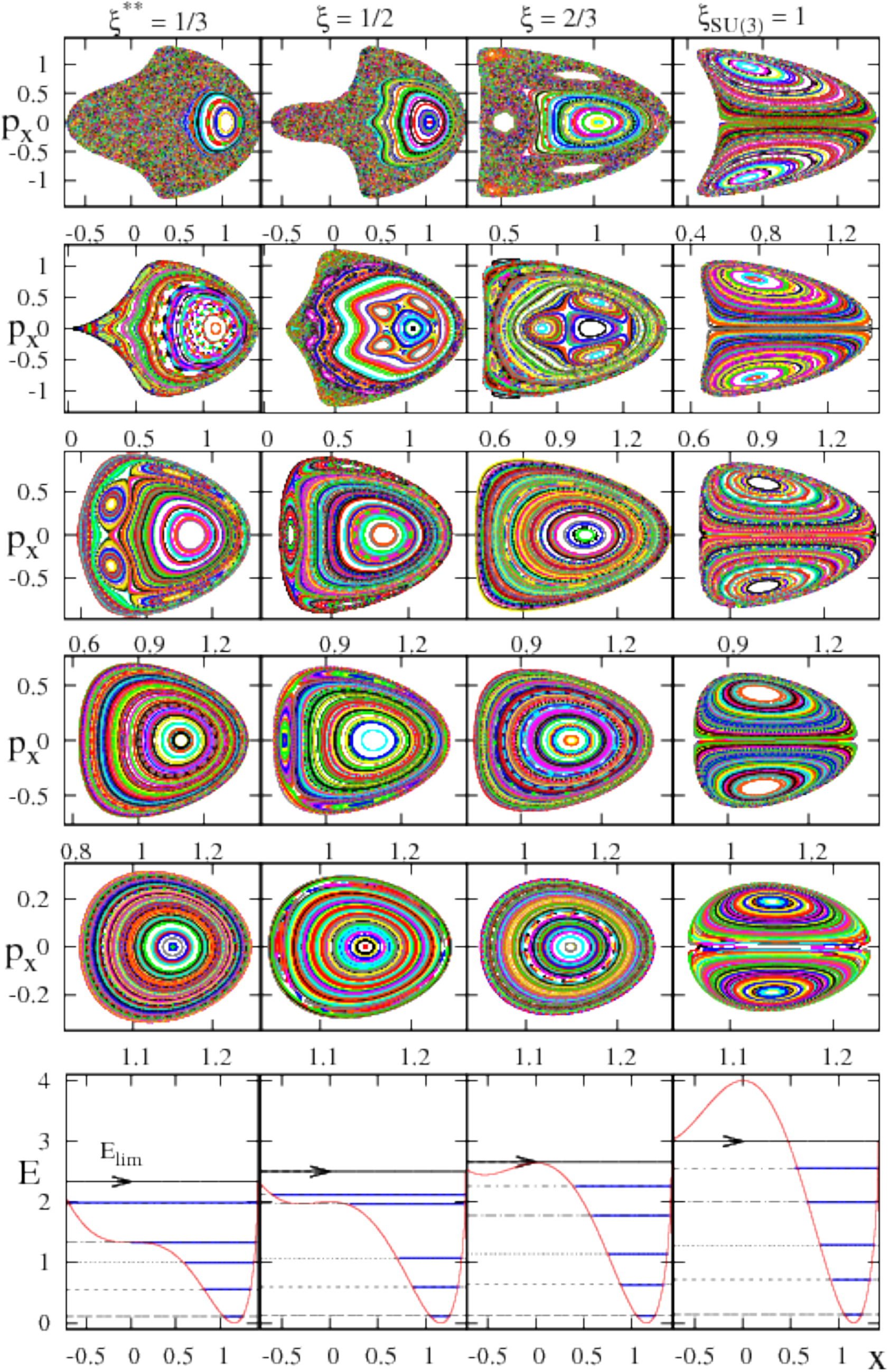,width=0.7\linewidth}
\end{center}
\caption{
Poincar\'e sections in the stable deformed phase (region~III)
The panels are as in Fig.~9, but for the classical intrinsic
Hamiltonian ${\cal H}_2(\xi)$~(\ref{eq:H2cl}) and potential
$V_{2}(\xi;x,y=0)$~(\ref{Vcl2xy}), with $\xi\geq\xi^{**}$ and
$E_{\rm lim} = h_2(2+\xi)$.
The left column ($\xi^{**}$) corresponds to the anti-spinodal point.
The right column ($\xi_{\rm SU(3)}=1$) corresponds to the integrable
SU(3) DS limit.}
\label{fig11}
\end{figure*}

We turn now to a comprehensive analysis of the
classical dynamics, constraint to $L=0$, evolving across the
first order QPT.
The evolution is accompanied by an intricate interplay of order
and chaos, reflecting the change in structure.
The shape-phase transition is induced by the intrinsic Hamiltonian
of Eq.~(\ref{eq:Hint})
with $\bz=\sqrt{2}$. The Poincar\'e surfaces of sections,
are shown in Figs.~9-10-11 for
representative energies, below the domain boundary ($V_\mathrm{lim}$),
and control parameters ($\rho,\xi$) in regions~I-II-III, respectively.
The surfaces record a total of 40,000 passages through the $y=0$ plane by
120 trajectories with randomly generated initial conditions,
in order to scan the whole accessible phase space at a given energy.
The bottom row in each figure displays the corresponding classical
potential $V(x,y=0)$, Eq.~(\ref{Vclxy}).

The classical dynamics of $L=0$ vibrations
in the stable spherical phase (region~I) is
governed by the Hamiltonian $\mathcal{H}_1(\rho)$, Eq.~(\ref{eq:H1cl}),
with $0\leq\rho\leq\rho^{*}$. The relevant potential $V_{1}(\rho)$,
Eq.~(\ref{Vcl1xy}), has a single minimum at $(x,y)=(0,0)$.
For $\rho=0$, the quantum Hamiltonian $\hat{H}_{1}(\rho=0)$,
Eq.~(\ref{H1u5}), has U(5) DS and its classical counterpart,
${\cal H}_1(\rho=0)= {\cal H}_{d,0}(2-{\cal H}_{d,0})$, involves
the 2D harmonic oscillator Hamiltonian,
${\cal H}_{d,0}=(p_{x}^2+p_{y}^2 + x^2 + y^2)/2$.
The system is completely integrable. The orbits are periodic
and, as shown in Fig.~9, appear in the surface of section
as a finite collection of points. As previously noted,
for small values of~$\rho$ ($\rho=0.03$
in Fig.~9), the sections are those of an anharmonic (quartic) oscillator,
weakly perturbed by the small $\rho (x^3-3x^2y)$ term.
The orbits are quasi-periodic and appear as smooth one-dimensional
invariant curves.
For larger values of $\rho$,
the importance of the latter perturbation increases.
The derived phase-space portrait near $x=0$, shown for $\rho\!=\!0.2$
in Fig.~9, is similar to the  H\'enon-Heiles system (HH)~\cite{Heno64}
with regularity at low energy and marked onset of chaos at higher energies.
The chaotic component of the dynamics increases with $\rho$ and
maximizes at the spinodal point $\rho^{*}\!=\!0.5$.
The chaotic orbits densely fill two-dimensional regions of the
surface of section.

The dynamics changes profoundly in the coexistence region (region~II).
Here the relevant classical Hamiltonians
are ${\cal H}_{1}(\rho)$, Eq.~(\ref{eq:H1cl}), with
$\rho^{*} < \rho \leq \rho_c$ and ${\cal H}_{2}(\xi)$, Eq.~(\ref{eq:H2cl}),
with $ \xi_c \leq \xi < \xi^{**}$.
The corresponding potentials $V_{1}(\rho)$, Eq.~(\ref{Vcl1xy}),
and $V_{2}(\xi)$, Eq.~(\ref{Vcl2xy}) have both
spherical and deformed minima, which become degenerate and cross
at the critical point $(\rho_c\!=\!1/\sqrt{2},\xi_c\!=\!0)$.
The Poincar\'e sections
before, at and after the critical point,
($\rho\!=\!0.6$, $\xi_c\!=\!0$, $\xi\!=\!0.1$) are shown in Fig.~10.
In general, the motion is predominantly regular at low energies and
gradually turning chaotic as the energy increases.
However, the classical dynamics evolves
differently in the vicinity of the two wells.
As the local deformed minimum develops, robustly regular dynamics
attached to it appears. The trajectories form a single island
and remain regular even
at energies far exceeding the barrier height $V_{\rm bar}$.
This behavior is in marked contrast to the HH-type of dynamics
in the vicinity of the spherical minimum, where a change with energy
from regularity to chaos is observed, until complete chaoticity is reached
near the barrier top.
The clear separation between regular and chaotic dynamics,
associated with the two minima, persists all the way to the
barrier energy, $E=V_{\rm bar}$, where the two regions just touch.
At $E > V_{\rm bar}$, the chaotic trajectories from the spherical region
can penetrate into the deformed region
and a layer of chaos develops,
and gradually dominates the surviving regular island
for $E\gg V_{\rm bar}$. As $\xi$ increases, the spherical minimum becomes
shallower, and the HH-like dynamics diminishes.

As seen in Fig.~11, the dynamics is robustly regular in the stable
deformed phase (region~III),
where the relevant classical Hamiltonian is ${\cal H}_{2}(\xi)$,
Eq.~(\ref{eq:H2cl}), with $\xi\geq \xi^{**}$.
The spherical minimum disappears
at the anti-spinodal point $\xi^{**}\!=\!1/3$
and the relevant potential $V_{2}(\xi)$, Eq.~(\ref{Vcl2xy}),
remains with a single deformed minimum.
Regular motion prevails for $\xi\!\geq\! \xi^{**}$ where
a single stable fixed point, surrounded by a family of
elliptic orbits, continues to dominate the Poincar\'e section.
In certain regions of the control parameter $\xi$ and energy,
the section landscape changes from a single to several regular
islands, reflecting the sensitivity of the dynamics
to local degeneracies of normal modes.
Such resonance effects will be elaborated in more detail in Section~5.3.
A notable exception to such variation is the SU(3) DS limit
($\xi=1$), for which the system is integrable and the phase
space portrait is the same for any energy.

\subsection{Resonance effects}\label{subsec:Res}

The preceding discussion has shown that even away from the integrable
SU(3) limit, the classical intrinsic dynamics associated with the deformed
well, remains robustly regular. In most segments of regions~II and ~III,
the Poincar\'e sections exhibit a single island, originating from
simple $\beta$ ($x$) and $\gamma$ ($y$) orbits, imprinting
the small amplitude vibrations of normal modes about the deformed minimum.
As noted, occasionally, resonances in these oscillations give rise
to additional chains of regular islands. In the present section we examine
in more detail this sensitivity of the classical motion and attempt to
demarcate the ranges of energy and control parameters where these
resonance effects occur.

The dynamical consequences of perturbing a classical integrable system,
are governed by the celebrated Kolmogorov-Arnold-Moser (KAM) and
Poincar\'e-Birkhoff (PB) theorems~\cite{Gutz90,Reic92,Licht92}.
According to the KAM theorem, most tori of the integrable system which
are sufficiently irrational,
get slightly deformed in the perturbed system but are not destroyed.
On the other hand, the resonant tori (the tori characterized by
a rational ratio of winding frequencies) of the integrable system,
disintegrate when the system gets perturbed and consequently,
according to the PB theorem, a chain of islands is formed on the
surface of section. The resonant tori decay into sets
of stable and unstable orbits, giving rise to
sequences of alternating  elliptic and hyperbolic fixed points.
The elliptic points lead to the emergence of regular islands,
inside which the trajectories are phase-locked and the ratio of the
corresponding frequencies remains equal
to the rational number of the corresponding initial resonant torus.
The hyperbolic points lie on separatrix intersections between the islands,
about which chaotic layers can develop.
\begin{figure}[!t]
\begin{center}
\epsfig{file=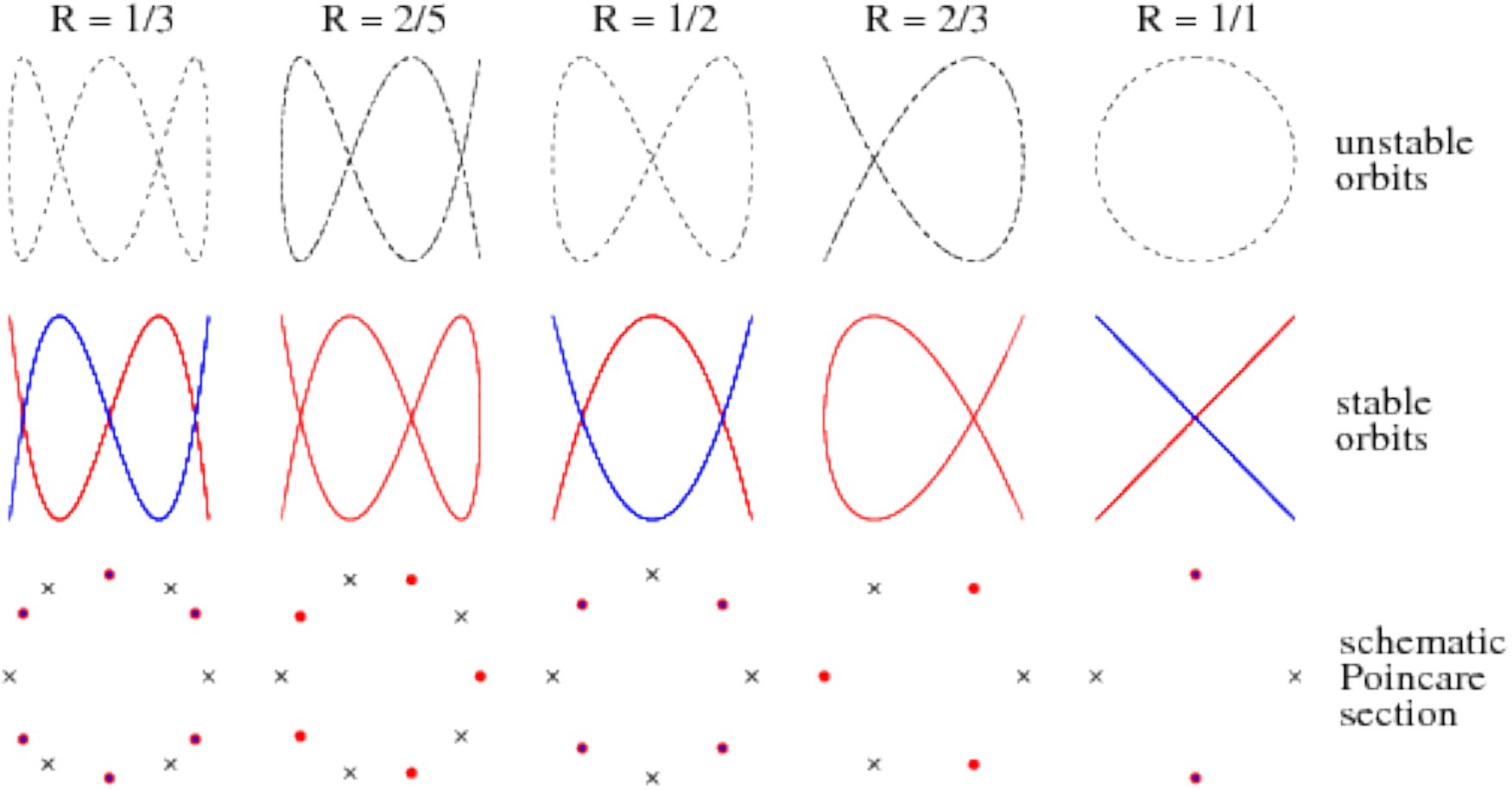,width=0.85\linewidth}
\end{center}
\caption{
Lissajous figures of the simplest stable (middle row)
and unstable (top row) resonant orbits of
${\cal H}_{2}(\xi)$, Eq.~(\ref{eq:H2cl}), with $\bz = \sqrt{2}$.
The bottom row illustrates the Poincar\'e-Birkhoff scenario
of the breakdown of resonant tori.
Sequences of alternating stable (dots) and
unstable (crosses) fixed points are seen in the
Poincar\'e sections due to the stable and unstable orbits
for each particular resonance.}
\label{fig12}
\end{figure}

For the considered classical intrinsic Hamiltonian
${\cal H}_{2}(\xi)$, Eq.~(\ref{eq:H2cl}),
in the deformed region $(\xi\geq\xi_c=0)$, the resonances are reached
when the ratio $R=\epsilon_{\beta}/\epsilon_{\gamma}=m/n$
of Eq.~(\ref{R}), is a rational number. The shape of the resonant orbits
resembles Lissajous figures with the same ratio of frequencies.
For more details on the topology of such orbits, the reader is referred
to~\cite{Cleary90}.
The most pronounced resonances (thicker PB islands)
correspond to small co-prime $(m,n)$ integers, and the number of
islands in a given chain is $2/R$. These features were observed in
Fig.~6, and are shown schematically in Fig.~12,
for $R=1,\,2/3,\,1/2,\,2/5,\,1/3$,
corresponding to $2,\,3,\,4,\,5,\,6$ islands, respectively.
\begin{figure}[!t]
\begin{center}
\epsfig{file=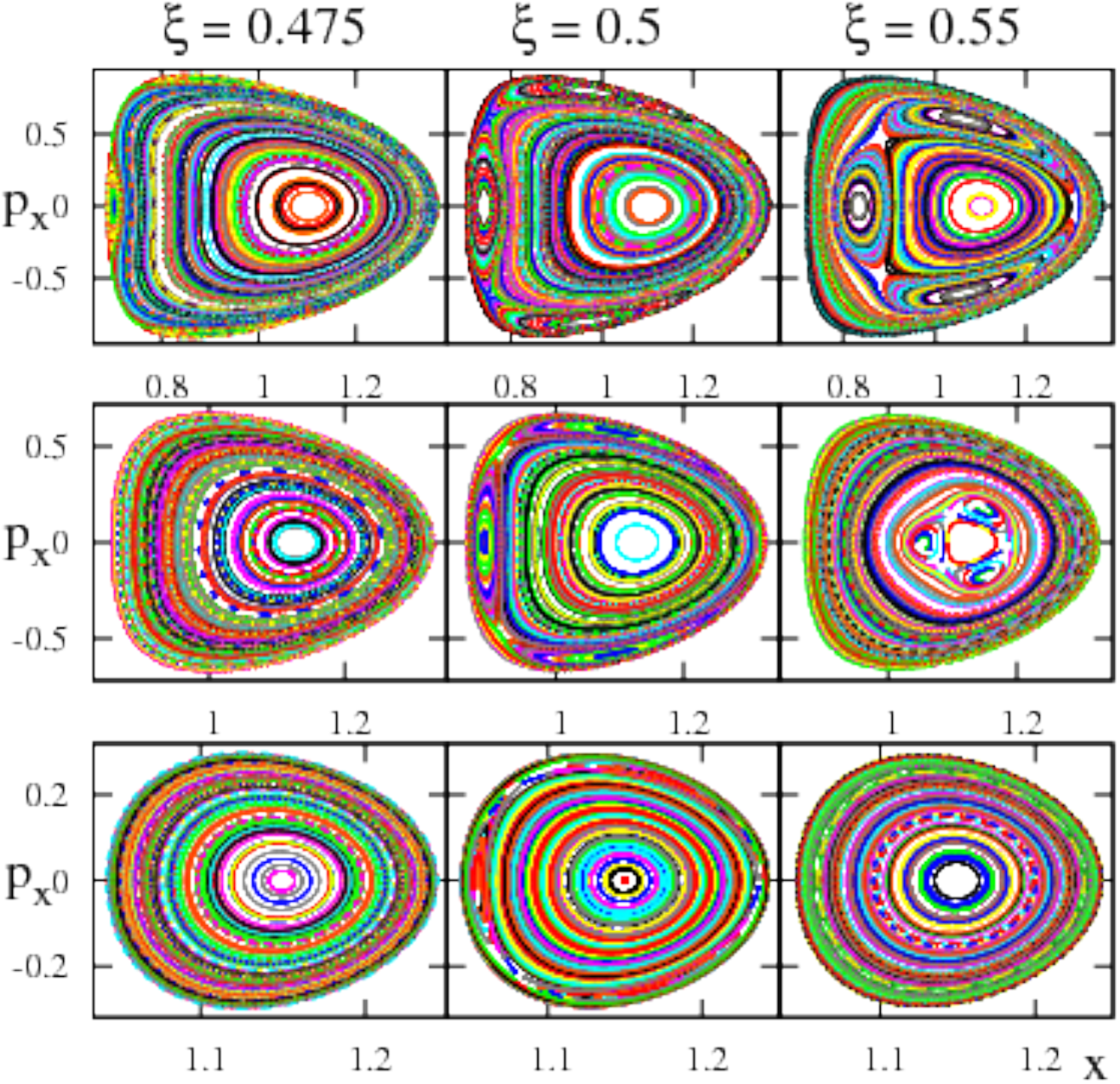,width=0.65\linewidth}
\end{center}
\caption{
Poincar\'e sections near the $R=2/3$ normal-mode resonance.
The three columns refer to values of $\xi=0.475,\,0.5,\,0.55$,
before, at, and after the resonance point, $\xi_R=0.5$, Eq.~(\ref{xiR}).
The bottom, center and top rows, correspond to energies
$E_1 = V_\mathrm{lim}(\xi)/21$, $E_2=5V_\mathrm{lim}(\xi)/21$,
$E_3=9V_\mathrm{lim}(\xi)/21$, respectively,
below the domain boundary $V_\mathrm{lim}(\xi)=(2+\xi)h_2$,
Eq.~(\ref{Vlim2}).}
\label{fig13}
\end{figure}
\begin{figure}[!t]
\begin{center}
\epsfig{file=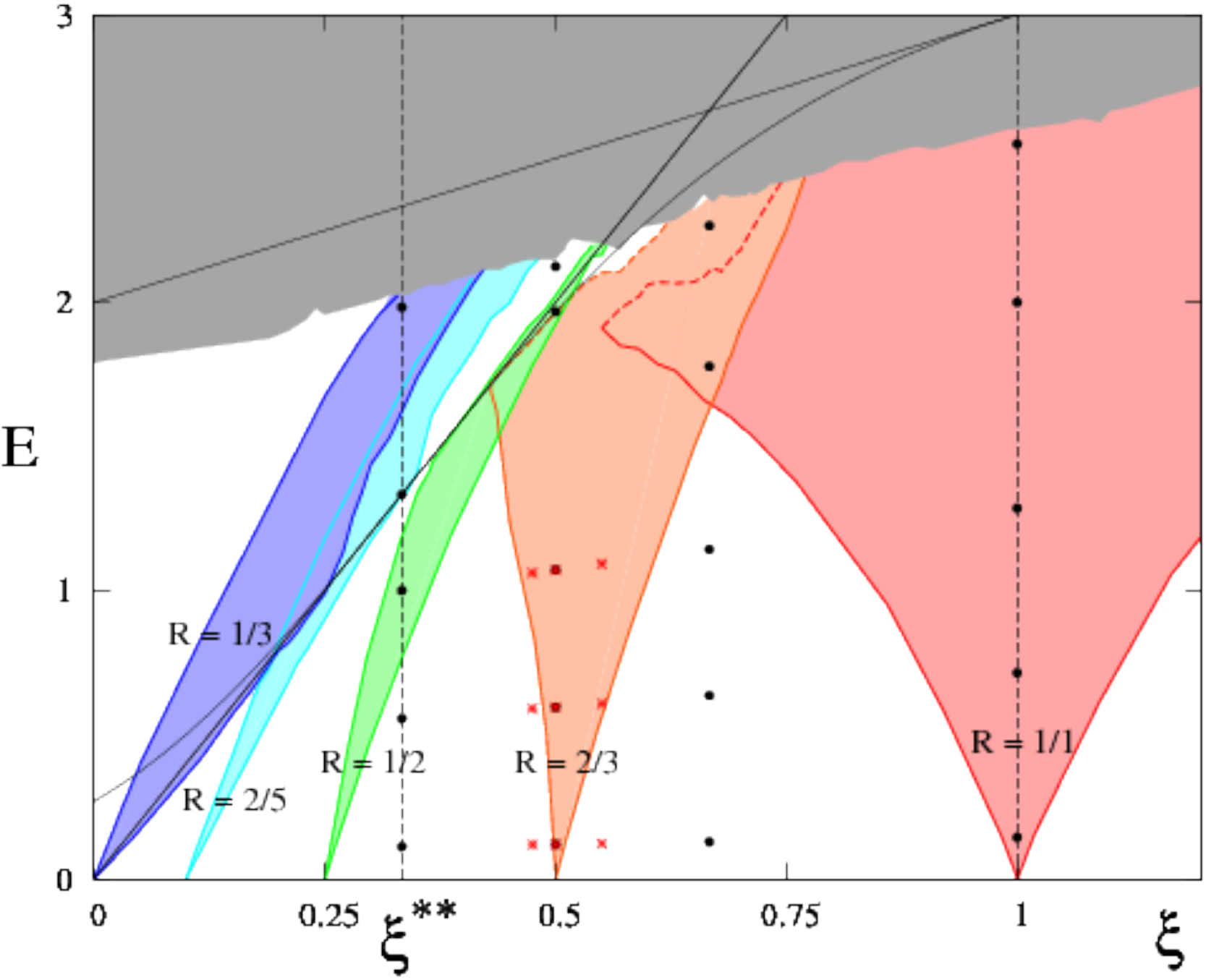,width=0.85\linewidth}
\end{center}
\caption{
Resonance map for the Hamiltonian ${\cal H}_2(\xi)$ with $\bz=\sqrt{2}$
and $h_2=1$, Eq.~(\ref{eq:H2cl}), on the deformed side of the QPT,
$\xi\geq 0$. The color-coded regions indicate the occurrence of major
Poincar\'e-Birkhoff chains of islands in the Poincar\'e sections due to
normal-mode resonances with frequency ratios
$R = 1/3$, $2/5$, $1/2$, $2/3$ and $1/1$,
giving rise to 6, 5, 4, 3, 2 islands, respectively.
As $E\!\rightarrow\! 0$, the colored regions tip exactly towards
the resonant values $\xi_R$ of Eq.~(\ref{xiR}).
White areas involve $(\xi,E)$ domains with a single island.
Black bullets (red stars) correspond to
individual panels in Figs.~10-11 (Fig.~13).
Stationary and boundary values of the potential surface
$V_{2}(\xi)$, Eq.~(\ref{eq:V2}), are marked by thin black lines
(compare with Fig.~3).
The gray area at high energies is inaccessible due to numerical
instability.}
\label{fig14}
\end{figure}

At low energy ($E\rightarrow 0$), where the harmonic approximation is valid,
one expects the resonances to occur
at discrete values of the control parameter $\xi\approx\xi_R$,
in a narrow interval around $\xi_R$,
\ba
\xi_{R} =
\frac{9}{2}(1+\bz^2)^{-1}R - \frac{1}{2} ~,
\label{xiR}
\ea
where the latter is obtained by inverting Eq.~(\ref{R}).
At a finite energy $(E>0)$, anharmonic effects in ${\cal H}_{2}(\xi)$
come into play and, consequently, a PB chain of
islands associated with a given rational $R$ ratio,
can occur in wider ranges of $\xi$ values.

The sensitivity of the classical dynamics to resonance effects is
demonstrated in Fig.~13 near $R=2/3$, where the PB chain consists
of three regular islands. The different columns show the
Poincar\'e sections for $\xi = 0.475,\,0.5,\,0.55$, at energies
$E_1 = V_\mathrm{lim}(\xi)/21$ (bottom row), $E_2 = 5E_1$
(center row), and $E_3 = 9E_1$ (upper row), where
$V_\mathrm{lim}(\xi) = (2+\xi)h_2$, Eq.~(\ref{Vlim2}).
At the resonance point, $\xi_R=0.5$ (middle column), one observes
at all chosen energies, the expected three regular islands
near the perimeter of the Poincar\'e sections,
indicating an instability with respect to the $\beta$-motion.
Their relative size compared to the total area of the section,
increases with energy.
In contrast, the PB islands are not seen at low $E$
neither at $\xi = 0.475$ (see panels for $E = E_1,\,E_2$),
nor at $\xi = 0.55$ (panel for $E=E_1$), where the
Poincar\'e sections display the usual pattern of a single island.
These islands, however, do appear at higher energies,
$E=E_3$ for $\xi=0.475$ and $E=E_2$ for $\xi = 0.55$.
In the latter case, the PB island-chain occurs
near the center of the Poincar\'e section, signaling an instability
with respect to the $\gamma$-motion.

Fig.~14 presents a detailed map of the (colored-coded)
regions in the $(\xi,E)$ plane, where PB chains with
$2,\,3,\,4,\,5,\,6$ islands occur. The latter are
associated with the most pronounced resonances having normal-mode
frequency ratios $R=1,\,2/3,\,1/2,\,2/5,\,1/3$, respectively.
For $E\rightarrow~0$, all the resonance regions end in a sharp tip
at $\xi_R = 0,\, 0.1,\, 0.25,\, 0.5,\,1$, in agreement with
Eq.~(\ref{xiR}). As the energy $E>0$ increases, the resonance regions
are either tilted away from $\xi_R$ (as for $R=1/3,\,2/5,\,1/2$) or
fan out and embrace $\xi_R$ (as for $R=2/3,\,1$).
At higher energies, pairs of regions [$(R=1/3,\,2/5)$, $(R=1/2,\,2/3)$,
$(R=2/3,\,1)$], can overlap, indicating that for a given Hamiltonian
${\cal H}_{2}(\xi)$, Eq.~(\ref{eq:H2cl}), two distinct PB island chains
can occur simultaneously in the Poincar\'e surface.
The white areas outside the color-coded resonance regions,
identify the $(\xi,E)$ domains where the Poincar\'e surfaces
exhibit a single island, without additional PB island chains.
The dominance of these areas for $E\leq 1$ explains why
this simple pattern prevails in most Poincar\'e sections at
low energies.

Fig.~14 is very instrumental for understanding the rich
regular structure arising from the classical intrinsic dynamics
in regions~II and~III of the QPT, for $\xi\geq 0$.
For orientation, a few black bullets are marked in some of the
color-coded resonance regions, corresponding
to particular Poincar\'e sections in Figs.~10-11.
For the critical point, the line $\xi_c=0$ is completely inside a white area
in Fig.~14 and no resonance regions are seen along it,
consistent with the single island observed in the panels of the
$\xi_c=0$ column in Fig.~10.
At the anti-spinodal point $(\xi^{**}=1/3)$, the lowest two bullets
marked in Fig.~14, are located in white areas and the remaining
bullets at higher energies reside inside the
$R=1/2,\,2/5,\,1/3$ resonance regions, consecutively.
This is consistent with the observed surfaces of the
$\xi^{**}=1/3$ column in Fig.~11,
where the lowest two panels display a single island and the remaining
panels in consecutive order, show PB chains with $4,\,5,\,6$ islands.
The Poincar\'e sections of the $\xi=2/3$ column in Fig.~11,
show a single island (lowest three panels) and a PB chain of three islands
in the remaining panels at higher energies. This is again in line with
the location of the bullets for $\xi=2/3$ in Fig.~14.
For the SU(3)-DS limit $(\xi=1)$, the Poincar\'e sections in Fig.~11
display two islands at all energies, consistent with the sole $R=1$ resonance
region embracing the $\xi=1$ line in Fig.~14.

Near the boundaries of each resonance region, the PB islands
are tiny in size. Upon varying $\xi$ and/or $E$ towards the center of a
given region, the islands migrate to the interior of the main regular
island in the respective Poincar\'e sections, and grow in relative size.
Such a scenario is seen clearly in the panels of Fig.~13.
The latter correspond to the red starred points near/inside the $R=2/3$
resonance region in Fig.~14.
The dashed lines marking the high-$E$ boundaries of the $R=2/3$ and $R=1$
resonance regions, indicate the location where the respective PB chains
disappear in the surrounding chaotic sea. Thus, for $\xi=0.5$ in
Fig.~14, the fourth black bullet lies on the dashed line marking
the boundary of the $R=2/3$ resonance region, where the three islands of
the PB chain just disappear in an emerging chaotic layer. Notice,
that the same black bullet lies simultaneously inside the $R=1/2$
resonance region and
indeed, we observe four additional pronounced islands in the
fourth panel from the bottom of the $\xi=0.5$ column in Fig.~11.
In contrast, the fifth black bullet at higher energy for $\xi=0.5$,
lies inside a white area in Fig.~14 and in the corresponding
fifth panel in Fig.~11, we observe just a single regular island,
without any PB island chains, embedded in a significant chaotic
environment.

\section{Quantum analysis}

The analysis of the classical dynamics, constraint to $L=0$,
has revealed a rich inhomogeneous phase space
structure with a pattern of mixed regular and chaotic dynamics, reflecting
the changing topology of the Landau potential across the QPT. It is
clearly of interest to examine the implications of this behavior
to the quantum treatment of the system.
In what follows, we consider the
evolution of levels in the corresponding quantum spectrum and examine
the regular and irregular features of these quantum states.

\subsection{Level evolution}

Fig.~15 shows the correlation diagrams for energies
of $(N=80,L=0)$ eigenstates of the intrinsic Hamiltonian,
Eq.~(\ref{eq:Hint}), with $\bz=\sqrt{2}$, as a function of the control
parameters, $0\leq\rho\leq\rho_c$ (upper portion) and
$\xi_c\leq\xi\leq 1$ (lower portion). The position of the spinodal
point ($\rho^{*}=1/2$) and the anti-spinodal point ($\xi^{**}=1/3$),
is indicated by vertical lines. In-between these points,
inside the coexistence region,
solid lines mark the energies of the barrier ($V_{\rm bar}$) at the
saddle point and of the local minima ($V_{\rm def}$ for
$\rho^{*}<\rho\leq \rho_c$ and $V_{\rm sph}$ for $\xi_c\leq\xi< \xi^{**}$)
in the relevant Landau potential (compare with Fig.~3).

On the spherical side of the QPT, outside of the coexistence region
($0\leq \rho\leq \rho^{*}$),
the spectrum of $\hat{H}_{1}(\rho)$ (\ref{eq:H1}) at low energy,
resembles the normal-mode expression of Eq.~(\ref{nmodesph}),
$E = \epsilon n_d$
($n_d = 0,2,3,\ldots$), with $\epsilon = 4\bar{h}_2N$
independent of $\rho$ (the missing $n_d=1$ state has $L=2$).
As seen in the upper portion of Fig.~15, this low-energy
behavior is observed
also inside the coexistence region ($\rho^{*} <\rho\leq \rho_c$)
at energies $E < V_{\rm def}$ below the local deformed minimum.
Anharmonicities are suppressed by $1/N$, as can be verified by
comparing the spectrum at $\rho=0$ with the U(5)-DS expression,
Eq.~(\ref{U5-DSspec}). At higher energies and $\rho>0$, there  are
noticeable level repulsion and (avoided) level crossing
occurring in the classical chaotic regime.
These effects become more pronounced
as $\rho$ increases and approaches the spinodal point~$\rho^{*}$, and
are due to the U(5) breaking $\rho$-term in Eq.~(\ref{Hrho}).

On the deformed side of the QPT,
outside of the coexistence region ($\xi^{**}\leq \xi\leq 1$),
the levels with $L=0$ serve as bandheads of rotational $K=0$ bands,
associated with the ground band $g(K=0)$ and multiple
$\beta^n\gamma^{2k}(K=0)$ excitations of the prolate-deformed shape.
The low energy spectrum of $\hat{H}_{2}(\rho)$~(\ref{eq:H2})
resembles the normal-mode expression of Eq.~(\ref{nmodedef}),
$E = \epsilon_{\beta}n_{\beta} + \epsilon_{\gamma}n_{\gamma}$
($n_{\beta}=0,1,2,\ldots$ and $n_{\gamma}=0,2,4,6,\ldots$)
with $\epsilon_{\beta} = 4\bar{h}_2N(2\xi+1)$ and
$\epsilon_{\gamma}= 12\bar{h}_2N$ (only bands with $n_{\gamma}$ even
support $L=0$ states).
In particular, bandhead energies involving pure
$\gamma$ excitations are independent
of $\xi$, while bandhead energies involving $\beta$ excitations
are linear in $\xi$, a trend seen in the lower portion of Fig.~15.
Local degeneracies of normal-modes
lead to bunching of energy levels and noticeable voids in the level density,
in the same regions of $(\xi,E)$ shown in the classical resonance map
of Fig.~14.
For $\xi=1$, one has $\epsilon_{\beta}=\epsilon_{\gamma}$
and the spectrum follows the SU(3)-DS expression,
Eq.~(\ref{SU3-DSspec}), with anharmonicities of order $1/N$.
This ordered pattern of levels is observed also inside the coexistence
region ($\xi_c \leq\xi<\xi^{**}$) at energies $E < V_{\rm sph}$ below the
local spherical minimum.

Dramatic structural changes in the level dynamics take place
in the coexistence region
$(\rho^{*} < \rho \leq \rho_c)$ and
$(\xi_c\leq \xi<\xi^{**})$.
As shown in Fig.~15,
at energies above the respective local minima,
($E > V_{\rm sph}$ or $E > V_{\rm def}$),
the spherical type and deformed type of levels
approach each other and their encounter results in marked modifications
in the local level density. In particular, there is
an accumulation of levels near the top of the barrier ($V_{\rm bar}$).
Such singularities in the evolution of the spectrum,
referred to as excited state quantum phase transition~\cite{caprio},
have been encountered in integrable models involving QPTs~\cite{CejStr08}.
In what follows, we plan to examine the regular and irregular features
of these quantum states, and explore how their properties echo
the mixed regular and chaotic dynamics observed in the classical analysis
of the first-order QPT.
\begin{figure}[!t]
\begin{center}
\epsfig{file=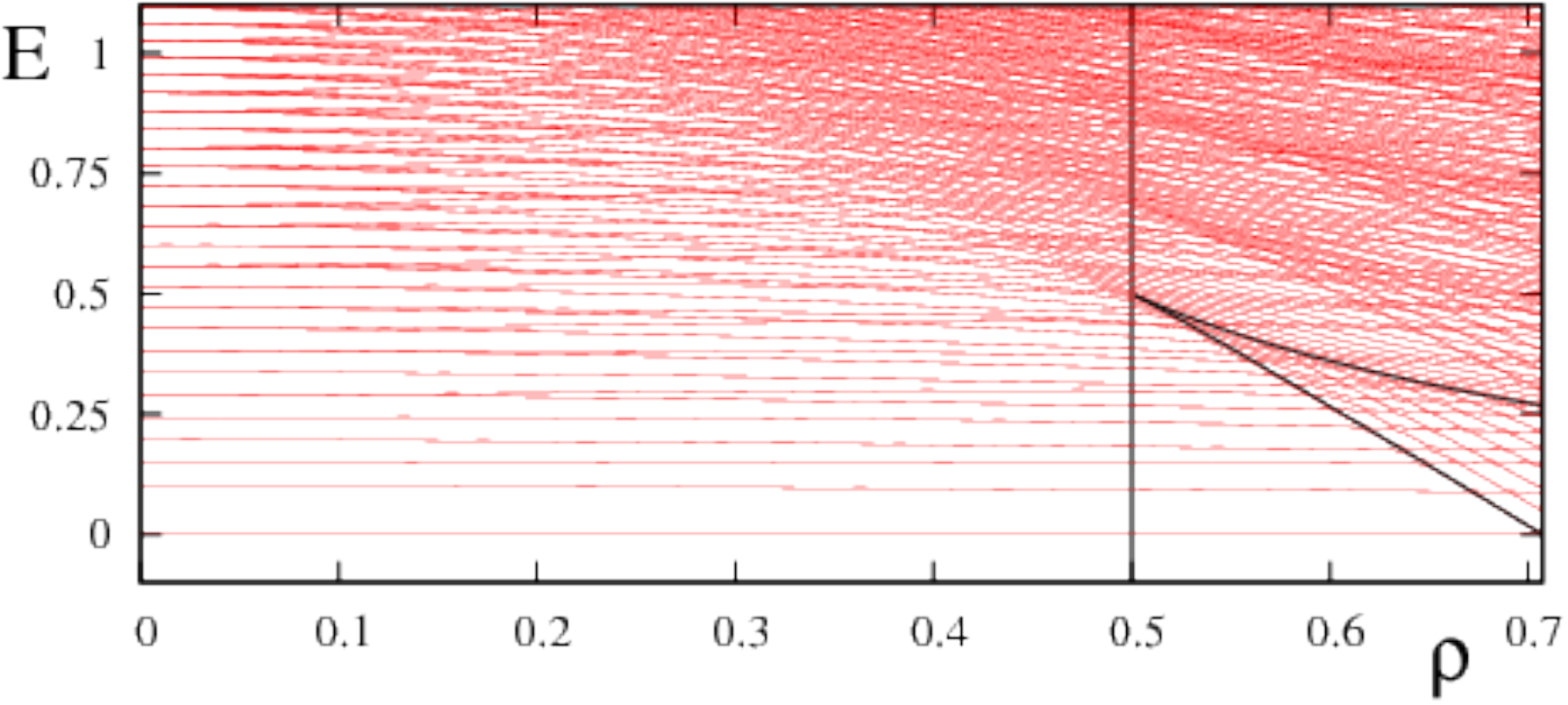,width=0.99\linewidth}\\
\vspace{12pt}
\epsfig{file=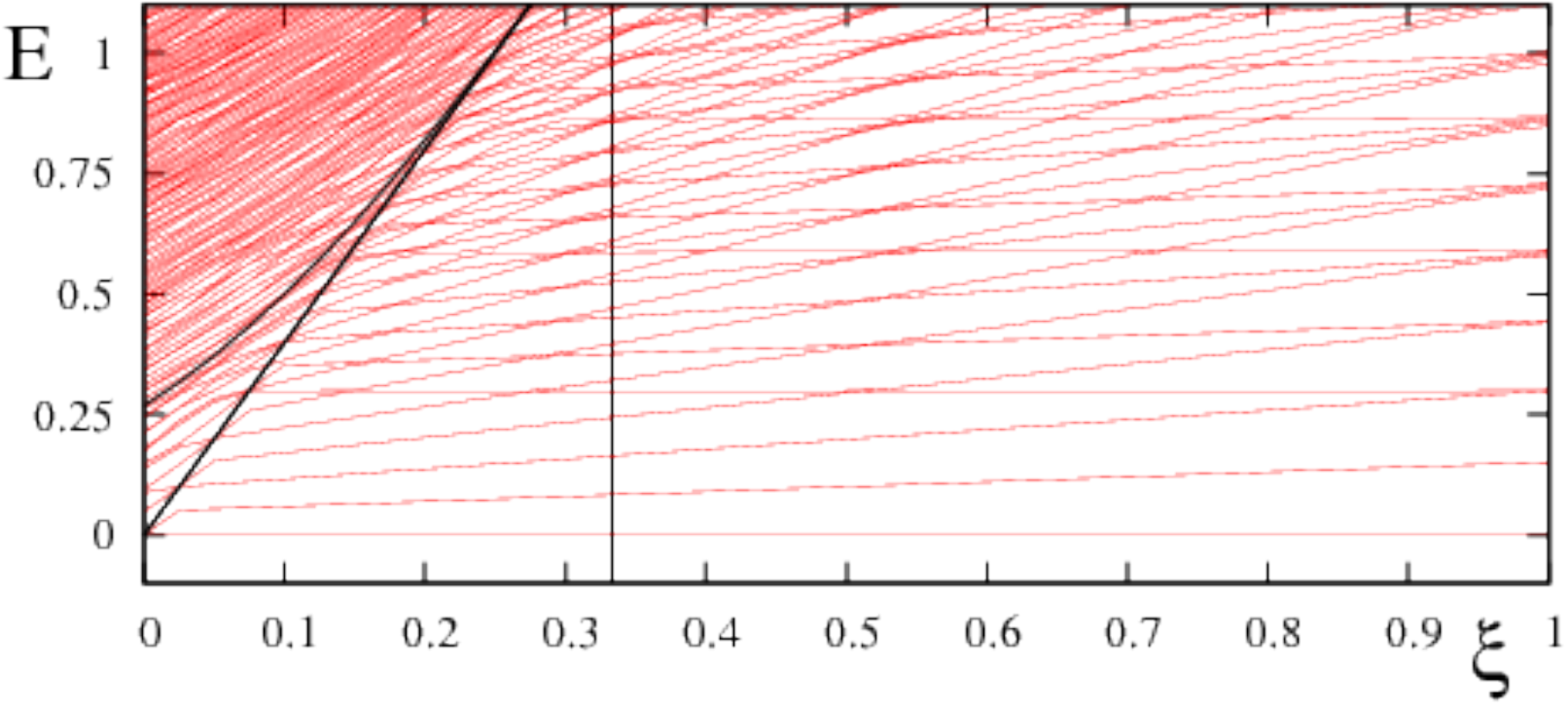,width=0.99\linewidth}
\end{center}
\caption{
The evolution of $L=0$ energy levels for the intrinsic Hamiltonian,
Eq.~(\ref{eq:Hint}), with $\bz=\sqrt{2}$, $h_2=1$ and $N=80$.
Upper portion: the $L=0$
spectrum of $\hat{H}_1(\rho)$, Eq.~(\ref{eq:H1}),
relevant to the spherical side of the QPT ($0\leq\rho\leq\rho_c$).
Here $\rho=0$ corresponds to the U(5)-DS limit, $\rho^{*}=0.5$ is the
spinodal point (marked by a vertical line)
and $\rho_c=1/\sqrt{2}$ is the critical point.
The energies of the barrier ($V_\mathrm{bar}$) at the saddle point
and of the local deformed minimum ($V_\mathrm{def}$) are marked by
solid lines inside the coexistence region ($\rho^{*}<\rho\leq\rho_c$).
Note that in this portion, the lowest $L=0$ level is the solvable U(5)
state with $n_d=0$, Eq.~(\ref{ePDSu5L0}).
Lower portion: the $L=0$
spectrum of $\hat{H}_2(\xi)$, Eq.~(\ref{eq:H2}),
relevant to the deformed side of the QPT ($\xi_c\leq\xi\leq 1$).
Here $\xi_c=0$ is the critical point, $\xi^{**}=1/3$
is the anti-spinodal point (marked by a vertical line)
and $\xi=1$ correspond to the SU(3)-DS limit.
The energies of the barrier ($V_\mathrm{bar}$) at the saddle point
and of the local spherical minimum ($V_\mathrm{sph}$) are marked by
solid lines inside the coexistence region ($\xi_c\leq\xi\leq\xi^{**}$).
Note that in this portion, the lowest $L=0$ level is the solvable SU(3)
state of Eq.~(\ref{solsu3g}).}
\label{fig15}
\end{figure}

\subsection{Peres lattices}

Quantum manifestations of classical chaos are
often detected by statistical analyses of energy
spectra~\cite{Gutz90,Reic92,Licht92}.
In a quantum system with mixed regular and irregular states, the
statistical properties of the spectrum are usually intermediate between
the Poisson and the Gaussian orthogonal ensemble (GOE) statistics.
Such global measures of quantum chaos are, however, insufficient to
reflect the rich dynamics of an inhomogeneous phase space structure
encountered in Fig.~9-11, with mixed but well-separated
regular and chaotic regions.
To do so, one needs to distinguish between regular and irregular
subsets of eigenstates in the same energy intervals.
For that purpose, we employ the spectral lattice method of
Peres~\cite{Peres84}, which
provides additional properties of individual energy eigenstates.
The Peres lattices are constructed by plotting the expectation
values $O_i = \bra{i}\hat{O}\ket{i}$ of an arbitrary operator,
$[\hat{O},\hat{H}]\neq 0$, versus the energy $E_i=\bra{i}\hat{H}\ket{i}$
of the Hamiltonian eigenstates $\ket{i}$.
The lattices $\{O_i,E_i\}$ corresponding to
regular dynamics can be shown to display a regular pattern, while chaotic
dynamics leads to disordered meshes of points.
The method has been recently applied to
the collective model of nuclei~\cite{Stran09a,Stran09b}
and to the IBM~\cite{Macek09,Macek10}.
The ability of the method to distinguish between regular and irregular
states, does not rely on the Peres operators $\hat{O}$ used,
and their choice can be made on physical grounds.
\begin{figure*}[!t]
\vspace{-2cm}
\begin{center}
 \epsfig{file=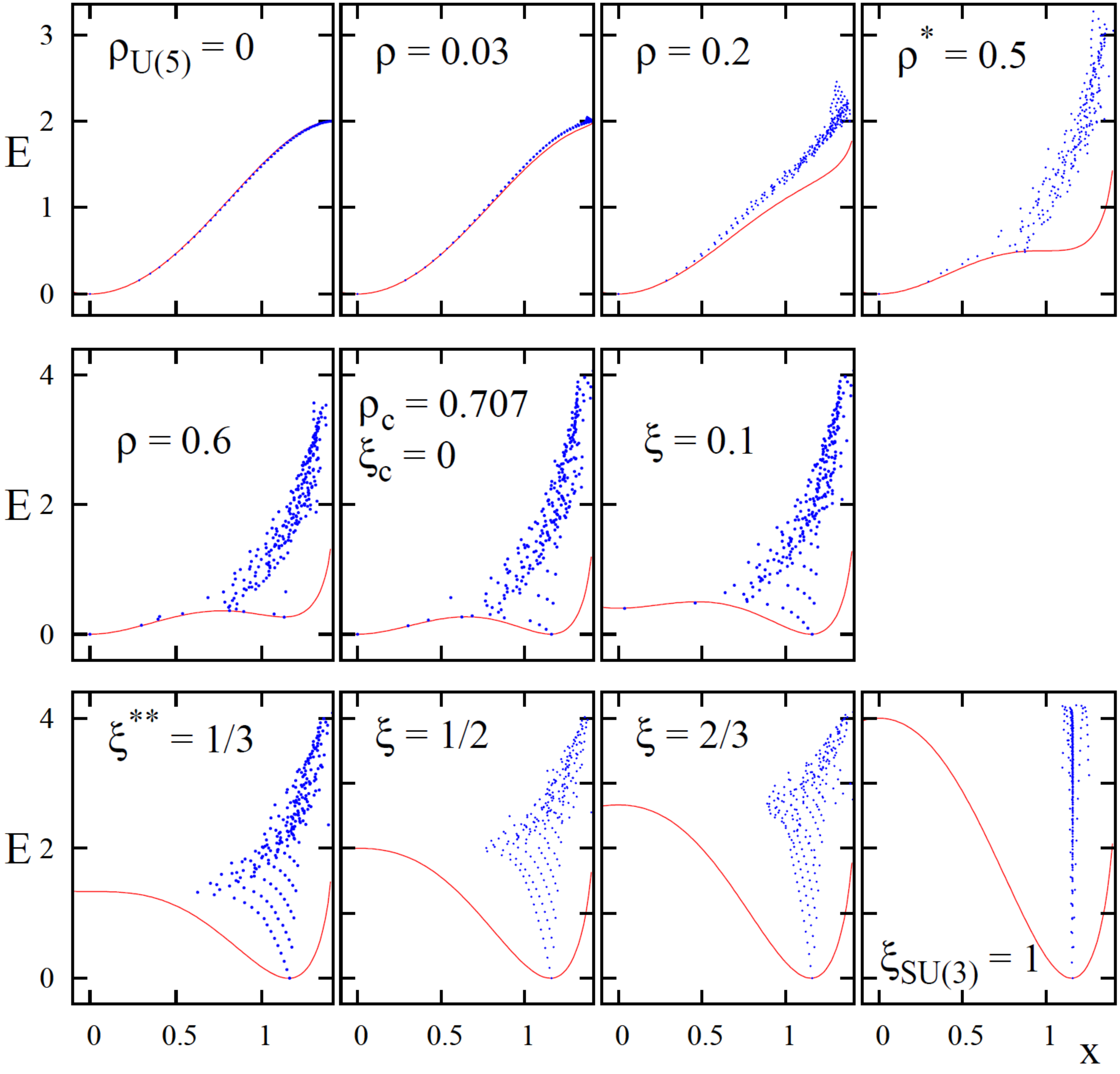,width=0.97\linewidth}
\end{center}
\vspace{-0.68cm}
\caption{
Peres lattices $\{x_i,E_i\}$, Eq.~(\ref{Peresnd}), of $(N=50,L=0)$
eigenstates of the intrinsic Hamiltonian, Eq.~(\ref{eq:Hint}),
with $h_2=1$, $\bz=\sqrt{2}$, for several values of
the control parameters in region~I (top row), region~II (center row)
and region~III (bottom row).
The lattices are overlayed on the classical potential $V(x,y=0)$,
Eq.~(\ref{Vclxy}).}
\label{fig16}
\end{figure*}

In the present analysis, in order
to highlight the classical-quantum correspondence,
we choose $\hat{O}=\hat{n}_d$
and define the Peres lattices 
as the set of points $\{x_i,E_i\}$, with
\ba
x_i \equiv \sqrt{\frac{2\bra{i}\hat{n}_d\ket{i}}{N}} ~,
\label{Peresnd}
\ea
and $\ket{i}$ being
the eigenstates of the IBM Hamiltonian.
The expectation value of $\hat{n}_d$ in the condensate
$\ket{\beta;N} \equiv \ket{\beta,\gamma=0;N}$ of Eq.~(\ref{condgen})
\ba
x = \beta =
\sqrt{\frac{2\bra{\beta;N}\hat{n}_d\ket{\beta;N}}{N}} ~,
\label{xbeta}
\ea
is related to the deformation $\beta$ (whose equilibrium value
is the order parameter of the QPT) and the coordinate
$x$ in the classical potential,
$V(x,y\!=\!0)\!=\!V(\beta,\gamma\!=\!0)$,
Eqs.~(\ref{eq:Vcl}) and (\ref{Vclxy}).
The spherical ground state is the $s$-boson condensate which
has $n_d\!=\!x_i=\!0$. Excited spherical states are obtained,
to a good approximation, by replacing
$s$-bosons in $\ket{\beta=0;N}$  with $d$-bosons,
hence $x_i\sim \sqrt{n_d/N}$ is small for $n_d/N<<1$.
Rotational members of the deformed ground band are obtained by
$L$-projection from  $\ket{\beta;N}$ and have $x_i\approx \beta$
to leading order in $N$.
This relation is still valid, to a good approximation, for states in
excited deformed bands, whose intrinsic states are obtained
by replacing condensate bosons in $\ket{\beta;N}$ with
the orthogonal bosons,
$\Gamma^{\dag}_{_\beta} \!=\! [\sqrt{2-\beta^2}d^{\dag}_0
- \beta s^{\dag}]/\sqrt{2}$ and
$\Gamma^{\dag}_{\gamma,\pm 2} \!=\! d^{\dag}_{\pm 2}$,
representing $\beta$ and $\gamma$
excitations~\cite{lev87,lev85}.
These attributes have the virtue that the
chosen lattices $\{x_i,E_i\}$ of Eq.~(\ref{Peresnd}),
can identify the regular/irregular quantum states and
associate them with a given region in the classical phase space.

\subsection{Evolution of the quantum dynamics across the QPT}

The Peres lattices for $(N\!=\!50,L\!=\!0)$ eigenstates of
the intrinsic Hamiltonian~(\ref{eq:Hint})
with $\bz\!=\!\sqrt{2}$ and $h_2\!=\!1$, are shown
in Fig.~16,
portraying the quantum dynamics across~the QPT
in regions I-II-III. To facilitate the
comparison of the quantum and classical analyses,
the Peres lattices $\{x_i,E_i\}$ of Eq.~(\ref{Peresnd})
are overlayed on the classical potentials $V(x,y=0)$ of Eq.~(\ref{Vclxy}).
These are the same potentials shown at the bottom rows in Figs.~9-10-11,
depicting the classical dynamics in these regions.

The top row of Fig.~16 displays the evolution of quantum Peres lattices
in the stable spherical phase (region~I) for the same values of
the control parameter $\rho\in [0,\rho^{*}]$
in $\hat{H}_{1}(\rho)$, Eq.~(\ref{eq:H1}), as in the classical Poincar\'e
sections of Fig.~9.
For $\rho\!=\!0$, the Hamiltonian~(\ref{H1u5})
has U(5) DS with a solvable spectrum
$E_i \!=\! 2\bar{h}_2[2N\!-\!1 - n_d]n_d$, Eq.~(\ref{U5-DSspec}).
For large $N$, and replacing $x_i$ by $\beta$, the Peres
lattice coincides with $V_{1}(\rho=0)$, Eq.~(\ref{V1rho0}),
a~trend seen for $\rho=0$ (full regularity) and $\rho=0.03$
(almost full regularity) in the top row of Fig.~16.
For $\rho=0.2$, at low energy a few lattice points still follow
the potential curve $V_{1}(\rho)$, but at higher energies
one observes sizeable deviations and disordered meshes of lattice points,
in accord with the onset of chaos in the classical
H\'enon-Heiles system considered in Fig.~9. The disorder in the
Peres lattice enhances at the spinodal point $\rho^{*}=0.5$,
where the chaotic component of the classical dynamics maximizes.

The center row of Fig.~16 displays the evolution of quantum Peres
lattices in the region of phase-coexistence (region~II) for
$\rho\in (\rho^{*},\rho_c]$ in $\hat{H}_{1}(\rho)$, Eq.~(\ref{eq:H1}),
and $\xi\in[\xi_c,\xi^{**})$ in $\hat{H}_{2}(\xi)$, Eq.~(\ref{eq:H2}).
The calculations shown are for the same values of control parameters
used in the classical analysis in Fig~10.
The occurrence of a second deformed minimum in the potential
is signaled by the occurrence of regular sequences of
states localized within and above the deformed well.
They form several chains of lattice points
close in energy, with the lowest chain
originating at the deformed ground state.
A close inspection reveals that the $x_i$-values
of these regular states, lie in the intervals of $x$-values
occupied by the regular tori in the Poincar\'e sections
in Fig.~10. Similarly to the classical tori, these regular
sequences persist to energies well above the barrier $V_{\rm bar}$.
The lowest sequence consists of
$L\!=\!0$ bandhead states of
the ground $g(K=0)$ and $\beta^n(K=0)$ bands.
Regular sequences at higher energy correspond to
multi-phonon $\beta^n\gamma^{2m}(K=0)$ bands.
In contrast, the remaining states,
including those residing in the spherical minimum,
do not show any obvious patterns and lead to
disordered (chaotic) meshes of points at high energy $E>V_{\rm bar}$.

The bottom row of Fig.~16 displays the Peres lattices
in the stable deformed phase (region~III) for $\xi\in [\xi_c,1]$,
and is the quantum counterpart of Fig.~11.
No lattice points are seen at small values of $x<0.5$, beyond
the anti-spinodal point $\xi^{**}$, where the spherical minimum disappears.
On the other hand, more and longer regular sequences of $K=0$ bandhead
states are observed in the vicinity of the single deformed
minimum ($x\approx 1$) as its depth increases.
These sequences tend to be more aligned above the center of the potential
well, as $\xi$ progresses from $\xi^{**}$ towards the SU(3) limit ($\xi=1$).
A close inspection reveals slight dislocations in the ordered pattern of
lattice points for those values of ($\xi,E$), mentioned in Section~5,
corresponding to a resonance.

\begin{figure*}[!t]
\begin{center}
\epsfig{file=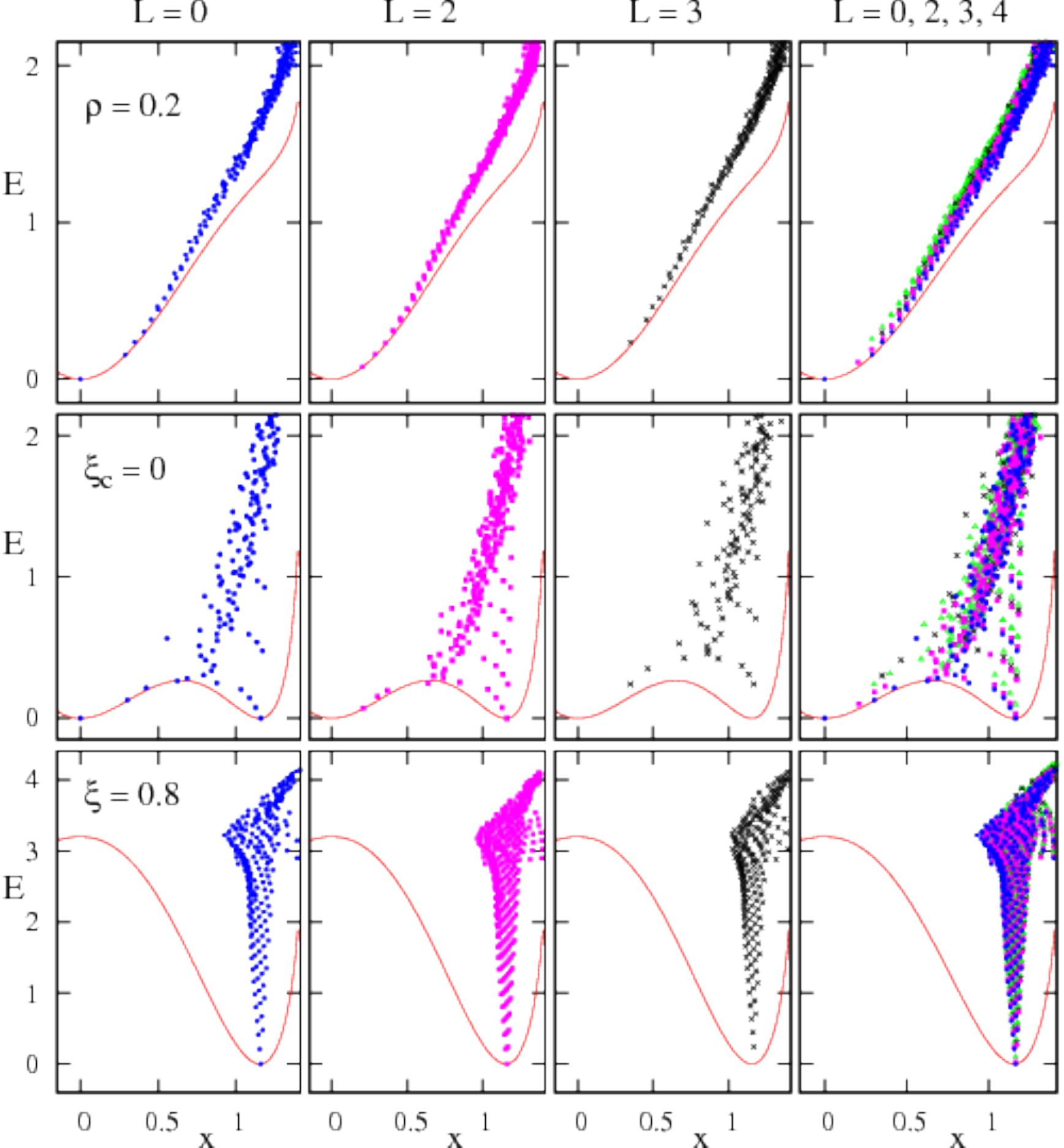,width=0.9\linewidth}
\end{center}
\caption{
Peres lattices $\{x_i,E_i\}$, Eq.~(\ref{Peresnd}),
for $L=0,2,3,4$, eigenstates
of the intrinsic Hamiltonian (\ref{eq:Hint})
with $h_2=1,\,\bz=\sqrt{2},\,N=50$. The lattices
in region~I $(\rho=0.2)$,
region~II $(\xi_c=0)$ and region~III $(\xi=0.8)$
are overlayed on the corresponding
classical potential $V(x,y=0)$, Eq.~(\ref{Vclxy}).
The right column combines
the separate-$L$ lattices.}
\label{fig17}
\end{figure*}
\begin{figure*}[!t]
\begin{center}
\epsfig{file=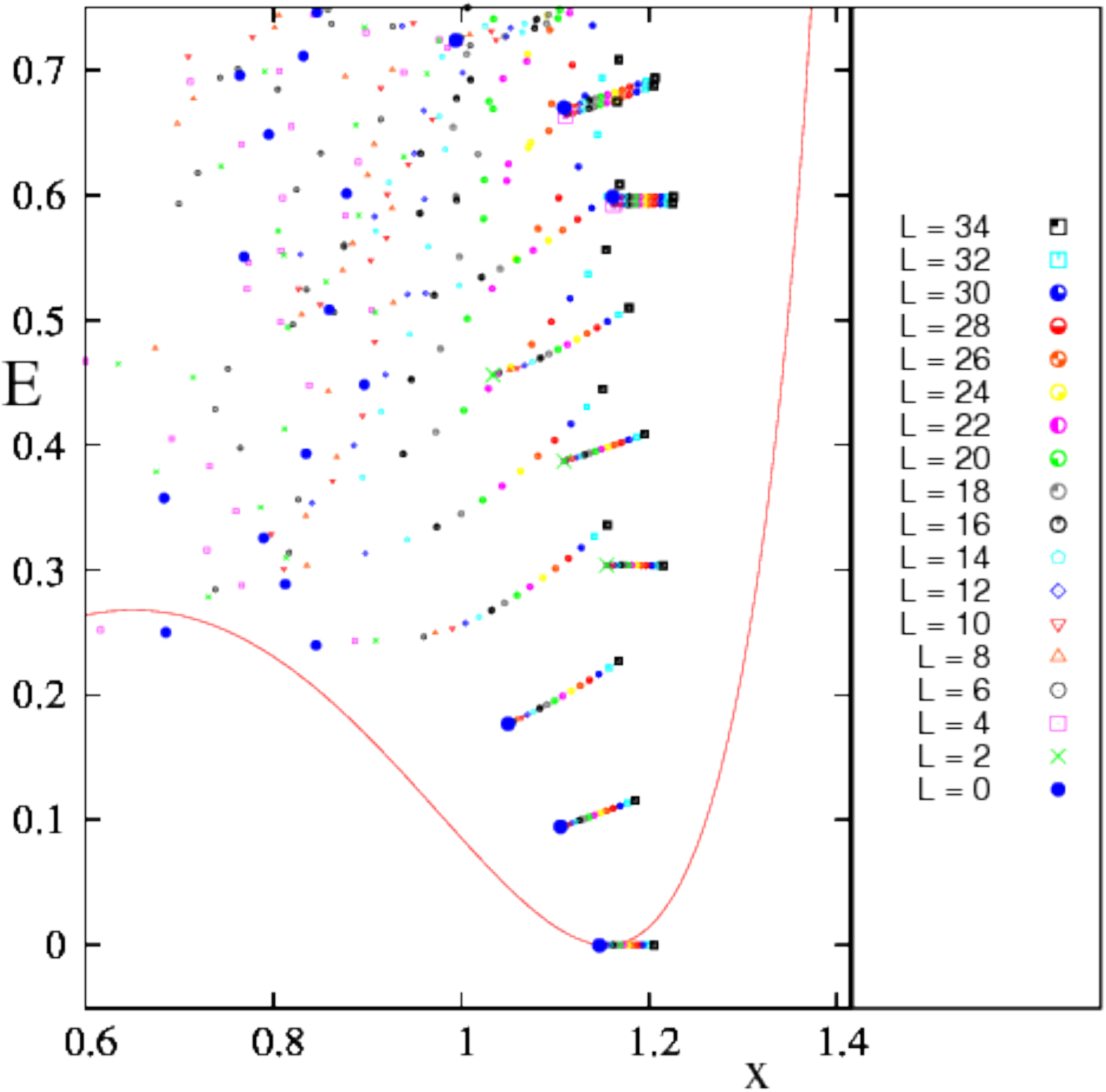,width=0.9\linewidth}
\end{center}
\caption{
Combined Peres lattices $\{x_i,E_i\}$, Eq.~(\ref{Peresnd}),
for $L=0,2,4,\dots,34$, eigenstates
of the intrinsic critical-point Hamiltonian, Eq.~(\ref{Hcri}), with
$h_2=1,\,\bz=\sqrt{2},\,N=40$.}
\label{fig18}
\end{figure*}

Unlike the Poincar\'e sections of the classical analysis,
the Peres spectral method can be used to visualize also the dynamics of
quantum states with non-zero angular momenta.
Examples of such Peres lattices of states with $L=0,\,2,\,3,\,4$
are shown in Fig.~17 for representative values of the control
parameters in region~I ($\rho=0.2$), region~II ($\xi_c=0$) and
region~III ($\xi=0.8$). The right column in the figure combines
the separate-$L$ lattices and overlays them on the relevant
classical potential. For $\rho=0.2$, at low energies typical
of the regular H\'enon Heiles (HH) regime, one can identify multiplets of
states with $L\!=\!0$, $L\!=\!2$, $L\!=\!0,2,4$,
similar to the lowest U(5) multiplets of Eq.~(\ref{u5mult}).
As will be discussed in Section~7,
their wave functions show the dominance of a single $n_d$ component
($n_d=0,1,2$, respectively), characteristic of a spherical vibrator.
No such multiplet structure can be detected at higher energy in
the chaotic HH regime.
Interestingly, a small number of low-energy U(5)-like multiplets
persists in the coexistence region, to the left of the barrier towards the
spherical minimum, as seen in the Peres lattice for the critical
point, $\xi_c=0$, in Fig.~17.

In regions II and III one can detect
the rotational states with $L=0,2,4,\ldots$, comprising the
regular $K\!=\!0$ bands mentioned above.
Additional $K$-bands with $L\!=\!K,K\!+\!1,K\!+\!2,\ldots$,
corresponding to multiple $\beta$ and $\gamma$ vibrations
about the deformed shape, can also be identified.
These ordered band structures show up in the vicinity of the deformed well
and are not present in the chaotic portions of the Peres lattice.
The panels for $\xi_c=0$ in Fig.~17 demonstrate the occurrence of such
regular $K\!=\!0,2,4$ bands inside the coexistence region (region~II),
alongside with other irregular states represented by the disordered
meshes of points in the Peres lattice.
The panels for $\xi=0.8$ in Fig.~17 indicate that
in region~III, as the single deformed minimum becomes deeper,
the regular $K$-bands exhaust larger portions of the Peres lattice.
Generally, the states in each regular band share a common intrinsic
structure as indicated by their nearly equal values of
$\langle \hat{n}_d \rangle$ and a similar coherent decomposition
of their wave functions in the SU(3) basis, to be discussed in Section~7.
The regular bands extend to high angular momenta as demonstrated
for the critical point in Fig.~18.
While it is natural to find regular rotational bands
in a region with a single well-developed deformed minimum,
their occurrence in the coexistence region,
including the critical point, is somewhat unexpected, in view of
the strong mixing and abrupt structural changes taking place.
Their persistence in the spectrum to energies well above the barrier
and to high values of angular momenta, amidst a complicated environment,
validates the relevance of an adiabatic separation of intrinsic and
collective modes~\cite{MDSC10}, for a subset of states.

To conclude, the classical and quantum analyses presented so far,
indicate that the variation of the control parameters ($\rho,\xi$) in
the intrinsic Hamiltonian, induces a change in the topology of the Landau
potential across the QPT which, in turn, is correlated with an
intricate interplay of order and chaos. For the considered Hamiltonian,
whenever a spherical minimum occurs in the potential,
the system exhibits an anharmonic oscillator (AO) type of
dynamics for small $\rho$, and a H\'enon Heiles (HH) type of
dynamics at larger values of~$\rho$. While the AO dynamics is
regular, the HH dynamics shows a variation
with energy from regular to chaotic character, which is reflected in
the Peres lattices by a change from ordered to disordered patterns.
Whenever a deformed minimum occurs in the potential, the Peres lattices
display regular rotational bands
localized in the region of the deformed well and corresponding
to the regular islands in the classical Poincar\'e sections.
In the coexistence region, these regular bands
persist to energies well above the barrier
and are well separated from the remaining states, which form
disordered meshes of lattice points in the classical chaotic regime.
The system in the domain of phase coexistence,
thus provides a clear cut demonstration of the classical-quantum
correspondence of regular and chaotic behavior, illustrating Percival's
conjecture concerning the distinct properties of regular and irregular
quantum spectra~\cite{Perc73}.

\section{Symmetry aspects}
\label{sec:SymAsp}

The intrinsic Hamiltonian, Eq.~(\ref{eq:Hint}), with $\bz=\sqrt{2}$,
interpolates between the U(5)-DS limit ($\rho=0$)
and the SU(3)-DS limit ($\xi=1$).
Away from these limits, ($\rho>0$ and $\xi<1$), both dynamical symmetries
are broken and the competition between terms in the Hamiltonian with
different symmetry character, drives the system through a first-order QPT.
It is of great interest to study the symmetry properties of the
Hamiltonian eigenstates and explore how they echo the observed
interplay of order and chaos accompanying the QPT.

The preceding quantum analysis has revealed regular
SU(3)-like sequences of states which persist in the deformed region
and, possibly, U(5)-like multiples which persist at low-energy in
the spherical region. It is natural to seek a symmetry-based
explanation for the survival of such regular subsets of states, in the
presence of more complicate type of states. In what follows, we show
that partial dynamical symmetry (PDS) and quasi-dynamical symmetry (QDS)
can play a clarifying role. They reflect, respectively, the enhanced
purity and coherence, observed in the wave functions of these
selected states.

\begin{figure}[!t]
\begin{center}
\epsfig{file=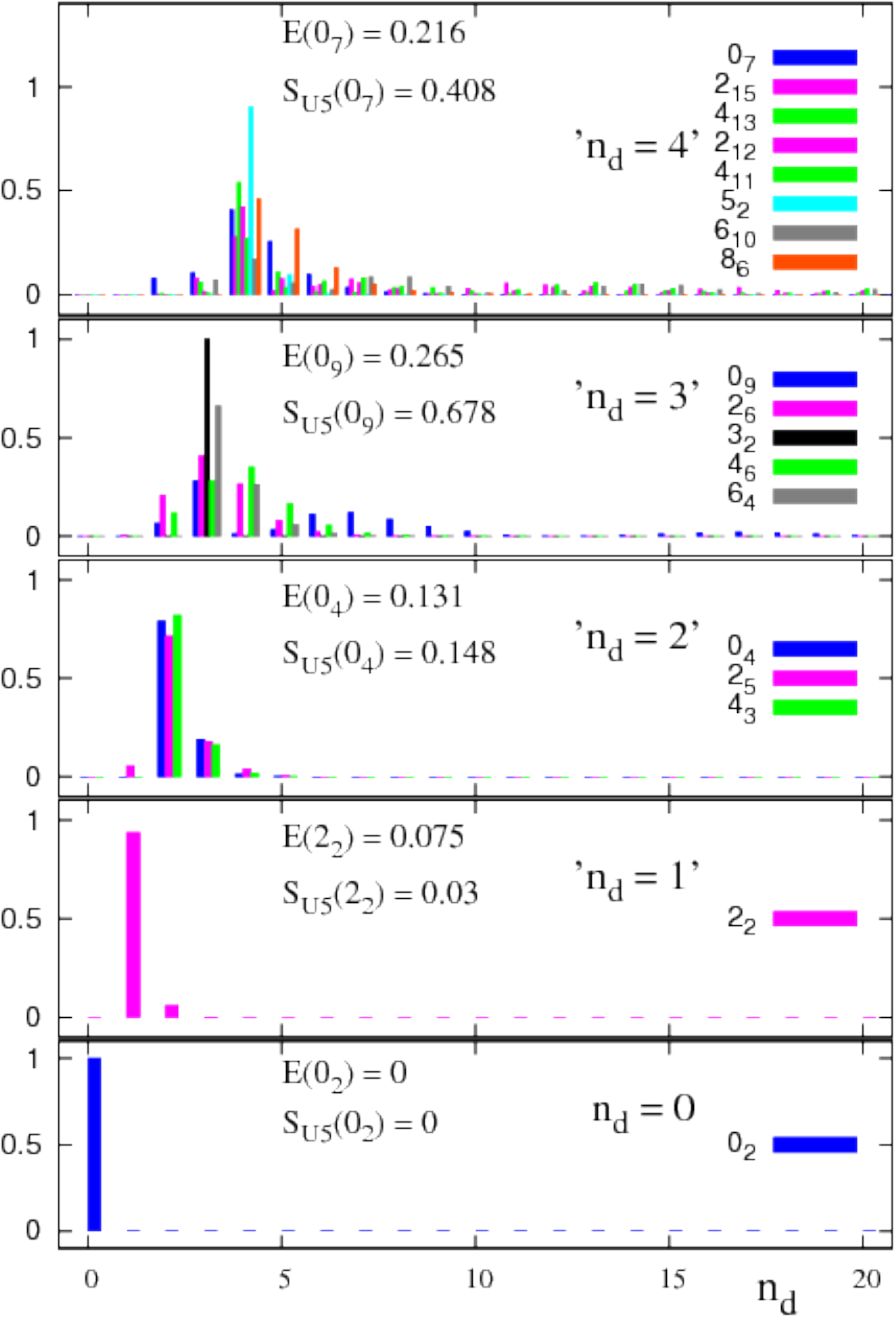,width=0.49\linewidth}
\epsfig{file=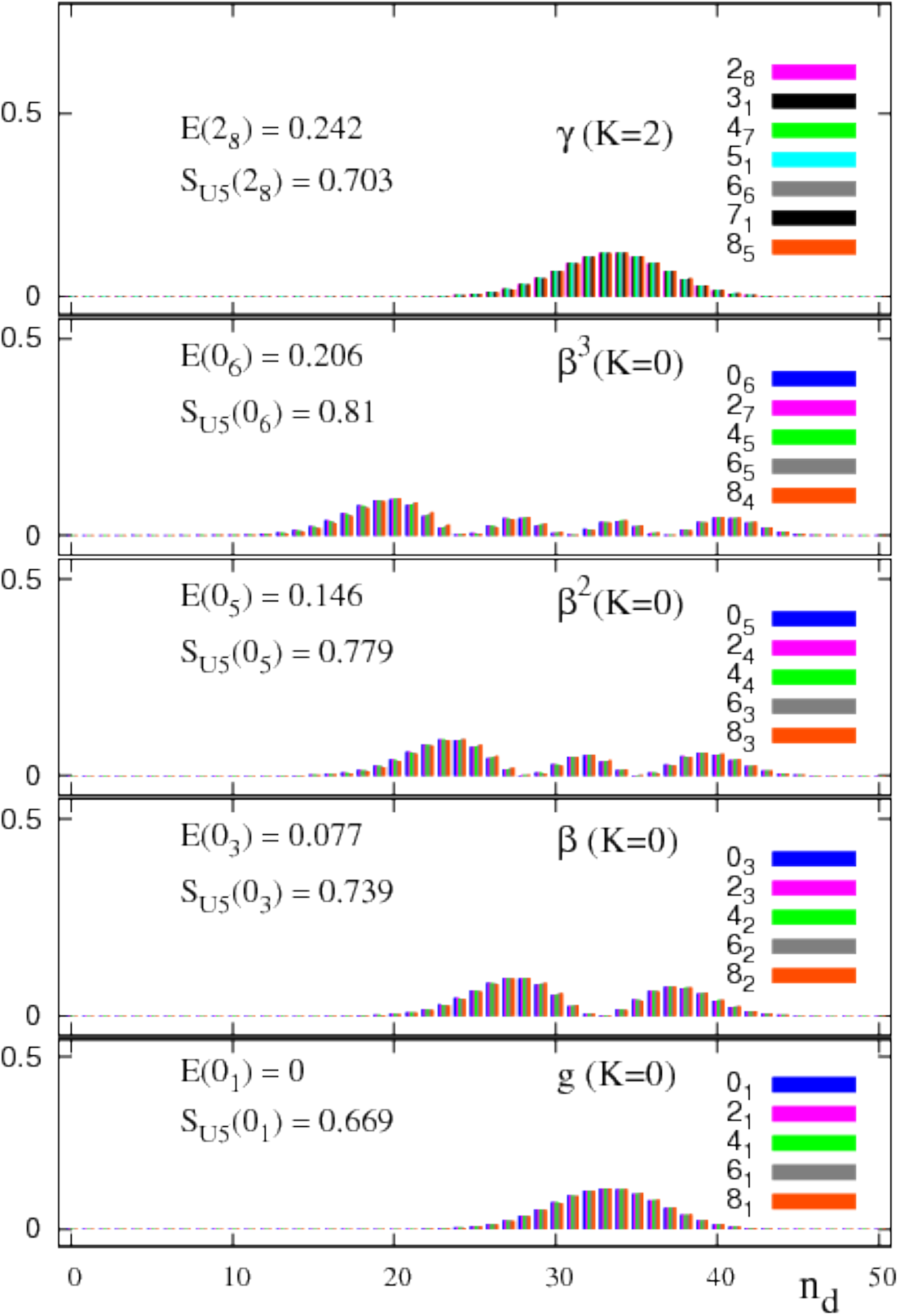,width=0.49\linewidth}
\end{center}
\caption{
U(5) ($n_d$) decomposition for selected eigenstates of the
intrinsic critical-point Hamiltonian, Eq.~(\ref{Hcri}), with
$h_2=1,\,\bz=\sqrt{2},\,N=50$.
Left column: $n_d$-probability distribution $P^{(L_i)}_{n_d}$,
Eq.~(\ref{Pnd}), for spherical-type of states arranged in
U(5)-like multiplets, `$n_d$', with a maximal $n_d=0,1,2,3,4$
component. Note the dominance of a single $n_d$-component
for states in the multiplets with `$n_d\leq 2$'.
Right column:  $P^{(L_i)}_{n_d}$ for deformed-type of states,
members of rotational $K$-bands, showing a broad $n_d$-distribution.
Each panel lists the energy and U(5) Shannon entropy $S_{\rm U5}(L_i)$,
Eq.~(\ref{Shannonu5}), of a representative state in the multiplet.}
\label{fig19}
\end{figure}

A number of works~\cite{WAL93,LevWhe96} have shown that
PDSs can cause suppression of chaos
even when the fraction of states which has the symmetry vanishes in
the classical limit.
SU(3) QDS has been proposed~\cite{Bona10} to underly
the ``arc of regularity''~\cite{AW91b},
a narrow zone of enhanced regularity in the
parameter-space of the IBM Hamiltonian, Eq.~(\ref{HndQQ}).
In conjunction with first-order QPTs, both U(5) and SU(3) PDSs
were shown to occur at the critical point~\cite{lev07}.
The QDS notion
was originally applied to properties of selected low-lying states
outside the coexistence region~\cite{RosRow05}.
Later works~\cite{Macek09,Macek10} have demonstrated the relevance of
SU(3) QDS not only to the ground band, but also to high-lying
bands in the stable deformed phase, with a single deformed minimum.
In what follows, we show that the PDS and QDS notions
can be used also inside the coexistence region
of the QPT and serve as fingerprints for structural
changes throughout this region. Their measures can uncover the survival of
order in the face of a chaotic environment.

\begin{figure}[!t]
\begin{center}
\epsfig{file=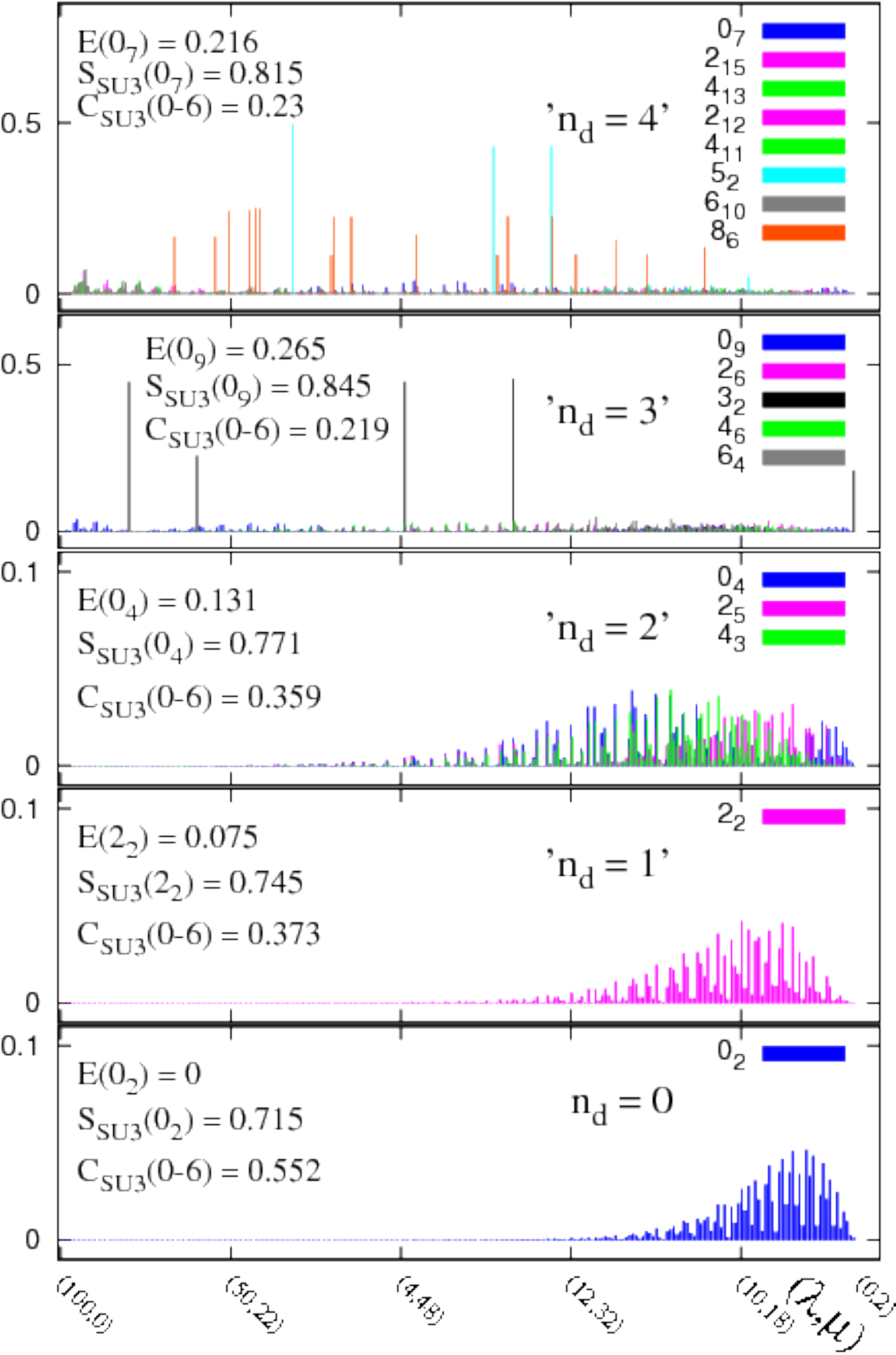,width=0.49\linewidth}
\epsfig{file=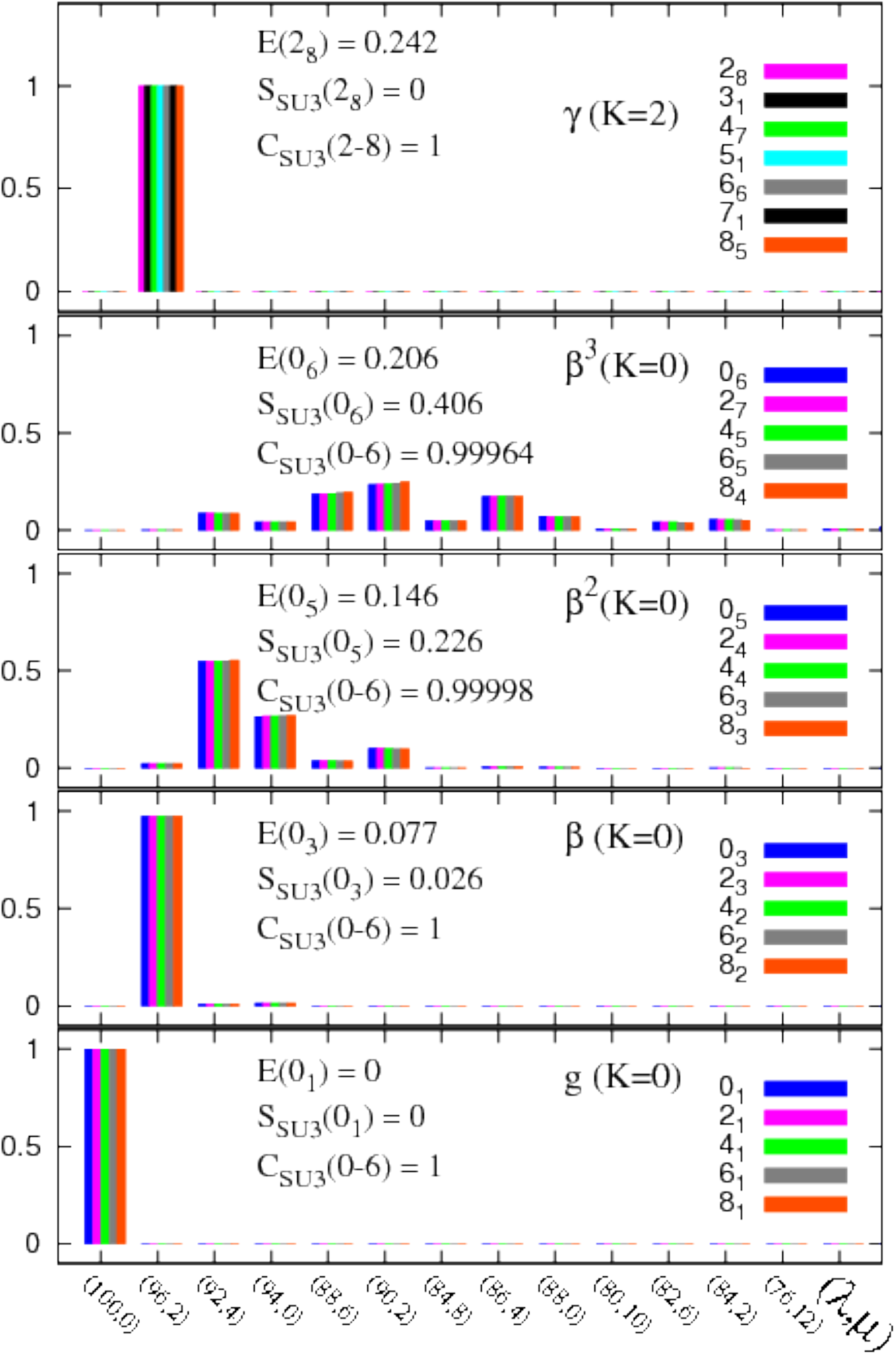,width=0.49\linewidth}
\end{center}
\caption{
SU(3) [$(\lambda,\mu)$] decomposition for the same eigenstates of the
intrinsic critical-point Hamiltonian as in Fig.~19.
Left column: $(\lambda,\mu)$-probability distribution
$P^{(L_i)}_{(\lambda,\mu)}$,
Eq.~(\ref{Plammu}), for spherical-type of states arranged in
`$n_d$' multiplets.
Right column: $P^{(L_i)}_{(\lambda,\mu)}$, for deformed-type of states
arranged in rotational bands. Note the coherent ($L$-independent)
mixing for states in the same band.
Each panel lists the energy, SU(3) Shannon entropy $S_{\rm SU3}(L_i)$,
Eq.~(\ref{Shannonsu3}) and SU(3) correlator $C_{\rm SU3}(0{\rm -}6)$,
Eq.~(\ref{Pearson}), for representative states in the multiplet.}
\label{fig20}
\end{figure}

\subsection{Decomposition of wave functions in the dynamical symmetry bases}

Consider an eigenfunction of the IBM Hamiltonian, $\ket{L_i}$, with
angular momentum $L$ and ordinal number $i$ (enumerating the
occurrences of states with the same $L$, with increasing energy).
Its expansion in the U(5) DS basis, $\ket{N,n_d,\tau,n_{\Delta},L}$,
of Eq.~(\ref{u5ds}) and in the SU(3) DS basis,
$\ket{N,(\lambda,\mu),K,L}$, of Eq.~(\ref{su3ds}) reads
\ba
\ket{L_i}  &=& \sum_{n_d,\tau,n_{\Delta}}C^{(L_i)}_{n_d,\tau,n_{\Delta}}
\ket{N,n_d,\tau,n_{\Delta},L_i\,} ~,
\nonumber\\
&=&
\sum_{(\lambda,\mu),K}C^{(L_i)}_{(\lambda,\mu),K}
\ket{N,(\lambda,\mu),K,L_i\,} ~,
\label{Li}
\ea
where, for simplicity, the dependence of $\ket{L_i}$ and
the expansion coefficients on $N$ is suppressed.
The U(5) ($n_d$) probability distribution, $P_{n_d}^{(L_i)}$,
and the SU(3) [$(\lambda,\mu)$] probability distribution,
$P_{(\lambda,\mu)}^{(L_i)}$, are calculated as
\bsub
\ba
P_{n_d}^{(L_i)} &=& \sum_{\tau,n_{\Delta}}
\vert C^{(L_i)}_{n_d,\tau,n_{\Delta}}\vert^2 ~,
\label{Pnd}\\
P_{(\lambda,\mu)}^{(L_i)} &=& \sum_{K}
\vert C^{(L_i)}_{(\lambda,\mu),K}\vert^2 ~.
\label{Plammu}
\ea
\label{Pndlammu}
\esub
The sum in Eq.~(\ref{Pnd}) runs over the O(5) labels
($\tau,n_{\Delta}$) compatible
with the ${\rm U(5)} \supset {\rm O(5)} \supset {\rm O(3)}$ reduction
and the sum in Eq.~(\ref{Plammu}) runs over the multiplicity label $K$,
compatible with the ${\rm SU(3)} \supset {\rm O(3)}$ reduction.

The quantity $P_{n_d}^{(L_i)}$ (\ref{Pnd}) provides
considerable insight on the nature of states.
This follows from the observation
that ``spherical'' type of states show
a narrow distribution, with a characteristic dominance of single $n_d$
components that one would expect
for a spherical vibrator. In contrast, ``deformed'' type of states show
a broad $n_d$-distribution typical of a deformed rotor structure.
This ability to distinguish different types of states, is illustrated
for eigenstates of the critical-point Hamiltonian in Fig~19.

The states shown on the left column of Fig.~19,
were selected on the
basis of having the largest components with $n_d=0,1,2,3,4$,
within the given $L$ spectra. States with different $L$ values
are arranged into panels labeled by `$n_d$' to conform with the
structure of the $n_d$-multiplets of the U(5) DS limit, Eq.~(\ref{u5mult}).
Each panel depicts the $n_d$-probability, $P_{n_d}^{(L_i)}$,
for states in the multiplet and lists the energy
of a representative eigenstate.
In particular,
the zero-energy $L=0^{+}_2$ state is seen to be a pure $n_d=0$ state
which is the solvable U(5)-PDS eigenstate of Eq.~(\ref{ePDSu5L0}).
The state $2^{+}_2$ has a pronounced $n_d=1$
component (96\%) and the states ($L=0^{+}_4,\,2^{+}_5,\,4^{+}_3$)
in the third panel, have a pronounced $n_d=2$ component.
All the above states with `$n_d\leq 2$' have a dominant single $n_d$
component, and hence qualify as `spherical' type of states.
These multiplets comprise the lowest left-most states shown in the
combined Peres lattices for $\xi_c=0$ in Fig.~17.
In contrast, the states in the panels `$n_d=3$' and `$n_d=4$' of Fig.~19,
are significantly fragmented. Notable exceptions are
the $L=3^{+}_2$ state, which is the solvable U(5)-PDS state of
Eq.~(\ref{ePDSu5L3})
with $n_d=3$, and the $L=5^{+}_2$ state with a dominant $n_d=4$ component.
The existence in the spectrum of specific spherical-type of states with
either $P_{n_d}^{(L_i)}=1$ or $P_{n_d}^{(L_i)}\approx1$, exemplifies
the presence of an exact or approximate U(5) PDS at the critical-point.

The states shown on the right column of Fig.~19, have a different
character. They belong to the five lowest regular sequences
seen in the combined Peres lattices for $\xi_c=0$ in Fig.~17.
The association of a set of $L_i$-states to a given sequence, is based
on a close proximity of their lattice points $\left\{x_i,E_i\right\}$,
and on having a similar decomposition in the SU(3) DS basis,
to be discussed below. The states shown,
exhibit a broad $n_d$-distribution, hence are qualified as
`deformed'-type of states, forming rotational bands:
$g(K\!=\!0),\,\beta(K\!=\!0),\,\beta^2(K\!=\!0),
\,\beta^3(K\!=\!0)$ and $\gamma(K\!=\!2)$.
The bandhead energy of each $K$-band is listed in each panel.
Note that the zero-energy deformed ground state, $L=0_1$, is degenerate
with the $(n_d=0,L=0_2)$ spherical state.
The $P_{n_d}^{(L_i)}$ probabilities for the $K=0$ bands in Fig.~19,
display an oscillatory behavior, reflecting the expected nodal structure
of these ground and multi $\beta$-phonon bands.

Fig.~20 shows the SU(3) $(\lambda,\mu)$-distribution,
$P^{(L_i)}_{(\lambda,\mu)}$ (\ref{Plammu}), for the same eigenstates
of the critical-point Hamiltonian as in Fig.~19.
The spherical-type of states, shown on the left column, involve
considerable mixing with respect to SU(3), without any obvious
common pattern among states in the same `$n_d$' multiplet,
and in marked contrast to their $n_d$-distribution shown in Fig.~19.
The states in the `$n_d\leq 2$' multiplets involve higher SU(3)
irreps, while those in the fragmented `$n_d\geq 3$' multiplets
are more uniformly spread among all $(\lambda,\mu)$-components.
The `rotational'-type of states, shown on the right column of Fig.~20,
show again a very different behavior.
First, the ground $g(K=0)$ and the $\gamma(K=2)$
bands are pure with $(\lambda,\mu) = (2N,0)$ and $(2N-4,2)$
SU(3) character, respectively. These are the solvable bands of
Eq.~(\ref{solsu3}) with SU(3) PDS.
Second, the non-solvable $K$-bands are mixed with respect
to SU(3), but the mixing is similar for the different $L$-states
in the same band. Such strong but coherent ($L$-independent) mixing is the
hallmark of SU(3) QDS. It results from the existence of a single
intrinsic state for each such band and imprints an adiabatic motion and
increased regularity~\cite{MDSC10}.

By comparing the right hand side panels in Fig.~20, with the
left hand side panels in Fig.~19, we find that the SU(3) QDS property
of the `deformed' states persists, while the U(5) PDS property of the
spherical states dissolves at higher energy.
This observation is in accord with the classical
and quantum analyses. As portrayed in the Poincar\' e
sections (Fig.~10) and Peres lattices (Figs.~17-18) at the critical
point ($\xi_c=0$),
the dynamics ascribed to the deformed well is
regular and persists to energies higher than the barrier.
In contrast, the dynamics ascribed
to the spherical well, shows a H\'enon-Heiles (HH) type of
transition from regular to chaotic motion as the energy increases.
A narrow chaotic layer in the classical phase space starts to occur
at $E\approx0.1$, while fully chaotic dynamics develops at $E\approx0.24$,
below the top of the barrier at $V_\mathrm{\rm bar}/h_2=0.268$.
For the boson number $N=50$ considered, the `$n_d= 0,1$' states
in Fig.~19, lie in the energy domain of the regular HH dynamics,
the `$n_d=2$' triplet resides in the relatively-regular
domain just above the appearance of the chaotic layer,
while the `$n_d = 3,4$' multiplets lie already near the barrier top,
in the highly chaotic domain.
Thus, the observed breakdown of the U(5)-character of the multiplets, can be
attributed to the onset of chaos at higher energy in the region of
the spherical well.

\subsection{Measures of purity (PDS) and coherence (QDS)}

The preceding discussion highlights the importance of U(5)-PDS
and SU(3)-QDS in identifying and characterizing the persisting
regular states.
These symmetry notions rely on the purity and coherence
of the states with respect to a DS basis. It is therefore of interest
to have at hand quantitative measures for these properties.

The Shannon state entropy is a convenient tool to evaluate
the purity of eigenstates with respect to a DS basis.
Given a state $\ket{L_i}$, with U(5) and SU(3) decomposition as in
Eq.~(\ref{Li}), its U(5) and SU(3) entropies are defined as
\bsub
\ba
S_\mathrm{U5}(L_i) &=& -\frac{1}{\ln D_{5}}
\sum_{n_d} P_{n_d}^{(L_i)} \ln P_{n_d}^{(L_i)} ~,
\label{Shannonu5}\\
S_\mathrm{SU3}(L_i) &=& -\frac{1}{\ln D_{3}}
\sum_{(\lambda,\mu)} P_{(\lambda,\mu)}^{(L_i)} \ln P_{(\lambda,\mu)}^{(L_i)} ~.
\label{Shannonsu3}
\ea
\label{eq:Shannon}
\esub
Here $P_{n_d}^{(L_i)}$ and $P_{(\lambda,\mu)}^{(L_i)}$ are the U(5) and SU(3)
probability distributions of Eq.~(\ref{Pndlammu}). The
normalization $D_5$ ($D_3$) counts the number of possible
$n_d$ [$\lm$] values for a given $L$ and, for simplicity, their
dependence on $L_i$ and $N$ is suppressed.
A Shannon entropy vanishes when the considered state
is pure with good $G$-symmetry
[$S_\mathrm{G}(L_i)=0$] and is positive for a mixed state.
The maximal value [$S_\mathrm{G}(L_i)=1$] is obtained when
the state $\ket{L_i}$ is uniformly spread among the irreps of $G$,
{\it i.e.} for $P_{G}^{(L_i)} = 1/D_G$. Intermediate values,
$0 < S_\mathrm{G}(L_i) < 1$, indicate partial fragmentation
of the state $\ket{L_i}$ in the respective DS basis.
The averaging of such quantities over all eigenstates
has been previously used to disclose the global DS content of
the IBM Hamiltonian, Eq.~(\ref{HndQQ}), and to correlate the implied
degree of the eigenfunction localization with chaotic
measures~\cite{CejJol98a,CejJol98b}.

The values of the U(5) entropy $S_{\rm U5}(L_i)$, Eq.~(\ref{Shannonu5}),
are listed for representative states in
Fig.~19. As expected, $S_{\rm U5}=0$ for the solvable U(5)-PDS states,
Eq.~(\ref{ePDSu5}), with $(n_d\!=\!0,L\!=\!0_2)$ and $(n_d\!=\!3,L\!=\!3_2)$.
Other spherical-type of states with `$n_d\leq 2$' have a low
value, $S_{\rm U5}<0.15$, while the more dispersed states with `$n_d=3,4$'
have $S_{\rm U5}>0.40$. The deformed-type of states,
shown on the right column of Fig.~19, have a large
U(5) entropy, $S_{\rm U5}>0.67$.
The values of the SU(3) entropy $S_{\rm SU3}(L_i)$, Eq.~(\ref{Shannonsu3}),
are shown for selected states in Fig.~20.
As expected, $S_{\rm SU3}=0$ for the solvable SU(3)-PDS states,
Eq.~(\ref{solsu3}), members of the $g(K=0)$ and $\gamma(K=2)$ bands.
The deformed $\beta^{n}(K=0)$ bands are mixed with respect to SU(3),
hence have non-zero values of $S_{\rm SU3}$, which increase with $n$.
The spherical-type of states, shown on the
left column of Fig.~20, are strongly mixed
with respect to SU(3) and have $S_{\rm SU3}>0.72$.

The coherent decomposition characterizing SU(3) QDS, implies
strong correlations between the SU(3) components of different $L$-states
in the same band. This can be used as a criterion for the identification
of rotational bands. We focus here on the $L=0,2,4,6$,
members of $K=0$ bands.
Given a $L=0^{+}_i$ state, among the ensemble of possible states,
we associate with it those $L_j>0$ states which show the maximum
correlation, $\max_{j}\{\pi(0_i,L_j)\}$.
Here $\pi(0_i,L_j)$ is a Pearson correlation coefficient whose
values lie in the range $[-1,1]$.  Specifically,
$\pi(0_i,L_j)=1,-1,0,$ indicate a perfect correlation,
a perfect anti-correlation, and no linear correlation, respectively,
among the SU(3) components of the $0_i$ and $L_j$ states.
More details on these coefficients in conjunction with the
present study, are discussed in Appendix~B.
To quantify the amount of coherence (hence of SU(3)-QDS) in the chosen
set of states, we adapt the procedure proposed in~\cite{Macek10}, and
consider the following product
of the maximum correlation coefficients
\ba
C_{\rm SU3}(0_i{\rm -}6) \equiv
\max_{j}\{\pi(0_i,2_j)\}\,
\max_{k}\{\pi(0_i,4_k)\}\,
\max_{\ell}\{\pi(0_i,6_{\ell})\} ~.
\label{Pearson}
\ea
We consider the set of states $\{0_i,\,2_j,\,4_k,\,6_{\ell}\}$ as
comprising a $K=0$ band with SU(3)-QDS,
if $C_{\rm SU3}(0_i{\rm -}6)\approx 1$.

The values of $C_{\rm SU3}(0_i{\rm -}6)$ for selected sets of states
are shown in Fig.~20. As expected, $C_{\rm SU3}(0_i{\rm -}6)\approx 1$ for
all the `deformed' $K$-bands. On the other hand, this quantity is
much smaller (but still non-zero) for the spherical-type of states.
Band structure based on SU(3) QDS thus
necessitates a value of $C_{\rm SU3}(0_i{\rm -}6)$ in very close
proximity to 1. It should be noted
that the coherence properly of a band of states, as measured by
$C_{\rm SU3}(0_i{\rm -}6)$, is independent of its purity, as measured
by $S_{\rm SU3}(L_i)$. Thus, in Fig.~20,
the pure $g(K=0)$ and $\gamma(K=2)$ bands with SU(3) PDS have
$C_{\rm SU3}(0_i{\rm -}6)= 1$ or $C_{\rm SU3}(2_i{\rm -}8)= 1$
and $S_{\rm SU3}=0$, while the mixed $\beta^3(K=0)$ band has
$C_{\rm SU3}(0_i{\rm -}6)= 0.9996$ and $S_{\rm SU3}=0.406$.

\subsection{Evolution of U(5)-PDS and SU(3)-QDS across the QPT}

We turn now to a detailed study of the evolution of
partial- and quasi-dynamical symmetries across the first order QPT,
induced by the intrinsic Hamiltonian $\hat{H}_{1}(\rho)$,
Eq.~(\ref{eq:H1}),
and $\hat{H}_{2}(\xi)$, Eq.~(\ref{eq:H2}), with $\bz=\sqrt{2}$.
For that purpose, we examine the change in the Shannon entropies,
$S_{\rm U5}(L_i)$ and $S_{\rm SU3}(L_i)$, Eq.~(\ref{eq:Shannon}),
and in the SU(3) correlation coefficient $C_{\rm SU3}(0_i{\rm -}6)$,
Eq.~(\ref{Pearson}), as a function of the control parameters,
$0\leq\rho\leq\rho_c$ and $\xi_c\leq\xi\leq 1$, for the entire spectrum.
\begin{figure}[!t]
\begin{center}
\epsfig{file=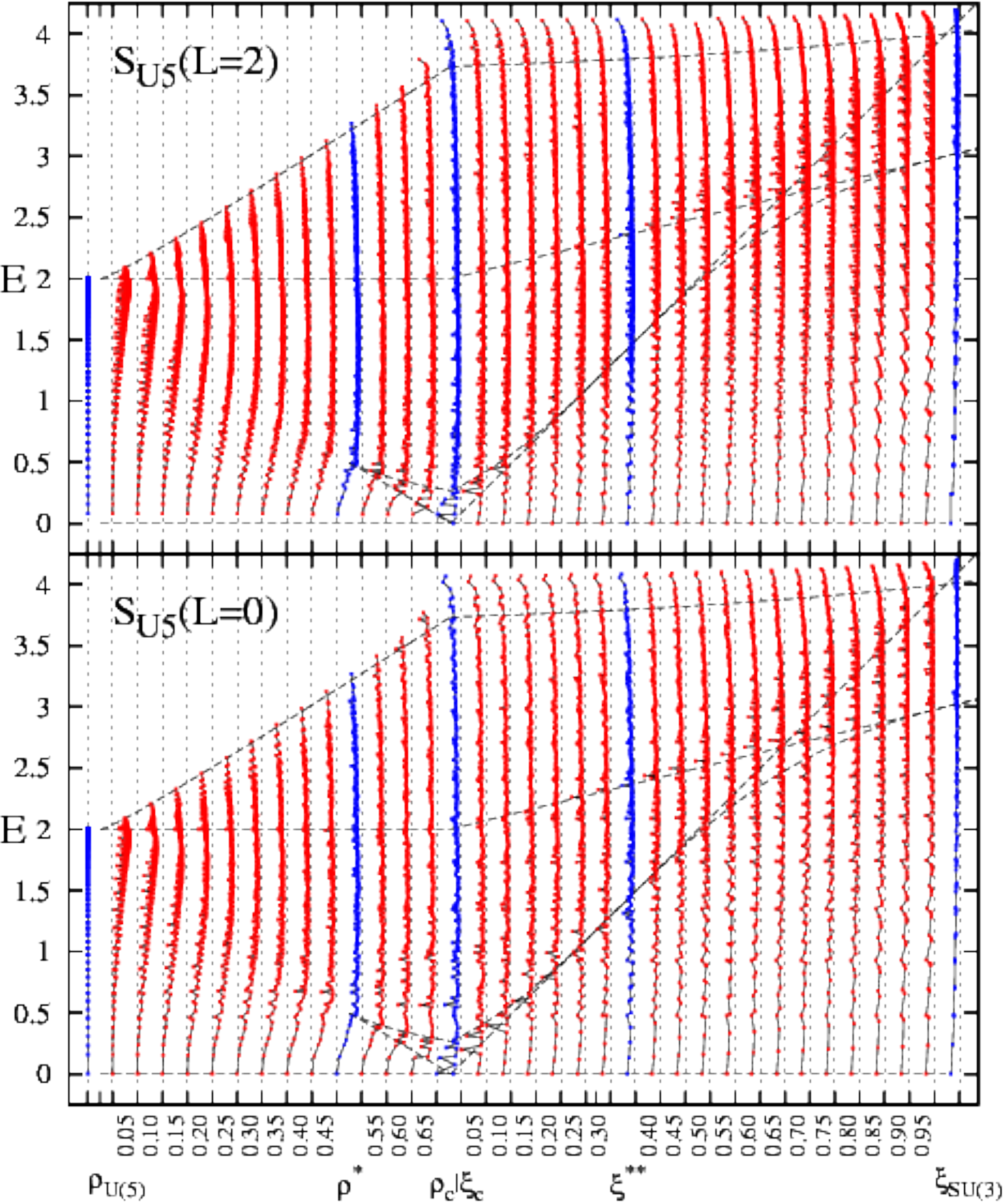,width=0.9\linewidth}
\end{center}
\caption{
U(5) Shannon entropy $S_{\rm U5}(L)$, Eq.~(\ref{Shannonu5}) for
$L=2$ (top) and $L=0$ (bottom) eigenstates of the intrinsic
Hamiltonian~(\ref{eq:Hint}), with $h_2=1,\,\bz=\sqrt{2},\,N=50$,
as a function of the control parameters $(\rho,\xi)$.
Vertical dashed lines to the left (right) of each numbered control parameter,
indicate the value $S_{\rm U5}(L)=0$ [$S_{\rm U5}(L)=1$].
Black dashed lines depict the stationary and asymptotic
values of the relevant classical potentials (compare with Fig.~3).
$S_{\rm U5}(L)\approx 1$ indicates an approximate U(5)-PDS.}
\label{fig21}
\end{figure}

Fig.~21 displays the values of the
U(5) Shannon entropy $S_{\rm U5}(L)$, Eq.~(\ref{Shannonu5}),
for $L=0$ (bottom) and $L=2$ (top) eigenstates of
$\hat{H}_{\rm int}$~(\ref{eq:Hint}), with $N=50$.
The vertical axis lists the energy $E$ of the states,
while the horizontal axis lists 35 values of the control parameters
$(\rho,\xi)$. Vertical dashed lines which
embrace each control parameter, correspond to the value
$S_{\rm U5}(L)=0$ (left) and $S_{\rm U5}(L)=1$ (right). Thus, states which
are pure with respect to U(5) (hence with $S_{\rm U5}(L)=0$),
are represented by points on the
vertical dashed line to the left of the given control parameter.
Departures from this vertical line, $0< S_{\rm U5}(L) \leq 1$,
indicate the amount of U(5) mixing.
For clarity, the values of $S_{\rm U5}(L)$ at
the U(5) DS limit ($\rho_{\rm U(5)}=0$), the spinodal ($\rho^*$),
the critical ($\rho_c,\xi_c$), and the anti-spinodal ($\xi^{**}$) points,
and the SU(3) DS limit ($\xi_{\rm SU(3)}=1$), are distinguished by
a different (blue) color. To gain further insight,
the stationary and asymptotic values
($V_{\rm sph}$, $V_{\rm max}$, $V_{\rm def}$, $V_{\rm bar}$,
$V_{\rm sad}$, $V_{\rm lim}$) of the relevant classical potentials,
are depicted by dashed black lines (compare with Fig.~3).

Starting on the spherical side of the QPT,
at the U(5) DS limit ($\rho_{\rm U(5)}=0$), the U(5) entropy,
$S_\mathrm{U5}(L) = 0$, vanishes for all states.
The spherical $L=0_1$ ground state
of $\Hone$ maintains $S_{\rm U5}(L=0_1) = 0$
throughout region~I ($0\leq\rho\leq\rho^{*}$)
and in part of region~II ($\rho^{*}< \rho\leq\rho_c$),
in accord with its U(5)-PDS property, Eq.~(\ref{ePDSu5L0}).
As seen in Fig.~21, for $\rho>0$, all other $L=0,2$ eigenstates of
$\hat{H}_{1}(\rho)$
have positive $S_{\rm U5}(L)>0$, reflecting their U(5) mixing.
$S_{\rm U5}(L)$ attains small positive values at low energy, corresponding
to spherical-type of states, and changes to moderate and high values
as the energy increases, indicating stronger U(5) mixing.
The departures of $S_{\rm U5}(L)$
from zero value start to occur at lower energy, as
$\rho$ approaches $\rho^{*}$. This behavior is consistent with
the H\'enon Heiles type of dynamics and the onset of chaos in this region.
In region~III ($\xi^{**}\leq\xi\leq 1$) all states, including
the ground state of $\hat{H}_{2}(\xi)$, have
$S_{\rm U5}(L)\in[0.7,0.9]$, exhibiting weaker variation with energy.
These large values reflect the deformed nature of the underlying
eigenstates, which are arranged in rotational bands.
In region~II of phase coexistence ($\rho^{*}<\rho\leq\rho_c$ and
$\xi_c\leq\xi<\xi^{**}$), $S_{\rm U5}(L)$ attains both low and
high positive values, reflecting the presence of both
spherical- and deformed-type of states.
This creates a zig-zag pattern, especially visible in the
triangular region bordered by the energies of the barrier ($V_{\rm bar}$)
and of the (deformed or spherical) local minima ($V_{\rm def}$ or $V_{\rm sph}$).

In spite of the U(5) mixing present in the overwhelming majority of
eigenstates, a subset of low-lying states in regions I and II
exhibit pronounced low-values of $S_{\rm U5}(L)$,
indicating an enhanced purity with respect to U(5).
Such states are members of U(5)-like multiplets,
of the form discussed in Fig.~19. Their
wave functions are dominated by a single $n_d$ component,
which has the largest (maximal) $n_d$-probability, $P_{n_d}^{(L_i)}$,
for a given $(n_d,L)$. These spherical type of states thus exemplify
the persistence of an (approximate) U(5) PDS.
\begin{figure}[!t]
\begin{center}
\epsfig{file=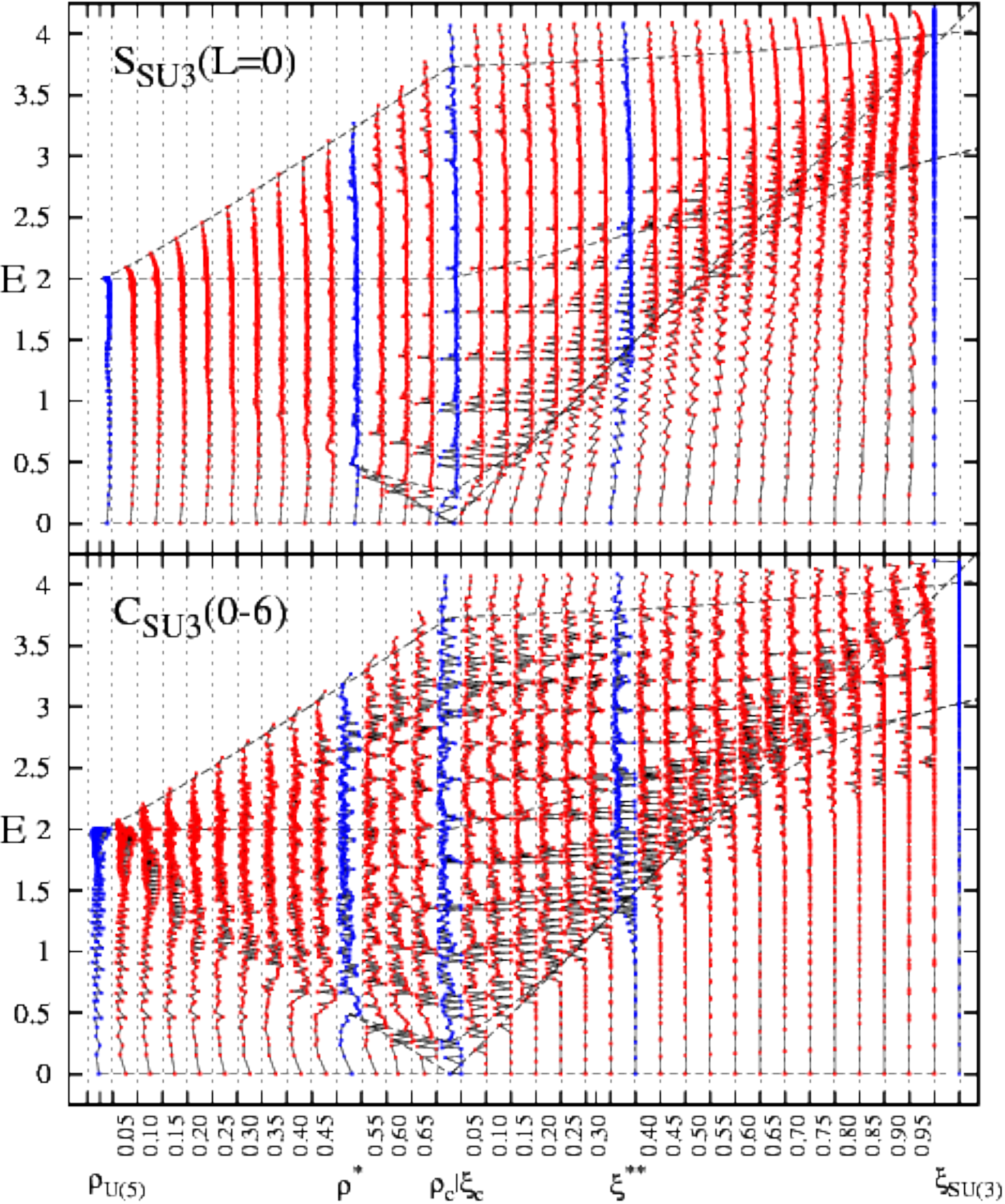,width=0.9\linewidth}
\end{center}
\caption{
As in Fig.~21, but for the
SU(3) Shannon entropy $S_{\rm SU3}(L=0)$, Eq.~(\ref{Shannonsu3}),
(top panel) and the SU(3)-correlation coefficient
$C_{\rm SU3}(0{\rm -}6)$, Eq.~(\ref{Pearson}), (bottom panel),
as a function of the control parameters $(\rho,\xi)$
and the energy $E$  of $L=0$ states.
Vertical dashed lines to the left (right) of each numbered control parameter,
indicate the value 0 (1) for both quantities.
$S_{\rm SU3}(L=0)\approx 1$ indicates an enhanced SU(3) purity, while
$C_{\rm SU3}(0{\rm -}6)\approx 1$ indicates a $K=0$ band
exhibiting SU(3)-QDS.}
\label{fig22}
\end{figure}

The top panel of Fig.~22 displays the values of the
SU(3) Shannon entropy $S_{\rm SU3}(L=0)$, Eq.~(\ref{Shannonsu3}),
for the entire energy spectrum of $L=0$ states.
The notation of lines is the same as in Fig.~21.
Thus, states which
are pure with respect to SU(3), are represented by points on the
vertical dashed line to the left of the given control parameter,
corresponding to $S_{\rm SU3}(L=0)=0$.
Departures from this vertical line, $0< S_{\rm SU3}(L) \leq 1$,
indicate the amount of SU(3) mixing.
The bottom panel of Fig.~22 displays the values of the
SU(3) correlation coefficient $C_\mathrm{SU3}(0{\rm -}6)$,
Eq.~(\ref{Pearson}), correlating sequences of $L=0,2,4,6$, states,
throughout the
entire spectrum. The energy $E$, listed on the vertical axis,
corresponds here to the energy of the
$L=0$ eigenstate in each sequence.
The vertical dashed lines which
embrace each control parameter $(\rho,\xi)$, correspond now to the value
$C_\mathrm{SU3}(0{\rm -}6)=0$ (left) and
$C_\mathrm{SU3}(0{\rm -}6)=1$ (right).
Thus, a highly-correlated sequence of $L=0,2,4,6$ states,
comprising a $K=0$ band and manifesting SU(3)-QDS,
are represented by points lying on or very close to the
vertical dashed line to the right of the given control parameter,
corresponding to $C_{\rm SU3}(0{\rm -}6)\approx 1$.
Even slight departures from this vertical line,
$C_{\rm SU3}(0{\rm -}6)<1$, indicate a reduction of SU(3) coherence.

Starting on the deformed side of the QPT,
at the SU(3) DS limit ($\xi_{\rm SU(3)}=1$), the SU(3) entropy,
$S_{\rm SU3}(L) = 0$, vanishes for all states.
In this case, the $L$-states in a given $K$-band belong
to a single SU(3) irrep, hence necessarily $C_{\rm SU3}(0{\rm -}6)=1$.
As one departs from the symmetry limit, ($\xi<1$),
$S_{\rm SU3}(L=0)>0$ acquires positive values, reflecting an SU(3) mixing.
The SU(3) breaking becomes stronger at higher energies and
as $\xi$ approaches $\xi_c=0$ from above,
resulting in higher values of $S_{\rm SU3}(L=0)$.
A notable exception to this behavior is
the deformed ground state ($L=0_1$)
of $\Htwo$, which maintains $S_{\rm SU3}(L=0_1) = 0$ throughout
region~III ($\xi^{**}\leq\xi\leq 1$) and in part of
region~II ($\xi_c\leq\xi<\xi^{**}$), in accord with
its SU(3)-PDS property, Eq.~(\ref{solsu3g}).
In contrast to the lack of SU(3)-purity in all excited
$L=0$ states, the SU(3) correlation function maintains a value close
to unity, $C_\mathrm{SU3}(0-6)\approx 1$. This indicates that the SU(3)
mixing is coherent and that these $L=0$ states serve as
bandhead states of $K=0$ bands with a pronounced SU(3) QDS.
This band-structure is observed throughout region~III
in extended energy domains. In particular, all such $K=0$ bands
show strong coherence up to the energy of the saddle point
$V_{\rm sad}$, Eq.~(\ref{Vsad}), for $\xi>\xi^{**}$, or of the spherical
local minimum $V_{\rm sph}$, Eq.~(\ref{bzero}), for $\xi<\xi^{**}$.
This observation is consistent
with the classical analysis, which revealed a robustly
regular dynamics in this region.
Coherent $K=0$ bands can also be seen seen
at high energy above $V_{\rm sph}$ in regions~III and II of Fig.~22.
One observes here numerous sequences of points
with $C_\mathrm{SU3}(0{\rm -}6)=1$,
alternating with other points for which $C_{\rm SU3}(0{\rm -}6)<0.7$.
The former correspond to the regular states, identified by the Peres
lattices in Fig.~16, while the latter correspond to irregular (chaotic)
states. In particular, at energies below $V_{\rm lim}$, Eq.~(\ref{Vlim2}),
there is a very sharp distinction between the two families,
corresponding to the sharp distinction between the regular and
chaotic states (dynamics) observed in the Peres lattices (Poincar\'e
surfaces). At very high energies, above $V_{\rm lim}$, some incoherence
appears, consistent with the onset of chaos in region~III.

In region~I ($0\leq\rho\leq\rho^{*}$), all states
exhibit high values of $S_{\rm SU3}(L=0)\approx 1$ and
$C_{\rm SU3}(0{\rm -}6)< 1$, indicating considerable SU(3)
mixing and lack of SU(3) coherence. This is in line with
the presence of spherical states
at low energy, and of more complex type of states at higher energy,
and the absence of rotational bands in this region.
In region~II of phase coexistence ($\rho^{*}<\rho\leq\rho_c$ and
$\xi_c\leq\xi<\xi^{**}$), one encounters
both, points with $C_{\rm SU3}(0{\rm -}6)\approx 1$, and points with
$C_{\rm SU3}(0{\rm -}6)< 1$. This reflects the presence of deformed states
arranged into regular bands, exemplifying SU(3) QDS and, at the same time,
the presence of spherical states and other more complicated type of
states of a different nature.

\subsection{Global features of U(5) PDS and SU(3) QDS
as fingerprints of the QPT}

The preceding discussion has demonstrated the relevance of
U(5) PDS and SU(3) QDS in characterizing the symmetry
properties of individual quantum eigenstates of the intrinsic Hamiltonian
across the QPT. The related measures of these quasi-symmetries,
the U(5) Shannon entropy, $S_{\rm U5}(L)$, Eq.~(\ref{Shannonu5}),
and the SU(3) correlation coefficient,
$C_{\rm SU3}(0{\rm -}6)$, Eq.~(\ref{Pearson}), quantify the U(5)-purity
and SU(3)-coherence in these states, respectively.
Considerable interest is drawn to subsets of regular states which
maintain a high degree of purity or coherence amidst a complicated
environment of other states.
In particular, a small value of $S_{\rm U5}(L)$ signals
an approximate U(5) PDS and identifies subsets of
spherical-type of states
which reflect a surviving regular dynamics in the vicinity
of the spherical minimum.
On the other hand, a large value of $C_{\rm SU3}(0{\rm -}6)\approx 1$
signals an SU(3) QDS and identifies rotational $K$-bands,
which reflect a persisting regular dynamics in the vicinity of the
deformed minimum. In the present Section, we wish to consider global
features of these measures which can shed light on the PDS and QDS
content of the entire system as a whole, and monitor its
evolution across the the QPT.

The presence or absence in the spectrum
of spherical or deformed type of regular states,
is intimately tied with the existence and depth of the corresponding
spherical or deformed wells in the classical potential.
As seen in Figs.~21-22, the number of such regular states is maximal
at the DS limits and it reduces as the control parameters approach
the values of the anti-spinodal or spinodal points, where the respective
local minimum disappears. The evolution with $(\rho,\xi)$
of the number of states having an approximate U(5) PDS or SU(3) QDS
reflects the change in the morphology of the underlying Landau potential
and can, therefore, serve as fingerprints of the QPT.

As a global measure of an approximate U(5) PDS, we consider the
quantity $\nu_{\rm U5}$, which denotes the
number of $L=0$ states satisfying $S_{\rm U5}(L=0) < 0.25$.
This quantity is an indicator of the amount of enhanced U(5)-purity
in the system. The choice of $0.25$ as an
upper limit is somewhat arbitrary, and is close to the value
of $S_{\rm U5}(L=0)\!=\!0.242$ at $\xi=0.17$
for which the maximal U(5) probability is $P_{n_d=0}^{(L=0)}\!=\! 0.8$.
Analogous quantities, $\nu_{\rm U5}(L)$,
can be calculated for states with other angular momentum $L$.
Henceforth, we continue to use the shorthand notation,
$\nu_{\rm U5}\equiv\nu_{\rm U5}(L=0)$.
In a similar spirit,
as a global measure of SU(3) QDS, we consider the
the quantity $\nu_{\rm SU3}$, which denotes the
number of $K=0$ bands whose $L=0,2,4,6$ members satisfy
$C_{\rm SU3}(0{\rm -}6)> 0.995$.
This quantity is an indicator of the amount of SU(3) coherence
in the system. The choice of $0.995$ as a
lower limit is again somewhat arbitrary. It is based on a detailed study
of the SU(3) correlator for the regular $K=0$ bands in Fig.~22,
which revealed a well-separated peak
in the range $C_\mathrm{SU3}(0-6) \in [0.995,1]$.
It should be pointed out that the chosen cutoff values
for $\nu_{\rm U5}$ and $\nu_{\rm SU3}$ apply to
eigenstates of the intrinsic Hamiltonian (\ref{eq:Hint}) with
$\bz=\sqrt{2}$ and $N=50$, and
in general these thresholds vary with $N$.
\begin{figure}[!t]
\begin{center}
\epsfig{file=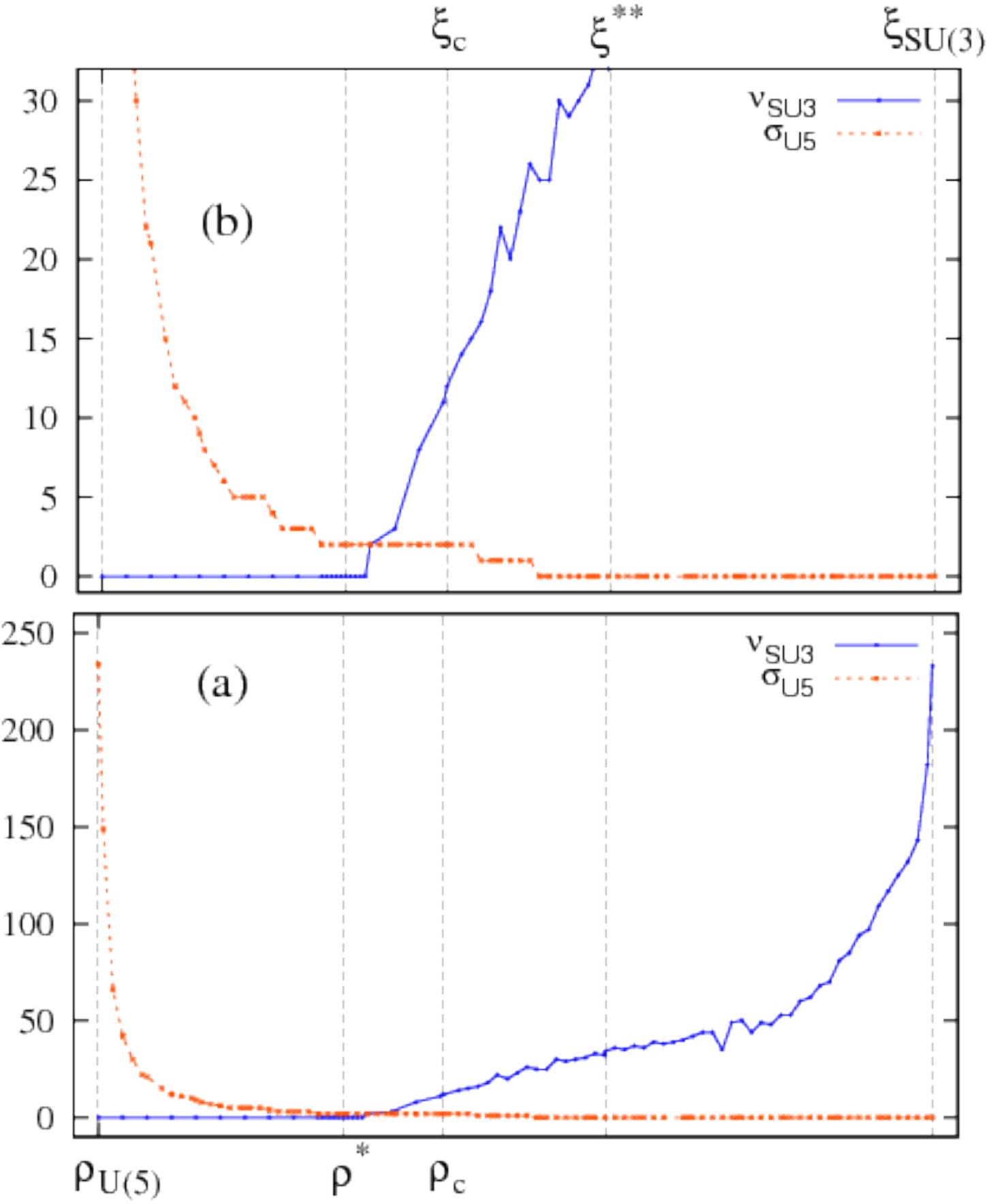,width=0.7\linewidth}
\end{center}
\caption{
Global measures of U(5) PDS [$\nu_{\rm U5}$: the number of
$L=0$ states with U(5) Shannon entropy $S_{\rm U5}(L=0) < 0.25$,
Eq.~(\ref{Shannonu5})]
and of SU(3) QDS [$\nu_{\rm SU3}$: the number of $K=0$ bands
with SU(3) correlator
$C_{\rm SU3}(0{\rm -}6)> 0.995$, Eq.~(\ref{Pearson})],
as a function of $(\rho,\xi)$.
(a)~Full evolution across the QPT. (b)~A detailed zoom.
At the critical point $(\rho_c,\xi_c)$,
$\nu_{\rm U5}=2$ and $\nu_{\rm SU3}=12$,
compared to a total of 234 $L=0$ eigenstates
of the Hamiltonian (\ref{eq:Hint}) with $\bz=\sqrt{2}$ and $N=50$.}
\label{fig23}
\end{figure}

Fig.~23 displays the quantities $\nu_{\rm U5}$ and $\nu_{\rm SU3}$
as a function of the control parameters $(\rho,\xi)$ across the QPT.
At the U(5) DS limit ($\rho_{\rm U(5)}=0$),
all states are pure with
respect to U(5) and hence, as seen in panel~(a) of Fig.~23,
$\nu_{\rm U5}=234$ equals the total number of $L=0$ states for $N=50$.
For $\rho >0$, the quantity
$\nu_{\rm U5}$ decreases, indicating a reduction in the U(5) PDS
of the system in region~I. This reduction in the U(5) purity
is accelerated for larger values of~$\rho$ (stronger U(5) mixing),
as the system enters region~II of phase-coexistence
($\rho^{*}<\rho\leq\rho_c$ and $\xi_c\leq\xi< \xi^{**}$).
Inside region~II, $\nu_{\rm U5}$ attains smaller values, it vanishes
as $\xi$ approaches the anti-spinodal $\xi^{**}$,
where the spherical minimum disappears,
and remains $\nu_{\rm U5}=0$ in region~III $(\xi\geq\xi^{**})$.
Similar trends are seen for $\nu_{\rm U5}(L)$
involving states with different angular momentum when scaled by the
number of states for each $L$.

At the SU(3) DS limit ($\xi_{\rm SU(3)}=1$), all states are pure
and coherent with respect to SU(3), in particular, all $L=0$ states
serve as bandheads of $K=0$ bands
and hence $\nu_{\rm SU3}=234$
in panel~(a) of Fig.~23.
For $\xi< 1$, the quantity $\nu_{\rm SU3}$ decreases,
indicating a reduction in the number of regular $K=0$ bands with
SU(3) QDS, as the deformed well becomes less deep in region~III.
This reduction in SU(3) coherence continues inside region~II of phase
coexistence where `spherical' states and chaotic-type of states
come into play. The quantity $\nu_{\rm SU3}$ vanishes
as $\rho$ approaches the spinodal point $\rho^{*}$,
where the deformed minimum disappears,
and remains $\nu_{\rm SU3}=0$ in region~I $(\rho\leq\rho^{**})$.

Panel~(b) of Fig.~23 zooms in and provides more details on the evolution of
$\nu_{\rm U5}$ and $\nu_{\rm SU3}$. As shown, both quantities are non-zero
throughout region~II, indicating the presence of both (approximate) U(5) PDS
and SU(3) QDS inside the coexistence region. These global measures
of purity and coherence in selected eigenstates, thus trace the crossing
of the spherical and deformed minima in the Landau potential by monitoring
the remaining regular dynamics associated with each of them.
The vanishing of $\nu_{\rm SU3}$ near $\rho^{*}$, appears to be
sharper and less gradual than the vanishing of
$\nu_{\rm U5}$ which occurs even before $\xi<\xi^{**}$.
This reflects the more abrupt disappearance of the deformed minimum
at $\rho^{*}$ compared to the disappearance of the spherical minimum
at $\xi^{**}$ [compare the behavior of $(V_{\rm bar} - V_{\rm def})$ near
$\rho^{*}$ with that of $(V_{\rm bar} - V_{\rm sph})$ at
$\xi\leq\xi^{**}$, in Fig.~3].

It is worthwhile emphasizing that
both (approximate) U(5) PDS and SU(3) QDS are present in region~II
of phase coexistence, including the critical point,
imprinting in a transparent manner, the evolution of the first-order QPT.
A number of factors have facilitated the exposure of such a
simple pattern in the present study;
(i)~a high barrier, (ii)~a wide coexistence region,
(iii)~invoking the resolution of the Hamiltonian, Eq.~(\ref{eq:H}),
and performing the analysis on its intrinsic part. The latter
does not contain rotation-vibration terms that can spoil the
simple patterns observed.
The effect of such collective kinetic terms will be considered in Section~8.
The rich symmetry structure uncovered in region~II and the
coexistence of PDS and QDS inside it, were not noticed in previous works
because the Hamiltonians employed did not meet the requirements
(i)-(iii).

\section{Collective effects}

The analysis presented so far, considered the evolution of the
dynamics associated with the intrinsic part of the Hamiltonian
across the QPT.
The intrinsic Hamiltonian determines the Landau potential and the
variation of its control parameters $(\rho,\xi)$ induces
the shape-phase transition.
In the present Section, we address the impact on the
order and chaos accompanying the QPT, of the
remaining collective part of the Hamiltonian.
For that purpose, we examine
the classical and quantum dynamics of the combined Hamiltonian
\ba
\hat{H} = \hat{H}_{\rm int}(\rho,\xi)
+ \hat{H}_{\rm col}(c_3,c_5,c_6) ~.
\label{Hfull}
\ea
The intrinsic Hamiltonian considered, $\hat{H}_{\rm int}(\rho,\xi)$, is that
of Eq.~(\ref{eq:Hint}) with $h_2=1$ and
$\bz=\sqrt{2}$, interpolating between the U(5) $(\rho=0$) and SU(3)
($\xi=1$) DS limits.
The collective Hamiltonian considered, $\hat{H}_{\rm col}(c_3,c_5,c_6)$,
is that of Eq.~(\ref{eq:Hcol}), composed of kinetic terms with couplings
$c_3,\, c_5$ and $c_6$, associated with collective O(3),
O(5) and $\overline{{\rm O(6)}}$ rotations in the Euler angles,
$\gamma$ and $\beta$ degrees of freedom, respectively.
By construction,
$\hat{H}$ and $\hat{H}_{\rm int}$ in Eq.~(\ref{Hfull})
have the same Landau potential which is not influenced
by $\hat{H}_{\rm col}$. The observed modifications in the dynamics
due to $\hat{H}_{\rm col}$, are thus kinetic in nature, arising from
momentum-dependent terms
which vanish in the static limit.
For simplicity, the impact of these rotational
$c_i$-terms are studied individually, by adding them one at a time to
$\hat{H}_{\rm int}(\rho,\xi)$, the latter taken at representative
values of $(\rho,\xi)$ in regions I-II-III of the QPT.
The results obtained indicate that,
although the collective Hamiltonian does not affect
the Landau potential,
it can have dramatic effects on the onset of classical chaos,
on the resonance structure
and on the regular features of the quantum spectrum.
\begin{figure}[!t]
\begin{center}
\epsfig{file=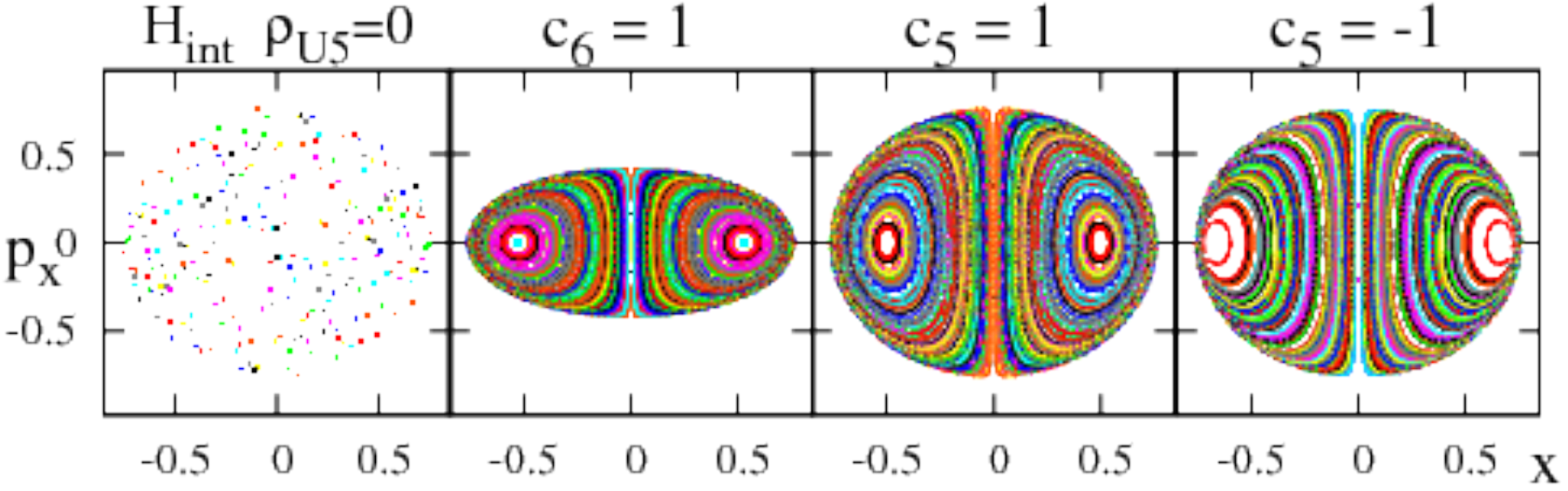,width=0.85\linewidth}
\end{center}
\caption{
Poincar\'e sections of the classical intrinsic
Hamiltonian ${\cal H}_{1}(\rho=0)$, Eq.~(\ref{eq:H1cl}),
at the U(5)-DS limit $(\rho=0)$,
with $h_2=1$ and $\bz=\sqrt{2}$ (left panel),
and of additional collective terms, Eq.~(\ref{eq:HColCl}),
with $c_6=1$ and $c_5=\pm 1$, involving $\overline{\rm O(6)}$ and
O(5) rotations. The energy is $E=V_\mathrm{lim}/2=1$.
The potential surface $V_{1}(\rho=0)$, Eq.~(\ref{V1rho0}),
is the same in all cases and is depicted in the
$\rho_{\rm U(5)}=0$ column of Fig.~9.}
\label{fig24}
\end{figure}
\begin{figure}[!t]
\begin{center}
\epsfig{file=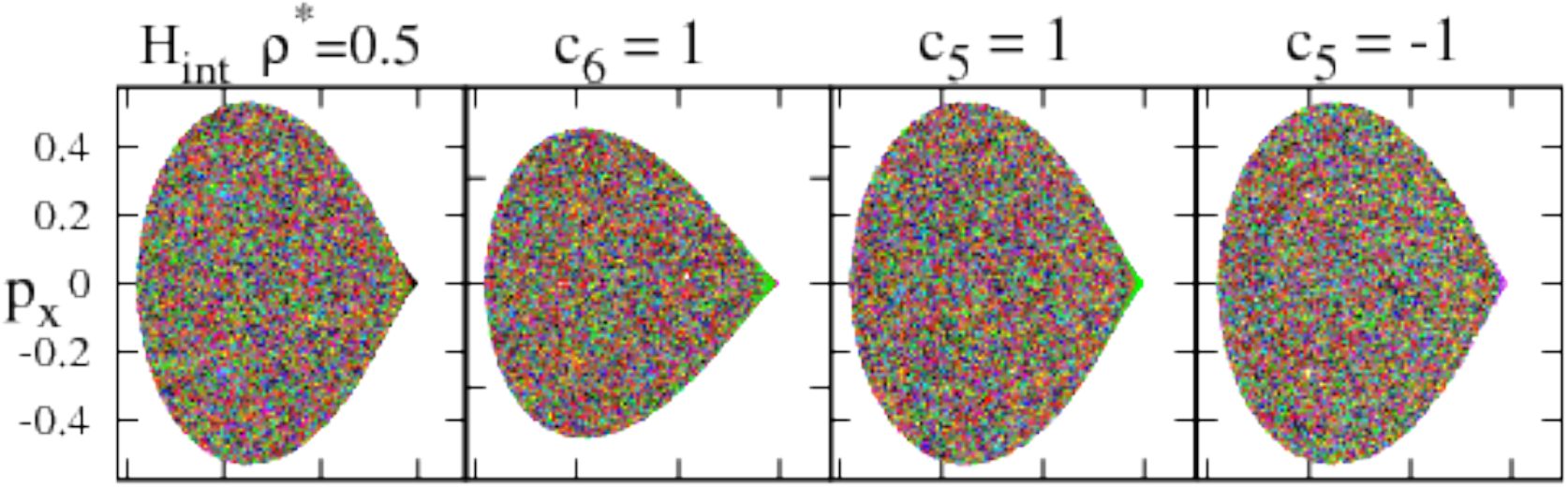,width=0.85\linewidth}
\epsfig{file=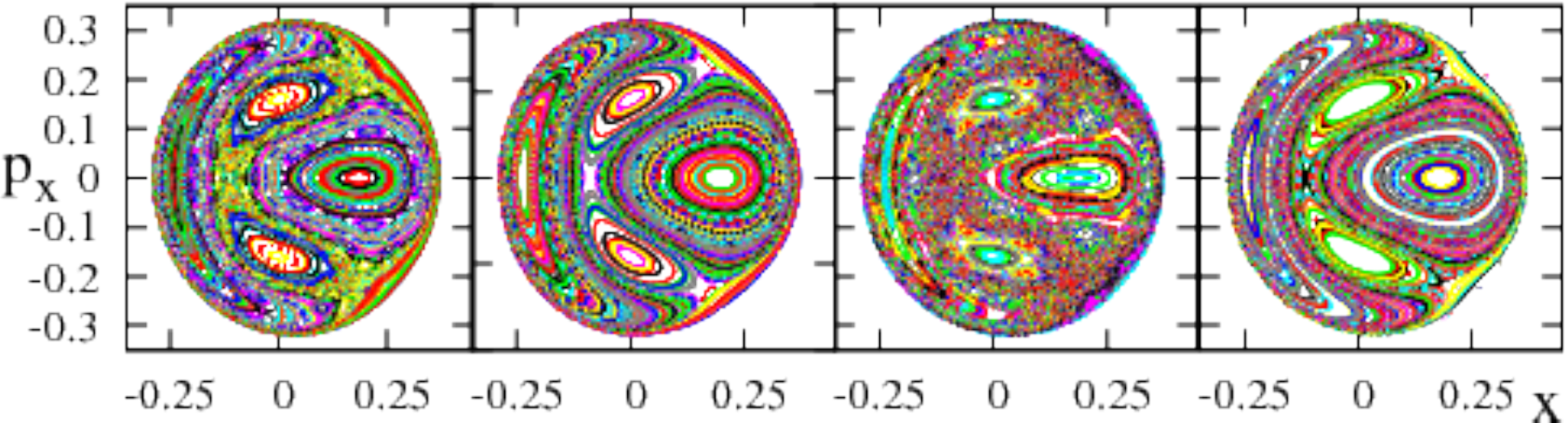,width=0.85\linewidth}
\end{center}
\caption{
Poincar\'e sections of the classical intrinsic
Hamiltonian ${\cal H}_{1}(\rho^{*})$, Eq.~(\ref{eq:H1cl}),
at the spinodal point $(\rho^{*})$,
with $h_2=1$ and $\bz=\sqrt{2}$ (left panel),
and of additional collective terms, Eq.~(\ref{eq:HColCl}),
with $c_6=1$ and $c_5=\pm 1$, involving $\overline{\rm O(6)}$ and
O(5) rotations. The energy is $E=V_\mathrm{lim}/10=0.2$ (bottom row)
and $E=V_\mathrm{lim}/4=0.5$ (top row).
The potential surface, $V_{1}(\rho^{*})$,
Eq.~(\ref{eq:V1}), is the same for all cases and is depicted in
the $\rho^{*}=0.5$ column of Fig.~9.}
\label{fig25}
\end{figure}
\begin{figure}[!t]
\begin{center}
\epsfig{file=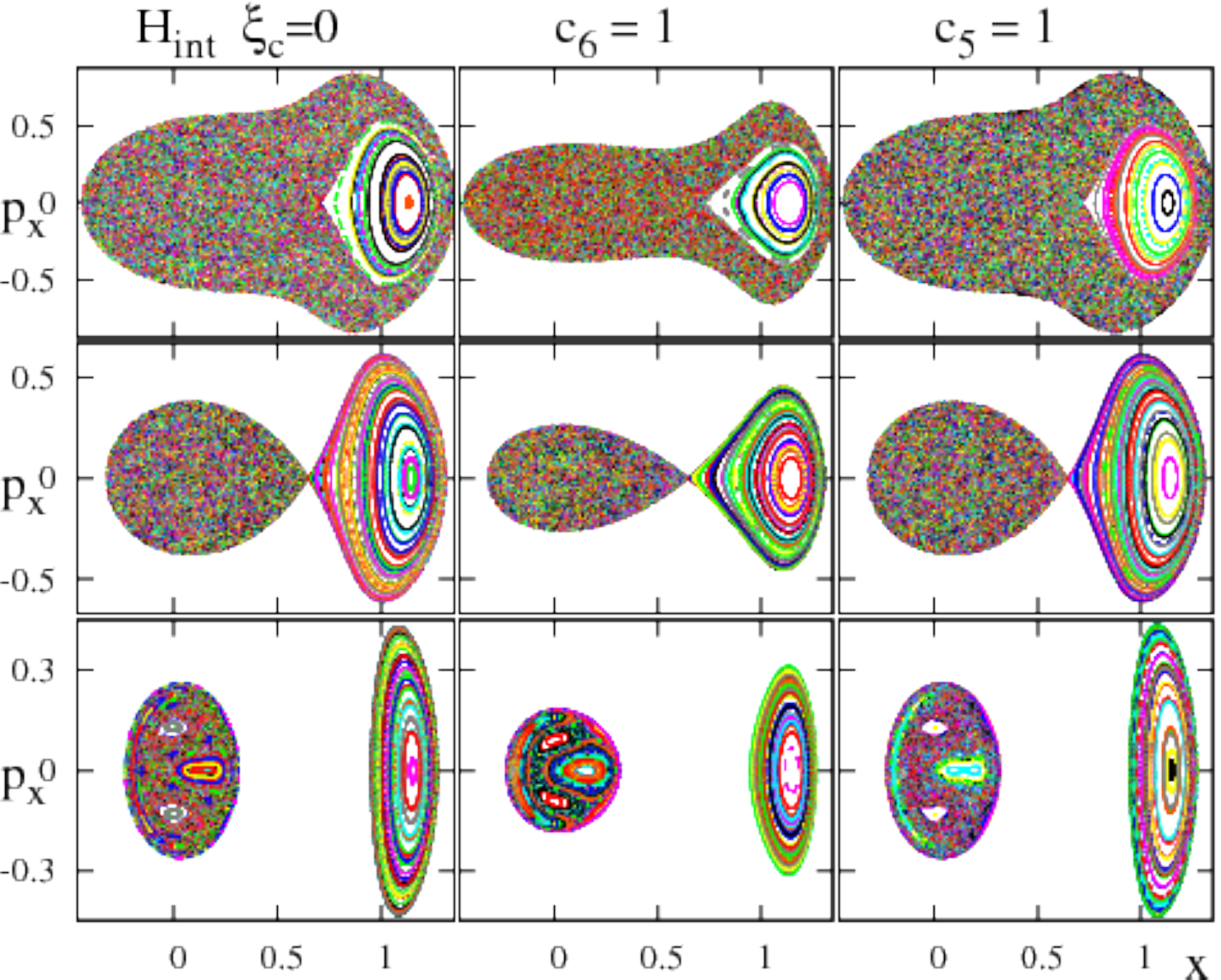,width=0.85\linewidth}
\end{center}
\caption{
Poincar\'e sections of the classical intrinsic
Hamiltonian ${\cal H}_{1}(\rho_c)= {\cal H}_{2}(\xi_c)$,
Eq.~(\ref{eq:Hintcl}),
at the critical point $(\rho_c,\xi_c)$,
with $h_2=1$ and $\bz=\sqrt{2}$ (left panel),
and of additional collective terms, Eq.~(\ref{eq:HColCl}),
with $c_6=1$ and $c_5=1$, involving $\overline{\rm O(6)}$ and
O(5) rotations.
The energy is $E=V_\mathrm{bar}/2=0.134$
(bottom row), $E=V_\mathrm{bar}=0.268$ (middle row) and
$E=2V_\mathrm{bar}=0.536$ (top row).
The potential surface $V_{1}(\rho_c)=V_{2}(\xi_c)$,
Eq.~(\ref{eq:Vcri}), is the same for all cases and is depicted in
the $\xi_c=0$ column of Fig.~10.}
\label{fig26}
\end{figure}

\subsection{Classical analysis in the presence of collective terms}
\label{subsec:CollC}

As previously done, we constrain the classical dynamics to zero angular
momentum and visualize it by means of Poincar\'e sections.
In such circumstances, the classical limit of the quantum Hamiltonian
of Eq.~(\ref{Hfull}) is given by
\ba
{\cal H} = {\cal H}_{\rm int}(\rho,\xi) + {\cal H}_{\rm col}(c_5,c_6) ~,
\label{Hfullcl}
\ea
where the first term is the classical intrinsic Hamiltonian of
Eq.~(\ref{eq:Hintcl}) and the second term is the classical
collective Hamiltonian of Eq.~(\ref{eq:HColCl}).
The O(3) $c_3$-term is absent from the latter since the
classical dynamics is constraint to $L=0$.
The O(5) $c_5$-term depends on $p_{\gamma}^2$ hence affects the
$\gamma$ motion, while the $\overline{O(6)}$
$c_6$-term depends on $T= p_{\beta}^2+ p_{\gamma}^2/\beta^2$, $T^2$ and
$\beta^2 p_{\beta}^2$, hence is the only collective term affecting
the $\beta$ motion.
The plane of the Poincar\'e section is chosen as before at  $y=0$ and its
envelope at a given energy $E$ is
defined by ${\cal H}(x,y=0,p_x,p_y=0)=E$, resulting in
the condition
\ba
{\cal H}_{\rm int}(x,y=0,p_x,p_y=0) + c_6[(2-x^2)p_x^2 - p_x^4] = E ~.
\label{envelope}
\ea
As seen, the envelope of the full classical Hamiltonian ${\cal H}$
is modified with respect to that of ${\cal H}_{\rm int}$, solely due to the
$\overline{O(6)}$ $c_6$-term.

Considering the classical dynamics of $L=0$ vibrations
in the stable spherical phase (region~I), the relevant
classical intrinsic Hamiltonian in Eq.~(\ref{Hfullcl}) is
$\mathcal{H}_1(\rho)$ of Eq.~(\ref{eq:H1cl}), with
$0\leq\rho\leq\rho^{*}$.
Fig.~24 shows for $\rho=0$
the Poincar\'e sections, at $E=1$,
of $\mathcal{H}_1(\rho=0)$
and of the added $c_5$ and $c_6$ collective terms.
The potential surface is $V_{1}(\rho=0)$, Eq.~(\ref{V1rho0}),
the same in all panels.
The $c_5$-term turns the exact U(5) symmetry of the intrinsic
Hamiltonian into a U(5) dynamical symmetry of the combined
Hamiltonian. The $c_6$-term breaks the U(5) symmetry but
maintains the reduced symmetry of the O(5) subgroup.
As a result, the system for $\rho=0$ remains integrable in the presence
of both terms. The main effect is that the
trajectories are no longer periodic but rather become quasi-periodic,
start to precess and densely cover the surfaces of the invariant tori.
In the Poincar\'e sections, instead of a finite collection of points,
we now see smooth curves organized into two
regular islands, forming a pattern typical of an anharmonic
(quartic) oscillator.
Fig.~25 shows similar sections, at $E=0.2$ (bottom)
and $E=0.5$ (top), for the spinodal-point $\rho=\rho^{*}$.
The relevant Landau potential is that depicted in the bottom panel
of the $\rho^{*}=0.5$ column in Fig.~9.
In general, the added collective terms maintain
the characteristic features of the intrinsic classical dynamics,
namely, a H\'enon-Heiles type of transition from
regular dynamics at low energy to chaotic dynamics at higher energy.
The classical dynamics in the spherical region is, to a large extent,
determined by the $\rho$-term in the intrinsic
Hamiltonian, Eq.~(\ref{eq:H1cl}).

The classical intrinsic Hamiltonian in Eq.~(\ref{Hfullcl}), appropriate
to the coexistence region (region~II), is $\mathcal{H}_1(\rho)$ of
Eq.~(\ref{eq:H1cl}), with $\rho^{*}< \rho\leq\rho_c$, and
$\mathcal{H}_2(\xi)$ of Eq.~(\ref{eq:H2cl}), with $\xi_c\leq\xi<\xi^{**}$.
Fig.~26 displays the Poincar\'e surfaces for the critical point
$(\rho_c,\xi_c)$ with energies below, at, and above the barrier,
arising from $\mathcal{H}_1(\rho_c)\equiv\mathcal{H}_2(\xi_c)$  and
from the added $c_5$ and $c_6$ rotational terms.
The potential surface in all panels is that of Eq.~(\ref{eq:Vcri}),
exhibiting a barrier separating the degenerate
spherical and deformed minima.
The $c_5$-term is seen to have a very little effect on the
Poincar\'e sections of the intrinsic Hamiltonian.
On the other hand, the sections with the $c_6$-term have a smaller size,
and are compressed for large $|p_x|$,
in accord with the properties of the envelope mentioned in
Eq.~(\ref{envelope}).
In addition, the regular island in the region of the deformed minimum
appears to be more elongated in the $x$-direction (see the middle
panel of the $c_6=1$ column in Fig.~26). This distortion can affect
the regular bands built on the deformed minimum, as will be discussed
in the subsequent quantum analysis.
The different impact of the two collective terms on the classical
dynamics can be attributed to the fact that in the coexistence region
the barrier at the saddle point is in the $\beta$-direction,
and hence is more sensitive to the $\beta$ motion. As mentioned,
the latter motion is affected by the $\overline{O(6)}$ term but not
by the O(5) term.
In general, throughout region~II, the presence of the collective terms
in the Hamiltonian does not destroy the simple pattern of robustly
regular dynamics confined to the deformed region and well-separated from
the chaotic dynamics ascribed to the spherical region.

The classical intrinsic Hamiltonian in Eq.~(\ref{Hfullcl}), relevant
to the stable deformed phase (region~III),
is $\mathcal{H}_2(\xi)$ of Eq.~(\ref{eq:H2cl}), with $\xi\geq\xi^{**}$.
For $\xi=1$, the intrinsic Hamiltonian has SU(3) symmetry and
the system is completely integrable.
As seen in Fig.~27, the inclusion of the $c_5$- and $c_6$- rotational
terms leads to substantial modifications in the phase space portrait,
showing chaotic layers and additional islands.
 \begin{figure}[!t]
\begin{center}
\epsfig{file=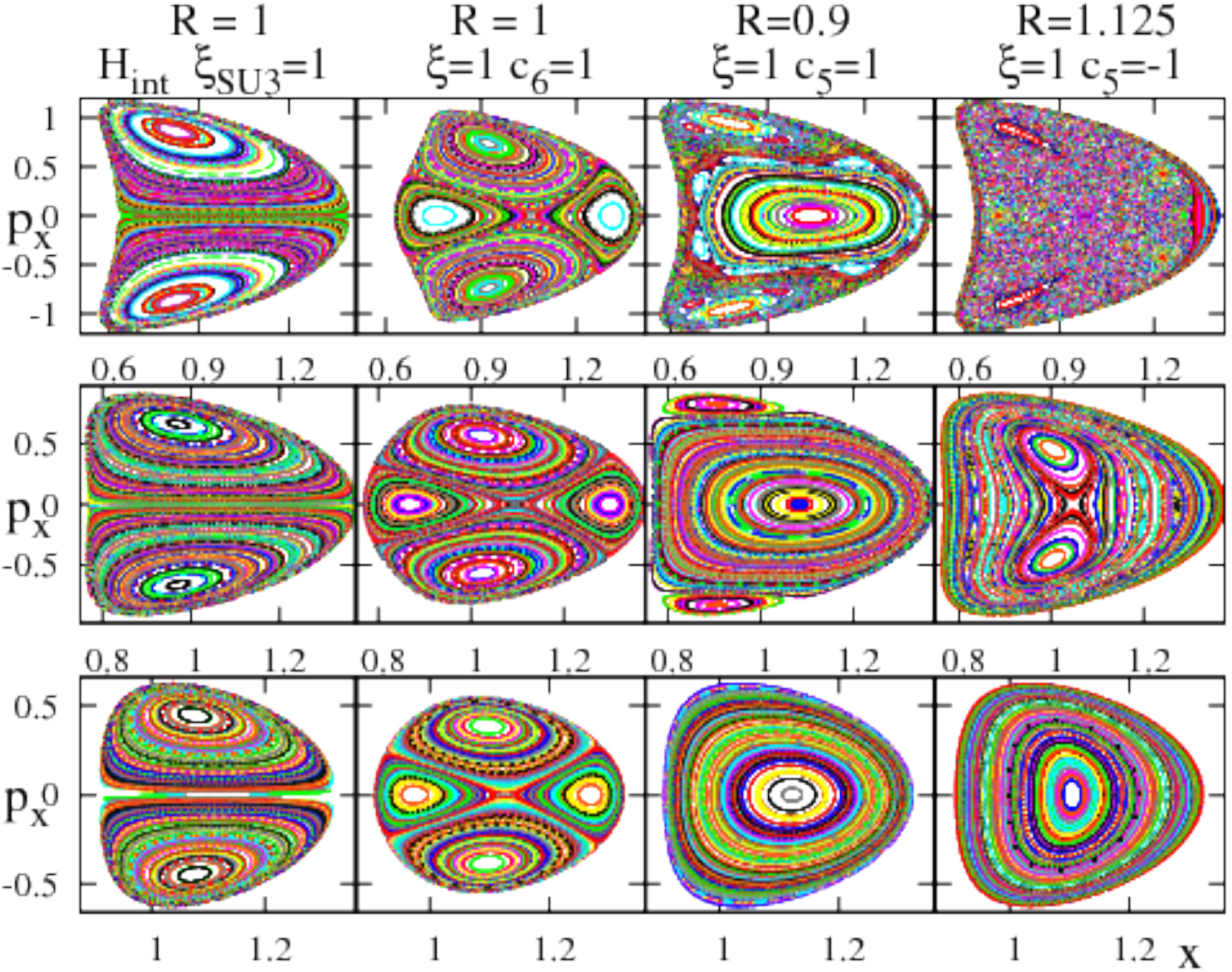,width=0.9\linewidth}
\end{center}
\caption{
Poincar\'e sections of the classical intrinsic
Hamiltonian ${\cal H}_{2}(\xi=1)$, Eq.~(\ref{eq:H2cl}),
at the SU(3)-DS limit $(\xi=1)$,
with $h_2=1$ and $\bz=\sqrt{2}$ (left panel)
and of additional collective terms, Eq.~(\ref{eq:HColCl}),
with $c_6=1$ and $c_5=\pm 1$, involving $\overline{\rm O(6)}$ and
O(5) rotations. The value of normal-mode frequency ratio $R$,
Eq.~(\ref{Rcol}), is indicated for each choice of parameters.
The energy is $E=V_\mathrm{lim}/4=0.75$ (bottom row),
$E=V_\mathrm{lim}/2=1.5$ (middle row) and
$E=3V_\mathrm{lim}/4=2.25$ (top row).
The potential surface $V_{2}(\xi=1)$, Eq.~(\ref{eq:V2}),
is the same in all cases and is depicted in the
$\xi_{\rm SU(3)}=1$ column of Fig.~11.}
\label{fig27}
\end{figure}

Both the O(5) and $\overline{{\rm O(6)}}$ symmetries
are incompatible with the SU(3) symmetry, hence the corresponding
added rotational terms break the integrability
of the intrinsic Hamiltonian at $\xi=1$. This can lead to the occurrence
of chaotic regions. The latter are more pronounced for the O(5)
term (see the panels for $c_5=\pm 1$ in Fig.~27), which can be attributed
to the fact that in region~III, the saddle point accommodates a barrier
in the $\gamma$-direction (see the contour plot in Fig.~2).
It should be stressed that, in this case, the onset of chaos
is entirely due to the kinetic terms of the collective Hamiltonian,
since the intrinsic Hamiltonian is integrable for $\xi=1$ and its
Landau potential, which has a single-deformed minimum, is kept intact
in all panels of Fig.~27.
This is a clear-cut demonstration that in the deformed side of the QPT,
chaos can develop from purely kinetic perturbations, without a change
in the morphology of the Landau potential.

\begin{figure}[!t]
\begin{center}
\epsfig{file=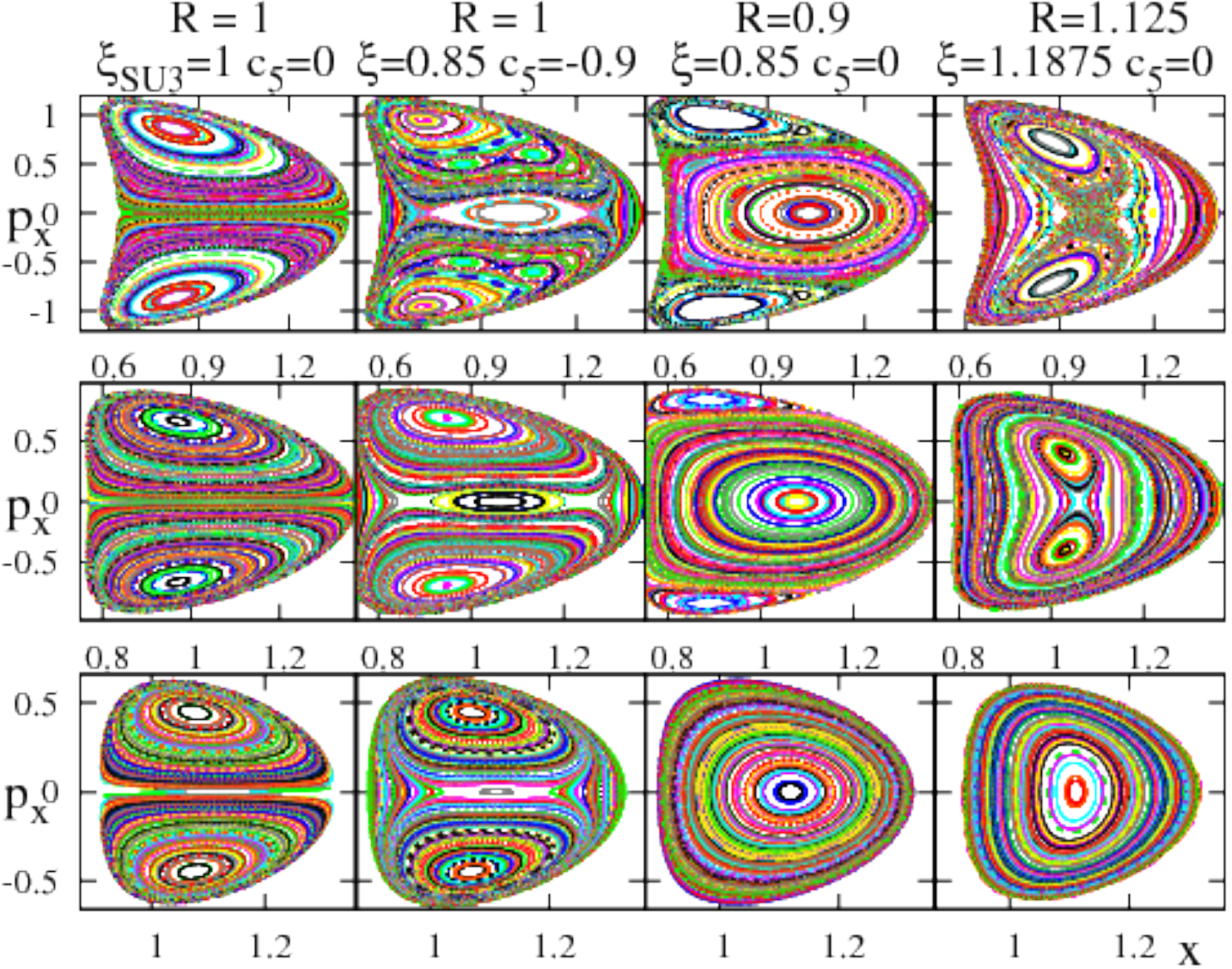,width=0.9\linewidth}
\end{center}
\caption{
Poincar\'e sections of the classical intrinsic
Hamiltonian ${\cal H}_{2}(\xi)$, Eq.~(\ref{eq:H2cl}),
with $h_2=1$ and $\bz=\sqrt{2}$ in region~III,
without and with a $c_5$-term,
Eq.~(\ref{eq:HColCl}).
The energies are $E=V_\mathrm{lim}(\xi)/4 $ (bottom row),
$E=V_\mathrm{lim}(\xi)/2$ (middle row), and
$E=3V_\mathrm{lim}(\xi)/4$ (top row), where
$V_\mathrm{lim}(\xi) = (2+\xi)h_2$.
Note the similarity of surfaces
in Figs. 27-28, with different values of $(\xi,c_5)$,
but with the same value of normal mode frequency ratio $R$,
Eq.~(\ref{Rcol}).}
\label{fig28}
\end{figure}

The phase space portrait of the integrable intrinsic Hamiltonian
at the SU(3) limit, shows the same pattern of two
regular islands at any energy (see the $\xi=1$ column in
Figs.~27 and~11).
The inclusion of the collective terms modifies this pattern.
As discussed in Section~5.3, the pattern of islands is affected by the
presence of resonances which, in turn, occur at low energy when the ratio
$R$ of normal mode frequencies is a rational number.
These resonance effects
are influenced by the presence of the collective Hamiltonian.
As seen in Eq.~(\ref{nmodescol}),
both the $c_5$- and $c_6$- terms contribute to the normal-mode frequencies
and hence change the ratio $R$ of Eq.~(\ref{R}), to
\ba
R \equiv \frac{\epsilon_{\beta}}{\epsilon_{\gamma}} =
\frac{h_2 \bz^2(2\xi+1) + c_6}{(9h_2 + c_5)\bz^2(1+\bz^2)^{-1}+c_6} ~.
\label{Rcol}
\ea
In the current study, we adapt the values $h_2=1$ and $\bz=\sqrt{2}$,
for which the above expression simplifies to
$R = [2 (2\xi+1) + c_6]/(6 + 2c_5/3 +c_6)$.

For $\xi=1$, the intrinsic Hamiltonian (with $\bz=\sqrt{2}$) has $R=1$.
The inclusion of the $\overline{O(6)}$ rotational term
does not alter this value,
but it stabilizes an additional family of orbits circulating around,
instead of passing through, the deformed minimum
(see the unstable orbit of ${\cal H}_{2}(\xi)$ for $R=1$ in Fig.~12).
As a result, two additional regular islands develop in the
Poincar\'e sections shown in the $c_6=1$ column Fig.~27,
compared to the SU(3) limit. The inclusion of the O(5) term changes
the ratio to $R=9/(9 + c_5)$, leading to $R<1$ ($R>1$) for $c_5>0$ ($c_5<0$).
The $\gamma$-motion, with $p_x\approx 0$, is stable for $R<1$
and is unstable for $R>1$.
In the latter case, the center of the Poincar\'e section
exhibits a hyperbolic fixed point
and chaos develops in its vicinity as the energy increases
(see the column with $c_5=-1$ in Fig.~27).
On the other hand, the $\beta$-motion, with large $|p_x|$, is stable
for $R>1$ and is unstable for $R<1$. In the latter case, chaos develops at
the perimeter of the Poincar\'e
section (see the column with $c_5=1$ in Fig.~27).

For $\xi<1$, the integrability associated with the SU(3)
limit is broken
due to the presence of the $(\xi-1)P^{\dagger}_0P_0$ in the
intrinsic Hamiltonian, Eq.~(\ref{H2sqrt2}). The effect of adding the
rotational $c_5$-term on the classical dynamics, is similar to that
of varying the control parameter $\xi$ in the intrinsic Hamiltonian.
This is illustrated in Fig.~28, where different combinations of
$\xi$ and $c_5$, which yield the same ratio $R$, give rise to similar
Poincar\'e surfaces. Specifically, the surfaces
obtained with $(\xi=0.85,c_5=-0.9; R=1)$ are similar (but not identical)
to the surfaces at the SU(3) limit $(\xi=1,c_5=0;R=1)$,
(compare the panels of the $R=1$ columns in Fig.~28). Similarly,
the surfaces with $(\xi=0.85,c_5=0; R=0.9)$ in Fig.~28
resemble the surfaces with $(\xi=1,c_5=1; R=0.9)$ in Fig.~27,
and the surfaces with $(\xi=1.1875,c_5=0; R=1.125)$ in Fig.~28,
resemble the surfaces with $(\xi=1,c_5=-1; R=1.125)$ in Fig.~27.
The slight differences are due to anharmonic effects beyond the
normal-mode approximation.

\subsection{Quantum analysis in the presence of collective terms}

The collective Hamiltonian $\hat{H}_{\rm col}(c_3,c_5.c_6)$ of
Eq.~(\ref{eq:Hcol}), is composed of the two-body parts of the Casimir
operators, $\hat{C}_{G}$, of the groups
$G={\rm O(3)},\,{\rm O(5)},\,\overline{{\rm O(6)}}$.
In studying their role in the quantum spectrum, $\hat{C}_{\rm O(3)}$
can be replaced by its eigenvalue $L(L+1)$, and has no effect on the
structure of wave functions.
For that reason, in the quantum analysis presented below, we focus on
the effect of the collective O(5) and $\overline{{\rm O(6)}}$
terms, when added to the intrinsic Hamiltonian in various regions of
the QPT.
As previously done, the regular and irregular nature of the quantum states
is revealed through the use of Peres lattices. Their
symmetry properties are examined by means of Shannon
entropies, Eq.~(\ref{eq:Shannon}), and the Pearson-based SU(3)
correlator, Eq.~(\ref{Pearson}). The former indicate the purity
with respect to U(5) and SU(3), and the latter indicates the
SU(3) coherence. These measures highlight
the (approximate) U(5)-PDS and SU(3)-QDS content of the quantum
eigenstates, across the QPT.

Figs. 29-32 portray the properties of quantum eigenstates
of the full Hamiltonian, Eq.~(\ref{Hfull}).
The intrinsic Hamiltonian $\hat{H}_{\rm int}(\rho,\xi)$, Eq.~(\ref{eq:Hint}),
has $h_2=1$ and $\bz=\sqrt{2}$, and the control parameters $(\rho,\xi)$
are taken at representative values in regions I-II-III of the QPT.
These values, as well as those of the coupling strengths, $c_5$ and $c_6$,
of the added collective terms, are identical to the ones used for
the classical analysis in Figs~24-27. The middle (third) column in
each of the Figures~29-32, displays the Peres lattices $\{x_i,E_i\}$,
Eq.~(\ref{Peresnd}), of eigenstates $\ket{i}$ with $N=50$ and $L=0,2,4,6$,
which are overlayed on the corresponding classical potentials
$V(x,y=0)$, Eq.~(\ref{Vclxy}). These potentials are unchanged when
the collective terms are included in the Hamiltonian.
As before, ordered (disordered) meshes of lattice points, identify regular
(irregular) type of states. The first, second and fourth columns
from the left, display for each energy eigenstate $E_i$,
the U(5) Shannon entropy $S_{\rm U5}(L_i)$, Eq.~(\ref{Shannonu5}),
for $L=0,2$, and the SU(3) Shannon entropy $S_{\rm SU3}(L_i)$,
Eq.~(\ref{Shannonsu3}) for $L=0$, respectively.
The range in these quantities is  $0\leq S_{\rm G}(L_i)\leq 1$,
with $S_{\rm G}(L_i)=0$ $(S_{\rm G}(L_i)>0)$ indicating
purity (mixing) with respect to $G=U(5),\,SU(3)$.
The right-most column displays the values of the
SU(3) correlation coefficient $0\leq C_\mathrm{SU3}(0{\rm -}6)\leq 1$,
Eq.~(\ref{Pearson}), correlating sequences of $L=0,2,4,6$ states
for each eigenstate $L=0_i$ with energy $E_i$.
The value $C_\mathrm{SU3}(0{\rm -}6)\approx 1$ indicates
a highly-correlated sequence, comprising a $K=0$ band
and manifesting SU(3)-QDS.
Slight departures, $C_{\rm SU3}(0{\rm -}6)<1$, indicate
a reduction in the SU(3) coherence.

The intrinsic Hamiltonian in region~I of the QPT is
$\hat{H}_{1}(\rho)$, Eq.~(\ref{eq:H1}), with $0\leq\rho\leq\rho^{*}$.
For $\rho\!=\!0$, it has U(5) DS and a solvable spectrum,
Eq.~(\ref{U5-DSspec}). The added collective $c_5$-term conforms with the
dynamical symmetry, the eigenstates remain the U(5) basis
states, $\ket{N,n_d,\tau,n_{\Delta},L}$ Eq.~(\ref{u5ds}),
and hence satisfy $S_{\rm U5}(L_i)=0$.
The combined spectrum becomes
$E_i \!=\! 2\bar{h}_2[2N\!-\!1 - n_d]n_d + c_{5}[\tau(\tau\!+\!3)- 4n_d]$,
which explains the observed spreading in the Peres lattice with $c_5=1$
in Fig.~29. The energies of the lowest $L=\tau=0$ states still follow the
potential curve $V_{1}(\rho=0)$, Eq.~(\ref{V1rho0}).
The $c_6$-term breaks the U(5) symmetry, inducing considerable
$\Delta n_d=\pm 2$ mixing, but retains the
O(5) symmetry and quantum number $\tau$. Accordingly, the U(5) Shannon
entropies are non-zero, as seen for $c_6=1$ in Fig.~29.
Nevertheless, a few low-lying (as well as high-lying) states exhibit
a low value of $S_{\rm U5}(L)$, indicating the persistence of an (approximate)
U(5)-PDS in the presence of the $\overline{O(6)}$ term.
The energies of the lowest $L=\tau=0$ states in the Peres lattice deviate
now from $V_{1}(\rho=0)$. In all cases considered in Fig.~29, with and
without the collective terms, the eigenstates in question are spherical
in nature, hence exhibit considerable SU(3) mixing
$(S_{\rm SU3}(L=0)\approx 1)$ and lack of SU(3) coherence
$(C_{\rm SU3}(0{\rm -}6)<1)$.
\begin{figure}[!t]
\begin{center}
\epsfig{file=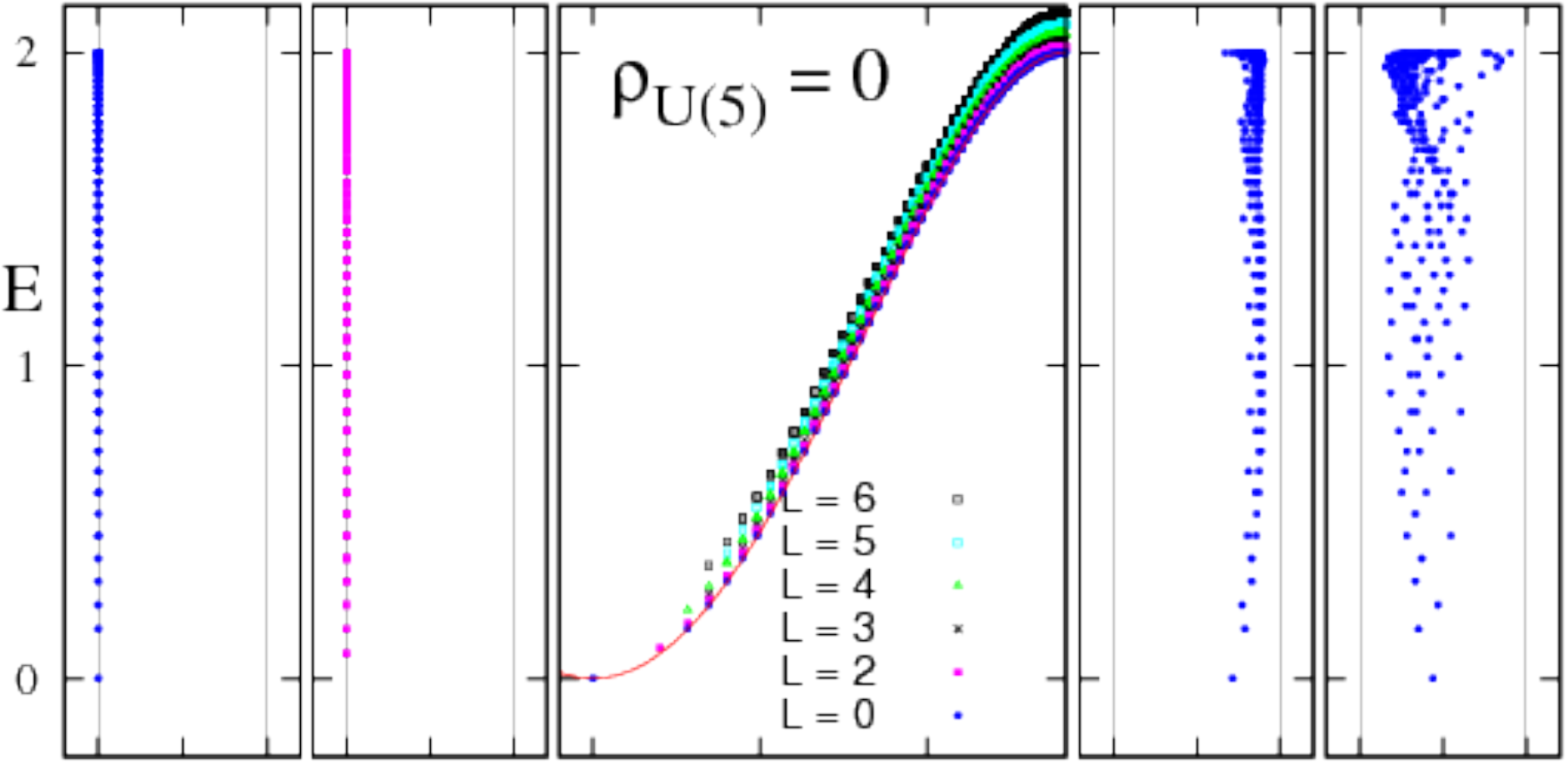,width=0.7\linewidth}
\epsfig{file=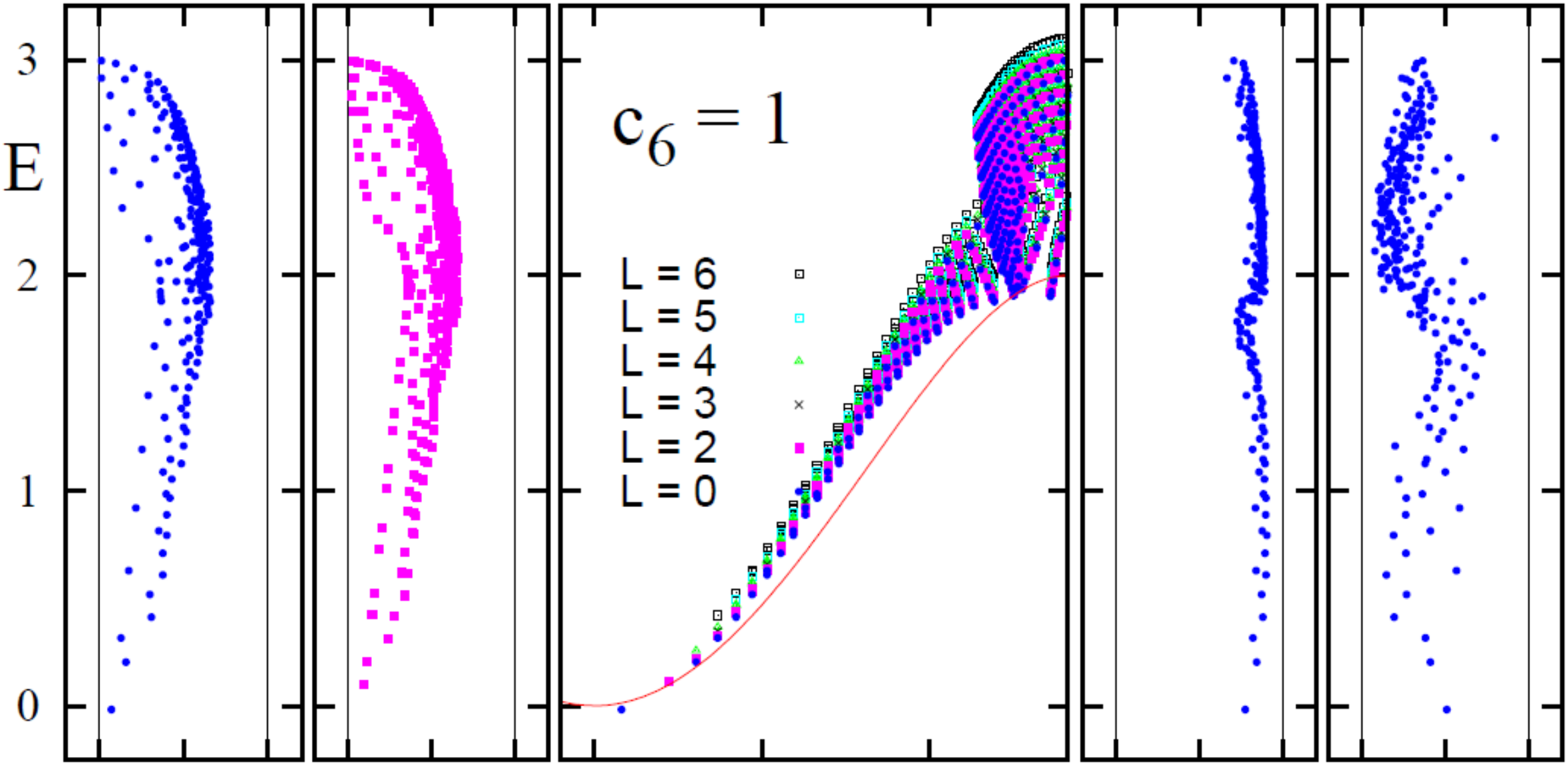,width=0.7\linewidth}
\epsfig{file=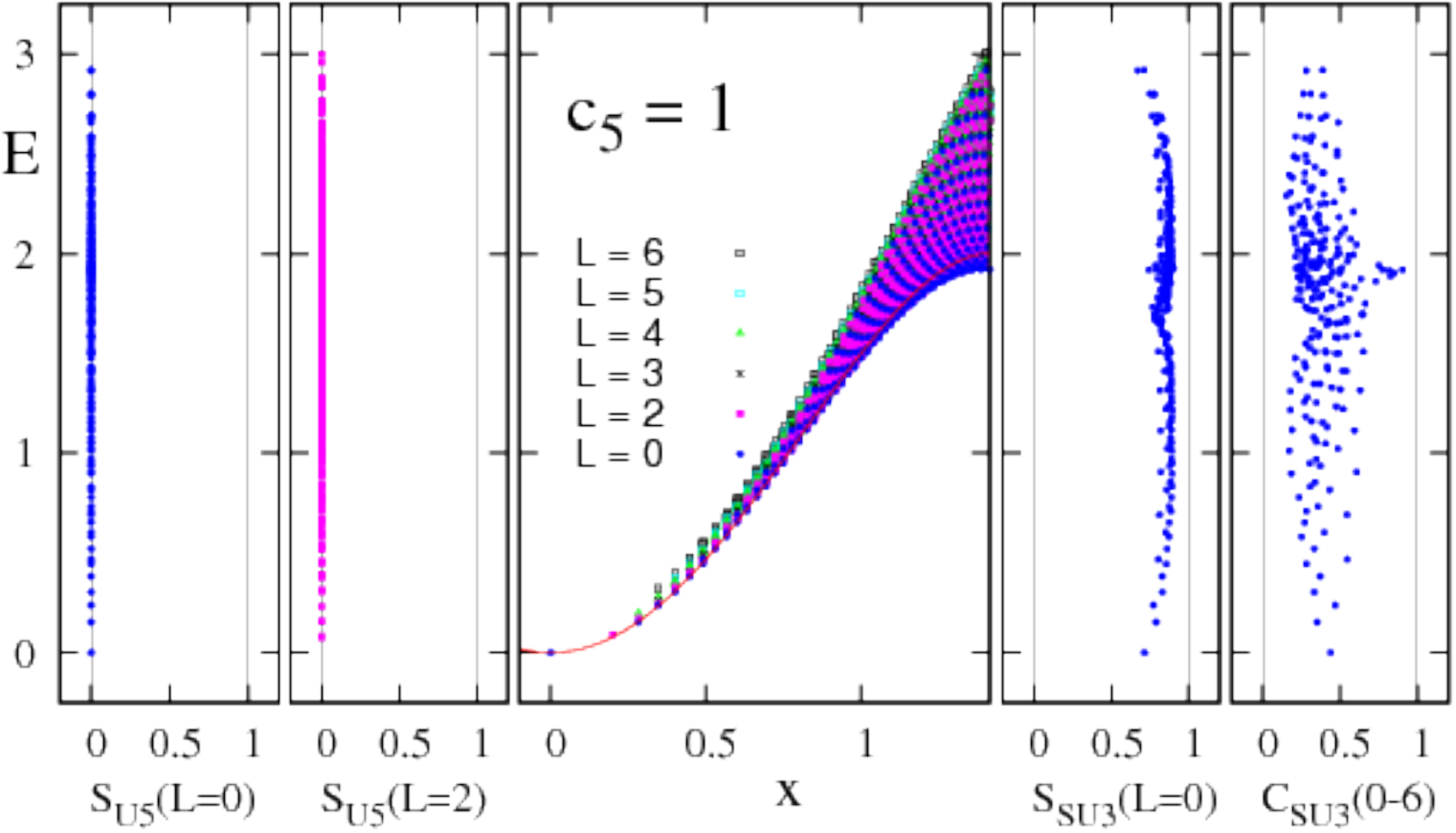,width=0.7\linewidth}
\end{center}
\caption{
U(5) Shannon entropy
$S_{\rm U5}(L=0)$ and $S_{\rm U5}(L=2)$, Eq.~(\ref{Shannonu5}),
Peres lattices $\{x_,E_i\}$, Eq.~(\ref{Peresnd}),
overlayed on the classical potential,
SU(3) Shannon entropy $S_{\rm SU3}(L=0)$, Eq.~(\ref{Shannonsu3}),
and SU(3) correlation coefficient $C_{\rm SU3}(0{\rm -}6)$,
Eq.~(\ref{Pearson}).
The $L=0,2,3,4,5,6$ eigenstates are those of the
intrinsic Hamiltonian, Eq.~(\ref{eq:Hint}),
with $h_2=1,\,\bz=\sqrt{2},\,N=50$,
at the U(5) DS limit ($\rho=0$, top row),
and of added $\overline{\rm O(6)}$ ($c_6=1$, middle row)
and O(5) ($c_5=1$, bottom row) collective terms, Eq.~(\ref{eq:Hcol}).}
\label{fig29}
\end{figure}
\begin{figure}[!t]
\begin{center}
\epsfig{file=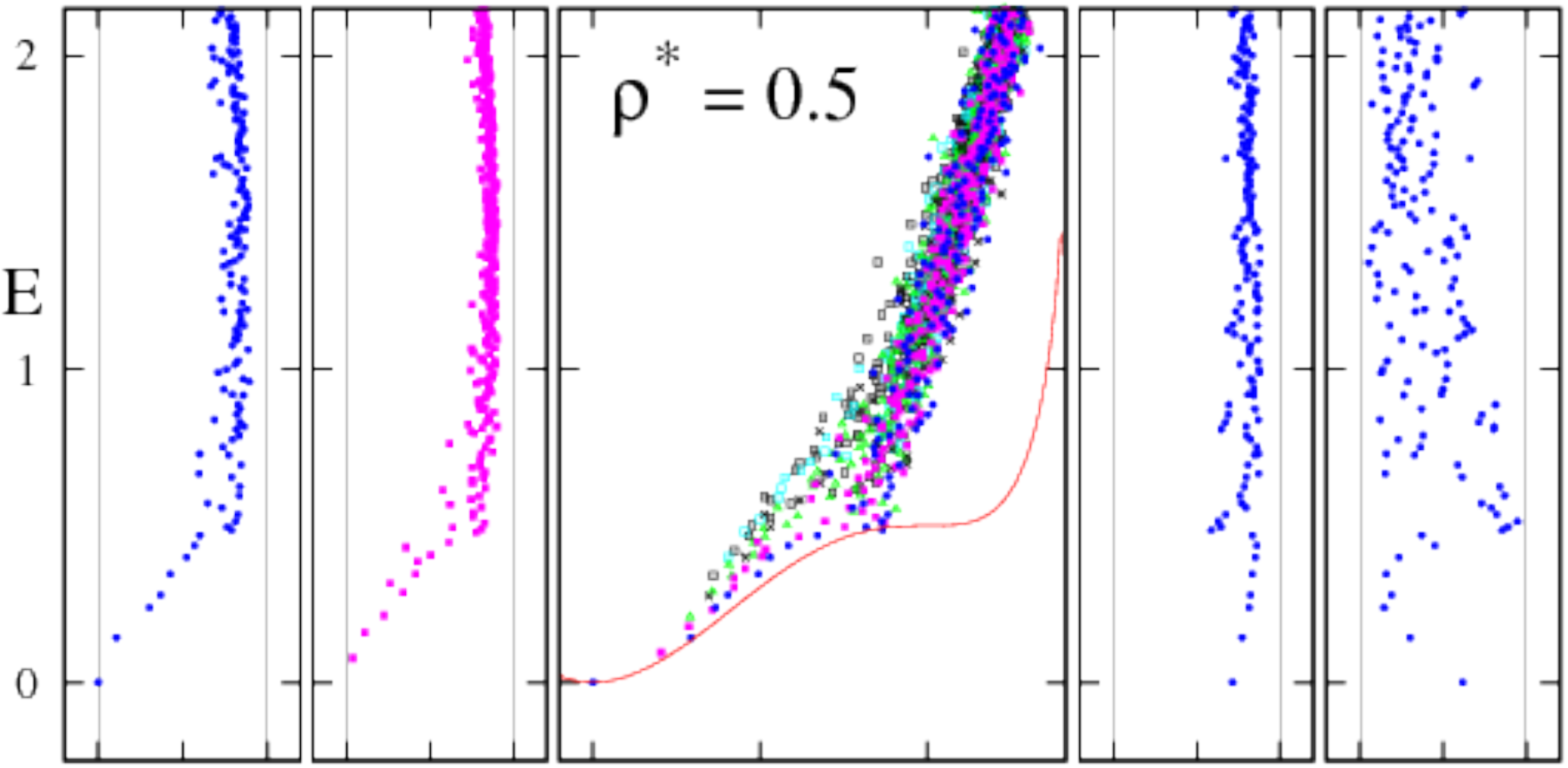,width=0.7\linewidth}
\epsfig{file=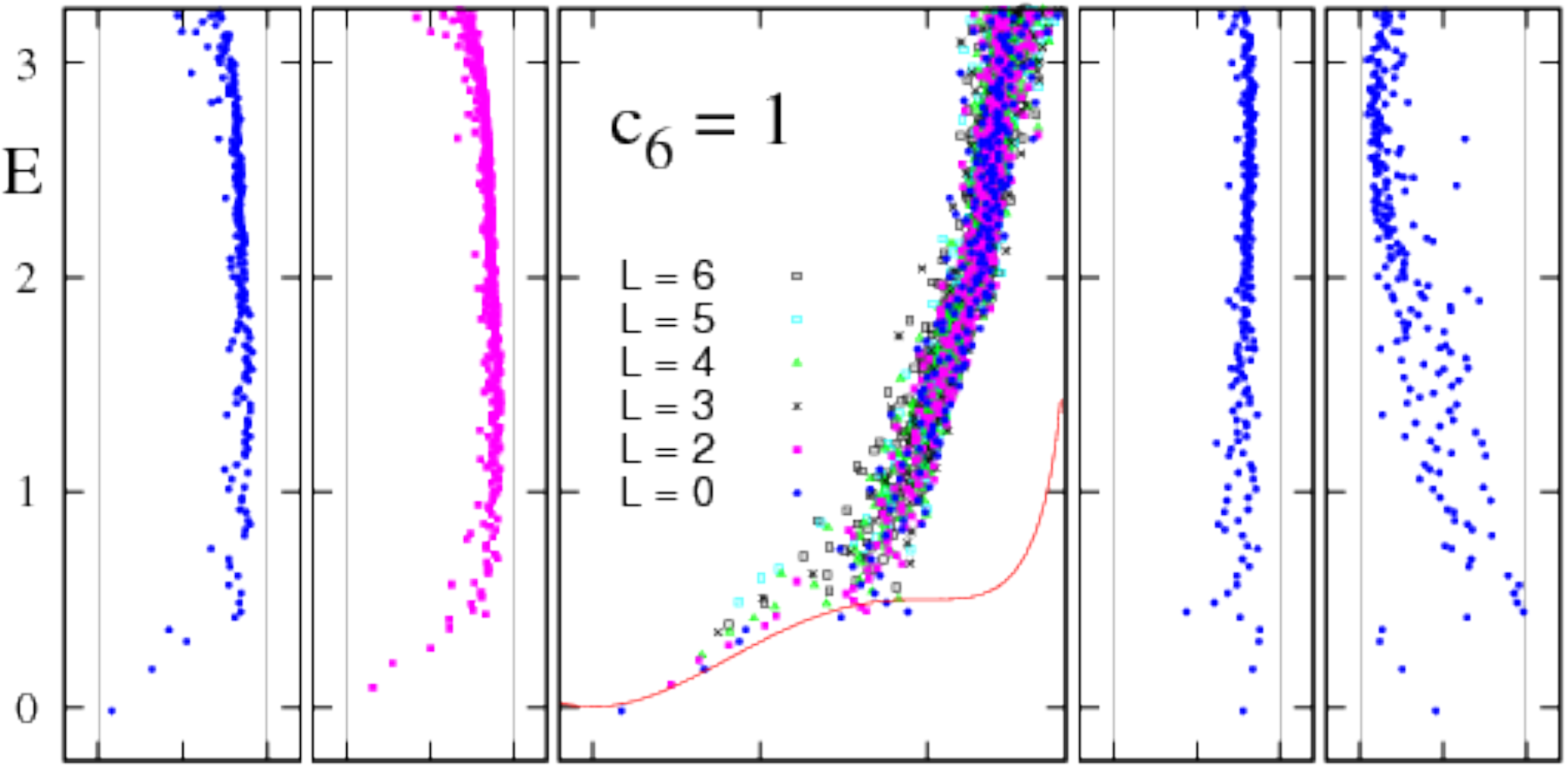,width=0.7\linewidth}
\epsfig{file=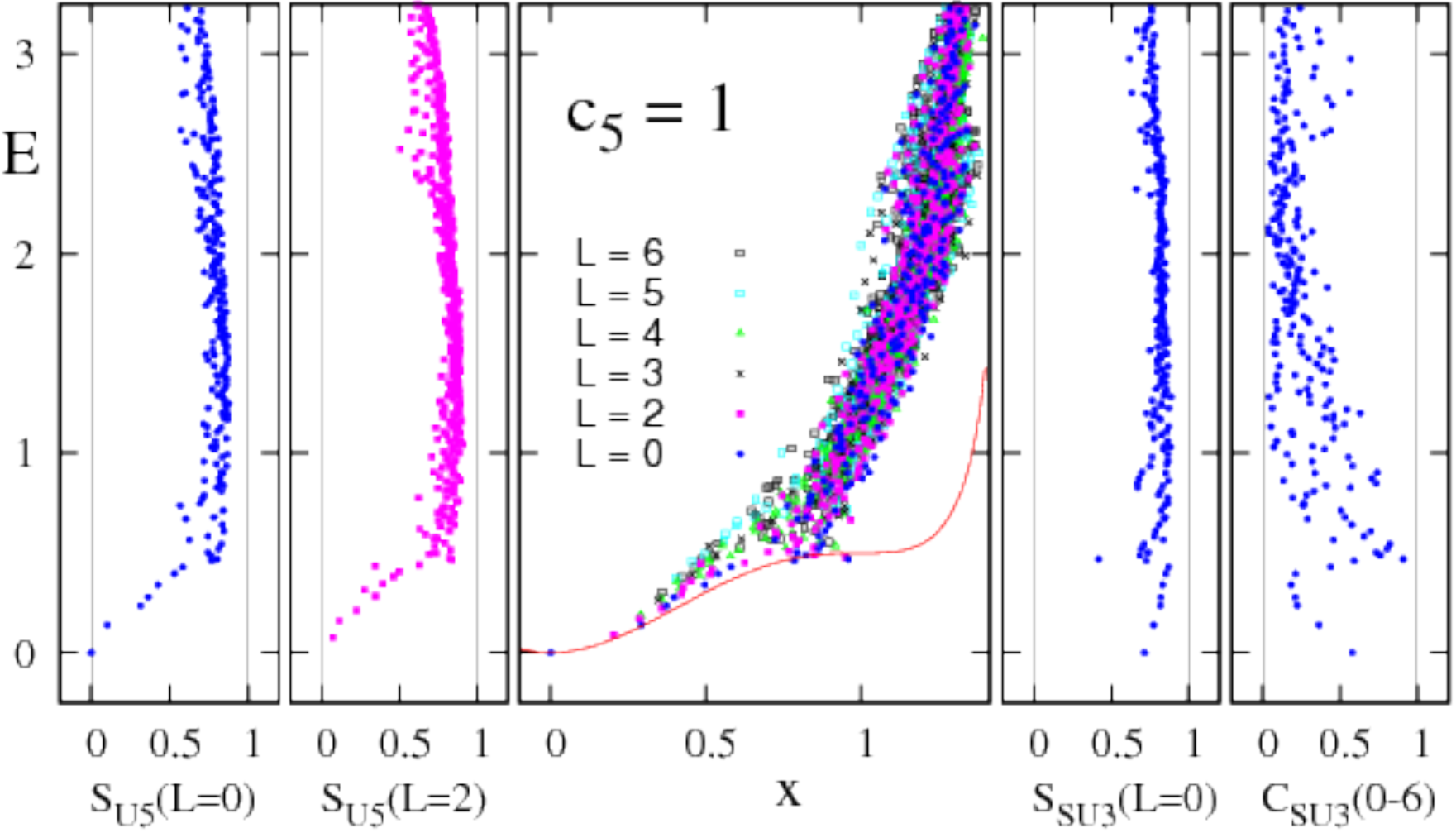,width=0.7\linewidth}
\end{center}
\caption{
Same as in Fig.~29, but at the spinodal point ($\rho^{*}=1/2$).}
\label{fig30}
\end{figure}
\begin{figure}[!t]
\begin{center}
\epsfig{file=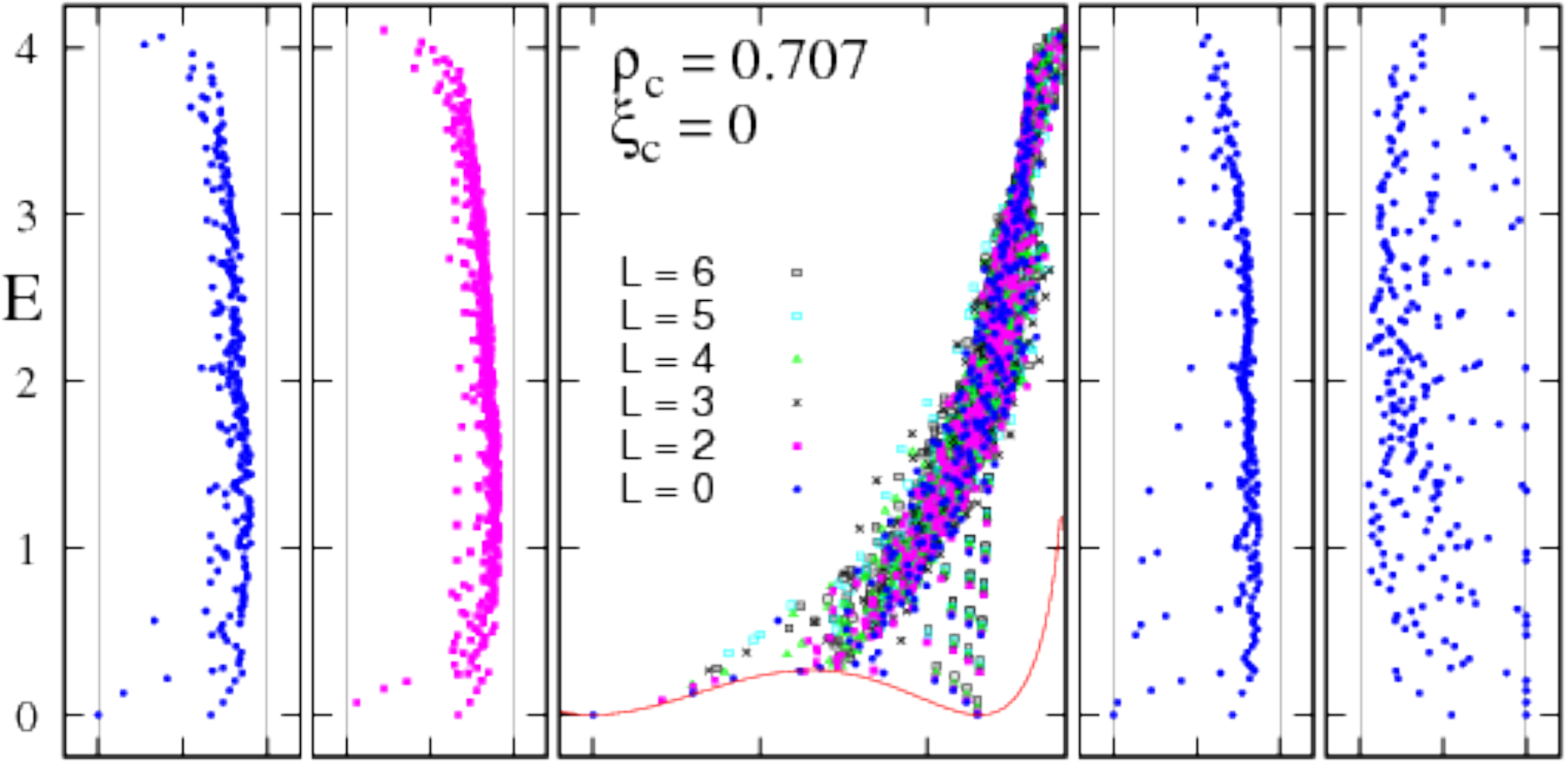,width=0.7\linewidth}
\epsfig{file=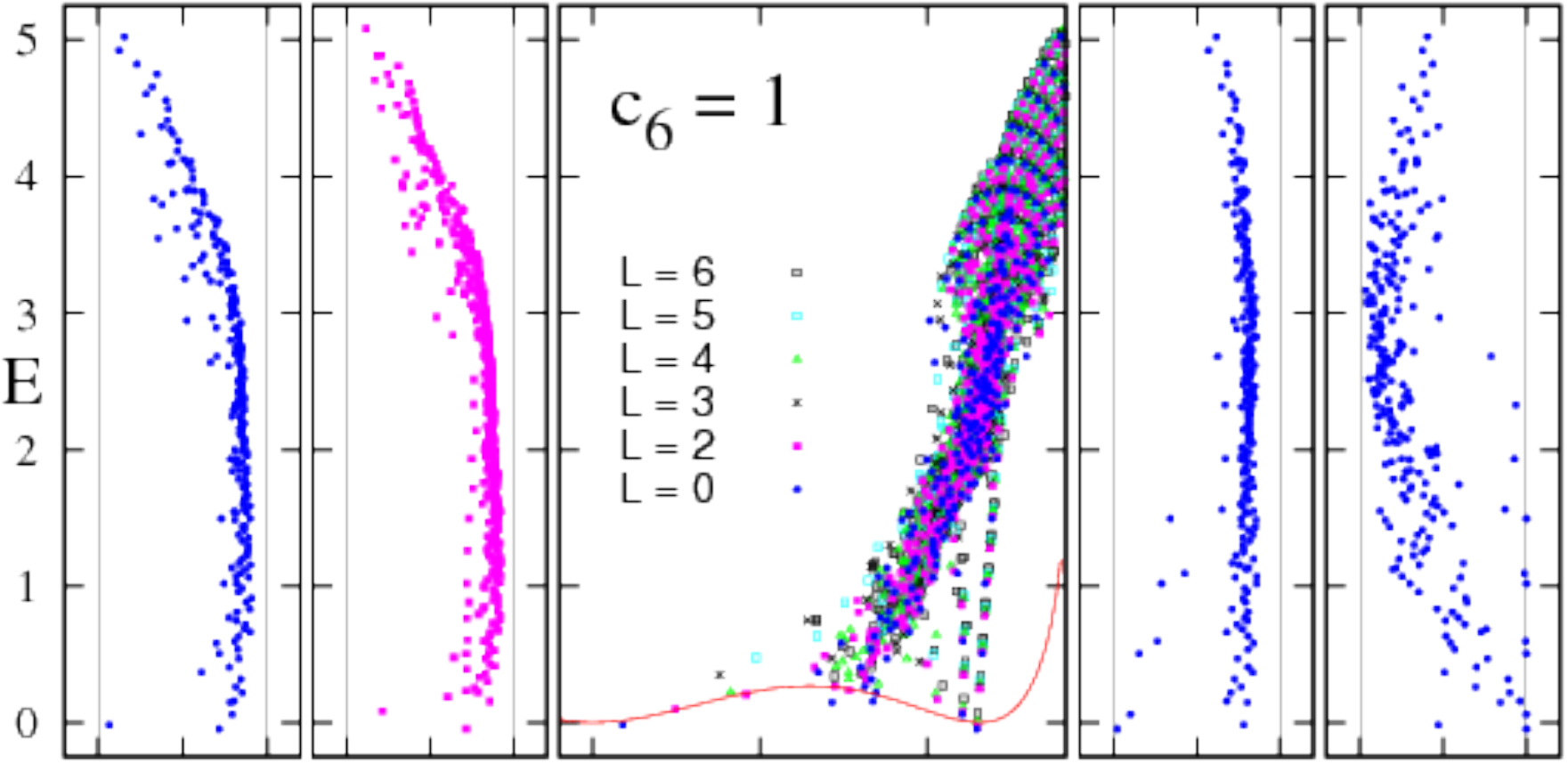,width=0.7\linewidth}
\epsfig{file=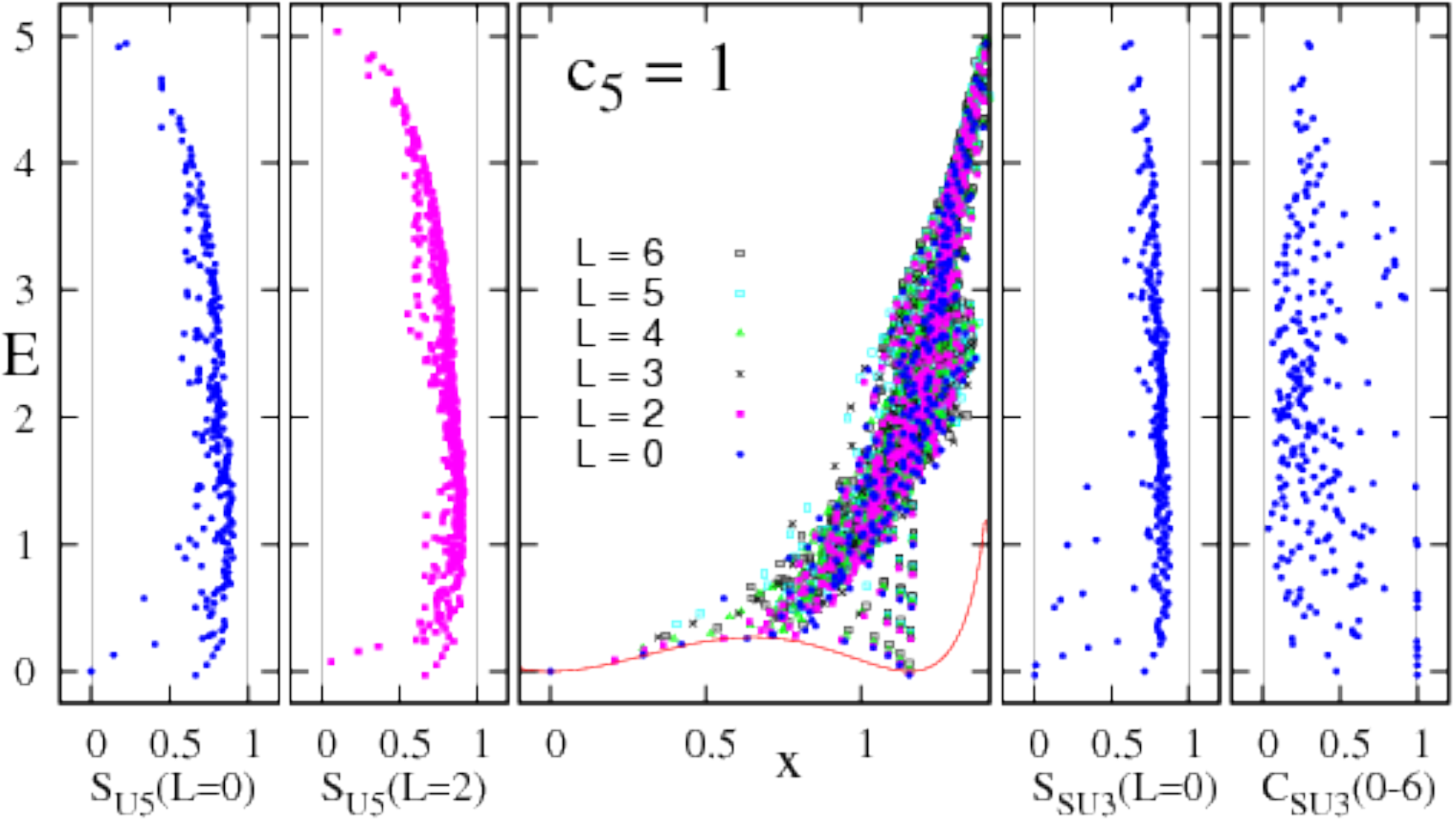,width=0.7\linewidth}
\end{center}
\caption{
Same as in Fig.~29, but at the critical point
($\rho_c=1/\sqrt{2},\xi_c=0$).}
\label{fig31}
\end{figure}
\begin{figure}[!t]
\vspace{-1cm}
\begin{center}
\epsfig{file=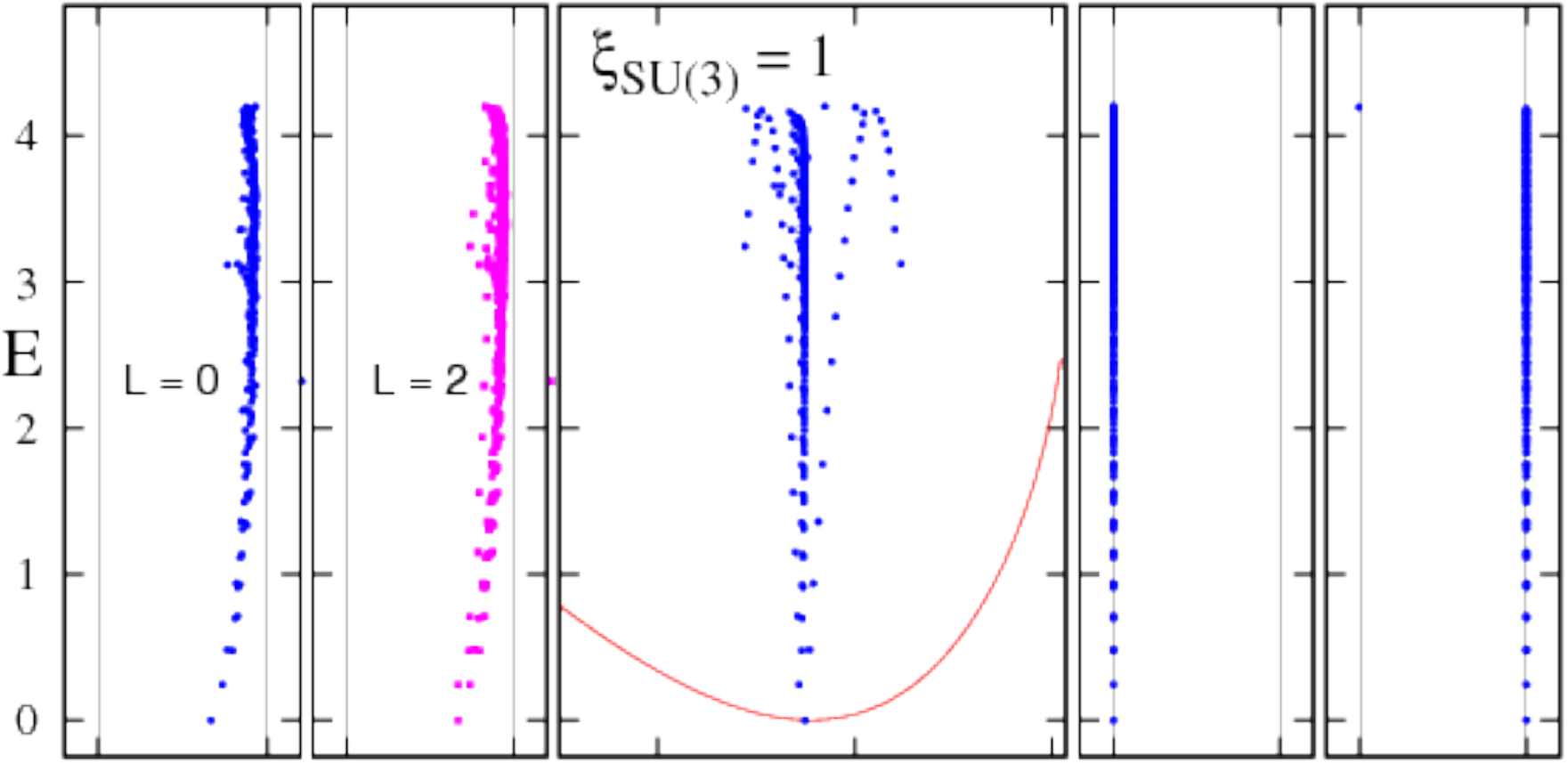,width=0.75\linewidth}
\epsfig{file=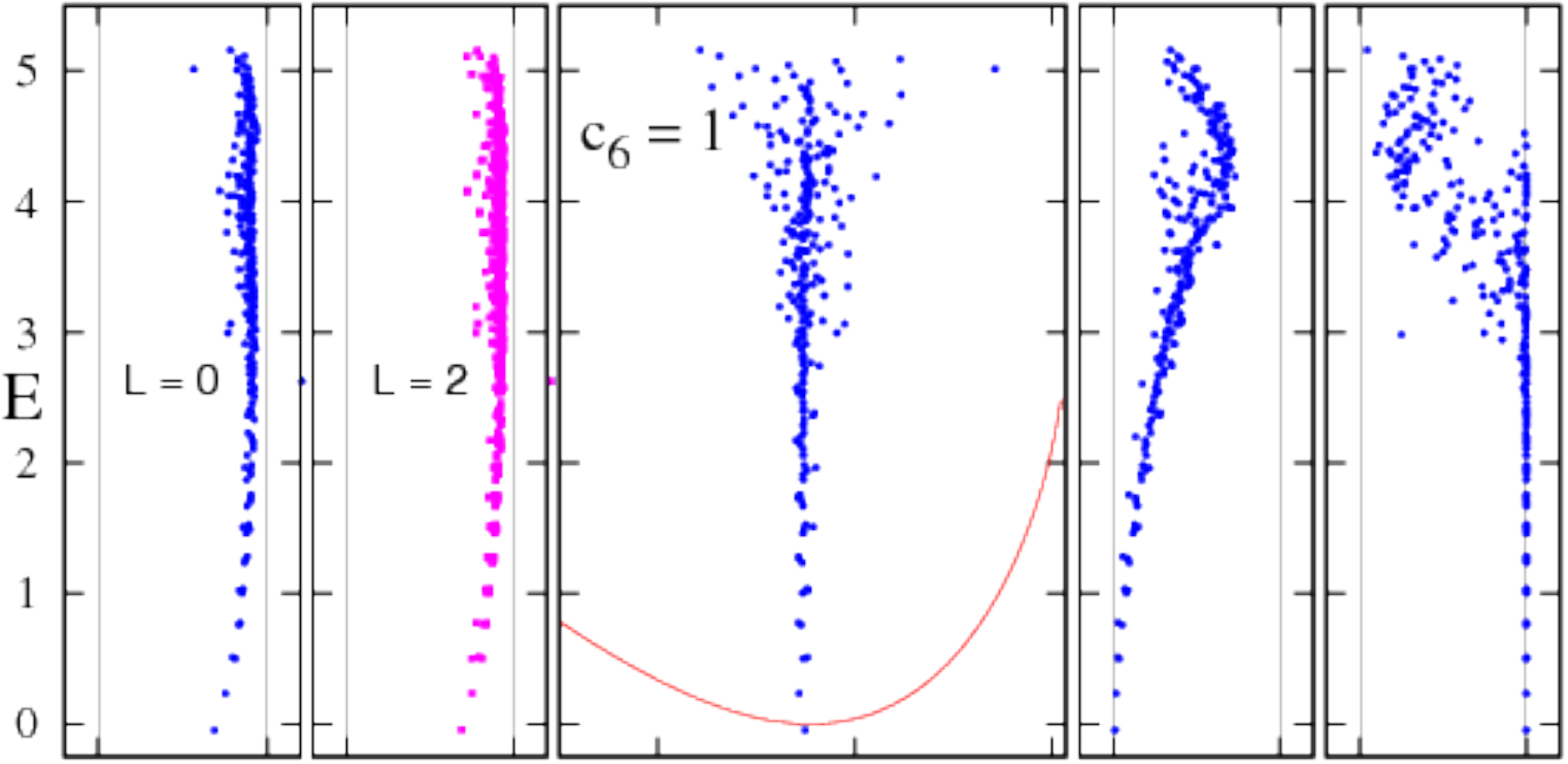,width=0.75\linewidth}
\epsfig{file=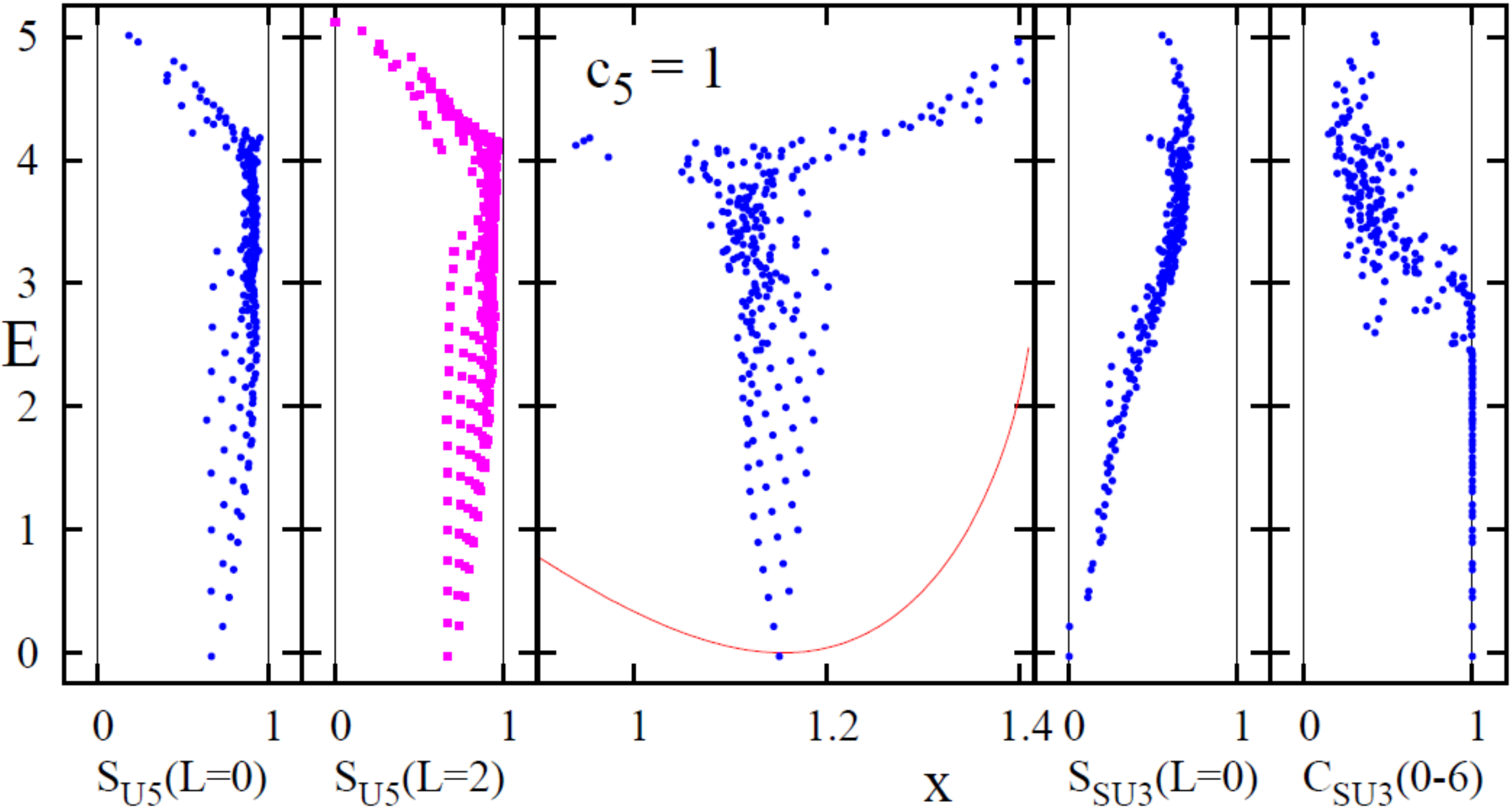,width=0.75\linewidth}\end{center}
\caption{
Same as in Fig.~29, but at the SU(3) DS limit ($\xi=1$).}
\label{fig32}
\end{figure}

For $\rho>0$, the intrinsic Hamiltonian $\hat{H}_1(\rho)$ itself breaks
the U(5) symmetry.
Most of its eigenstates are mixed with respect to U(5)
except for the U(5)-PDS states of Eq.~(\ref{ePDSu5}),
with $n_d=\tau=L=0,\,3$.
The U(5)-PDS property still holds when the $c_5$-term is included,
but is violated by the $c_6$-term. This can be seen in Fig.~30
for the spherical ground state, $L=0_1$, which has $S_{\rm U5}=0$
($S_{\rm U5}>0$) for the $c_5$ ($c_6$) term.
In general, the added collective terms maintain the characteristic
features of the intrinsic quantum dynamics in region~I, namely,
the presence of spherical-type of states at low energy,
with an approximate U(5)-PDS, of more complex-type of states
at higher energy, and the absence of rotational bands, hence
$S_{\rm SU3}(L=0)\approx 1$ and $C_{\rm SU3}(0{\rm -}6)<1$ for all states.
The quantum dynamics in the spherical region is, to a large extent,
determined by the O(5)-breaking
$\rho$-term in the intrinsic Hamiltonian, Eq.~(\ref{Hrho}).
These observations are exemplified in Fig.~30 by the
Peres lattices, Shannon entropies and SU(3) correlator
at the spinodal point $\rho^{*}$, and are
consistent with the classical analysis of Fig~25,
showing the H\'enon-Heiles scenario for the onset of chaos in this region.

The intrinsic Hamiltonian in region~II is
$\hat{H}_{1}(\rho)$, Eq.~(\ref{eq:H1}), with $\rho^{*}< \rho\leq\rho_c$,
and $\hat{H}_2(\xi)$ of Eq.~(\ref{eq:H2}), with $\xi_c\leq\xi<\xi^{**}$.
As discussed in Section~7, the new element entering the
intrinsic quantum dynamics in the shape-coexistence region,
is the occurrence of deformed-type of states forming rotational
$K$-bands, associated with the deformed minimum, coexisting with
low-energy spherical-type of states, associated with the spherical minimum,
in the background of more-complicated type of states at higher-energies.
The regular rotational $K$-bands exhibit coherent SU(3) mixing,
and for $K=0$ bands are signaled by $C_{\rm SU3}(0{\rm -}6)\approx 1$.
As shown for the critical point $(\xi_c=0)$ in Fig~31,
the inclusion of the collective $c_5$-term
maintains these features. In contrast,
the regular band-structure is
disrupted by the inclusion of the $c_6$-term.
The number of quasi-SU(3) bands for which $C_{\rm SU3}(0{\rm -}6)>0.995$,
is now reduced from 12 to 6.
Thus, most of the reduction of SU(3)-QDS is due to the collective
$\overline{O(6)}$ rotations which couple the deformed and
spherical configurations and mix strongly the regular
and irregular states. This disruption of band-structure
is consistent with the $\beta$-distortion of the regular island,
observed in the classical analysis of Fig.~26.
It highlights the importance for QPTs of the coupling of the order
parameter fluctuations with soft modes~\cite{Belitz05}.

The intrinsic Hamiltonian in region~III is
$\hat{H}_2(\xi)$ of Eq.~(\ref{eq:H2}), with $\xi\geq\xi^{**}$.
For $\xi=1$, it has SU(3) DS and a solvable spectrum,
Eq.~(\ref{SU3-DSspec}). The added collective $c_5$- and $c_6$ terms
both break the SU(3) symmetry and consequently, as seen in Fig.~32,
the SU(3) Shannon entropy in both cases is positive, $S_{\rm SU3}(L)>0$.
At low and medium energies, ($E\leq 3$ for the $c_5$ term
and $E\leq 4.5$ for $c_6$ term),
the SU(3) mixing is coherent and the $L$-states are still arranged in
rotational bands. The number of such regular $K=0$ bands is smaller for the
$c_5$-term, consistent with the classical analysis of Fig.~27, showing
a more pronounced onset of chaos in the $\gamma$-motion due to the
O(5) rotational term. At higher energies, the SU(3)-QDS property is
dissolved due to mixing with other types of states.
In general, there are no spherical-type of states in region~III, and
the U(5) entropy is positive,
$S_{\rm U5}(L)>0$, in all panels of Fig.~32. This is in line
with the fact that the classical Landau potential
has a single deformed minimum in this region.

\section{Height of the barrier}\label{sec:BarHeight}

All calculations presented so far, were performed at a fixed value of
$\bz=\sqrt{2}$,
ensuring a high barrier $V_{\rm bar}/h_2 = 0.268$, Eq.~(\ref{Vbarcri}),
at the critical point $(\xi_c=0)$. For this choice, the intrinsic
Hamiltonian $\hat{H}_{2}(\xi;\bz=\sqrt{2})$, Eq.~(\ref{H2sqrt2}),
attains the SU(3) limit for $\xi=1$ and
exhibits SU(3)-PDS for states in the ground and selected gamma bands
of Eq.~(\ref{solsu3}), throughout the deformed region, $\xi\geq\xi_c$.
A variation of the  parameter $\bz$ in the intrinsic Hamiltonian,
Eq.~(\ref{eq:Hint}), affects the symmetry properties of quantum
states and the morphology of the classical potential, in
particular, the height of the barrier.
In the present Section we examine the implied changes in the dynamics in
the coexistence region of the QPT, reflecting the impact of different
barrier heights.
\begin{figure}[!t]
\vspace{-0.5cm}
\begin{center}
\epsfig{file=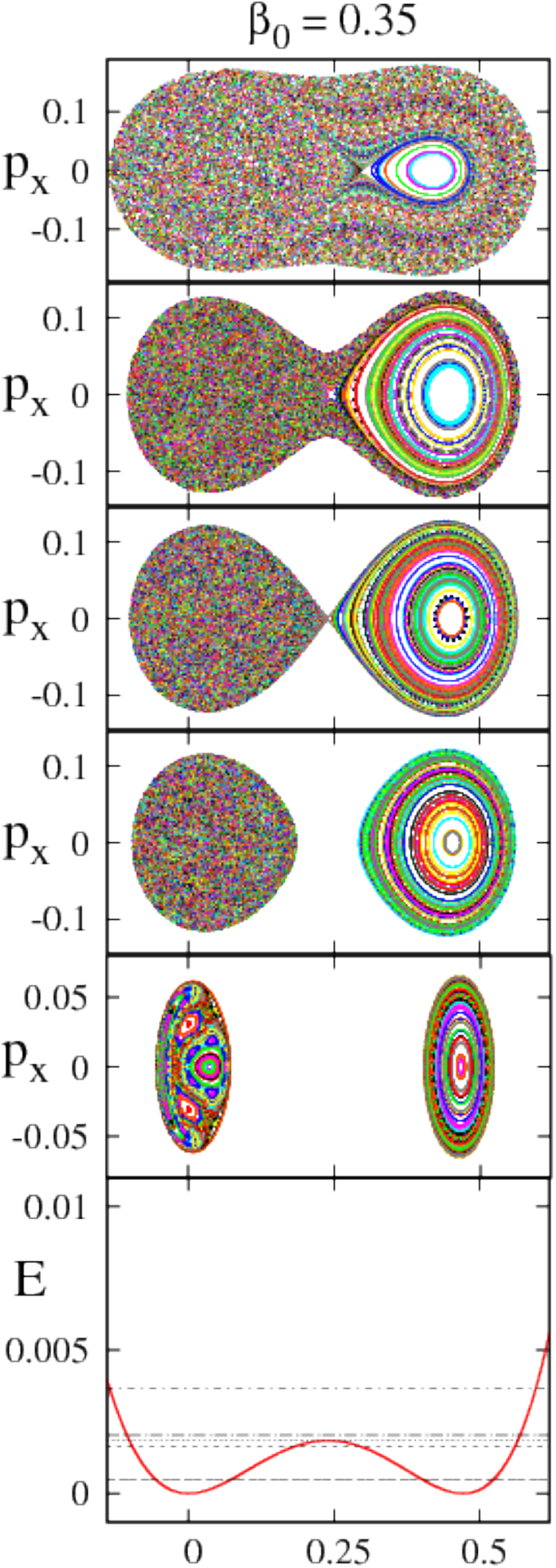,width=0.305\linewidth}
\epsfig{file=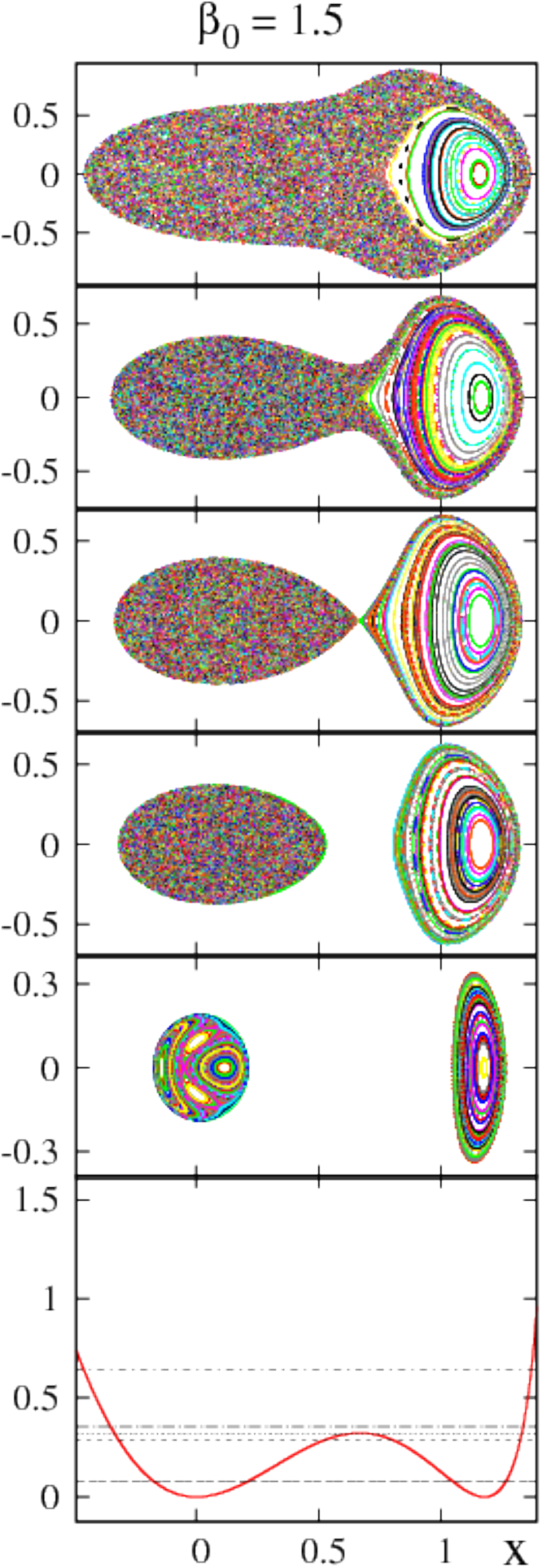,width=0.3000\linewidth}
\epsfig{file=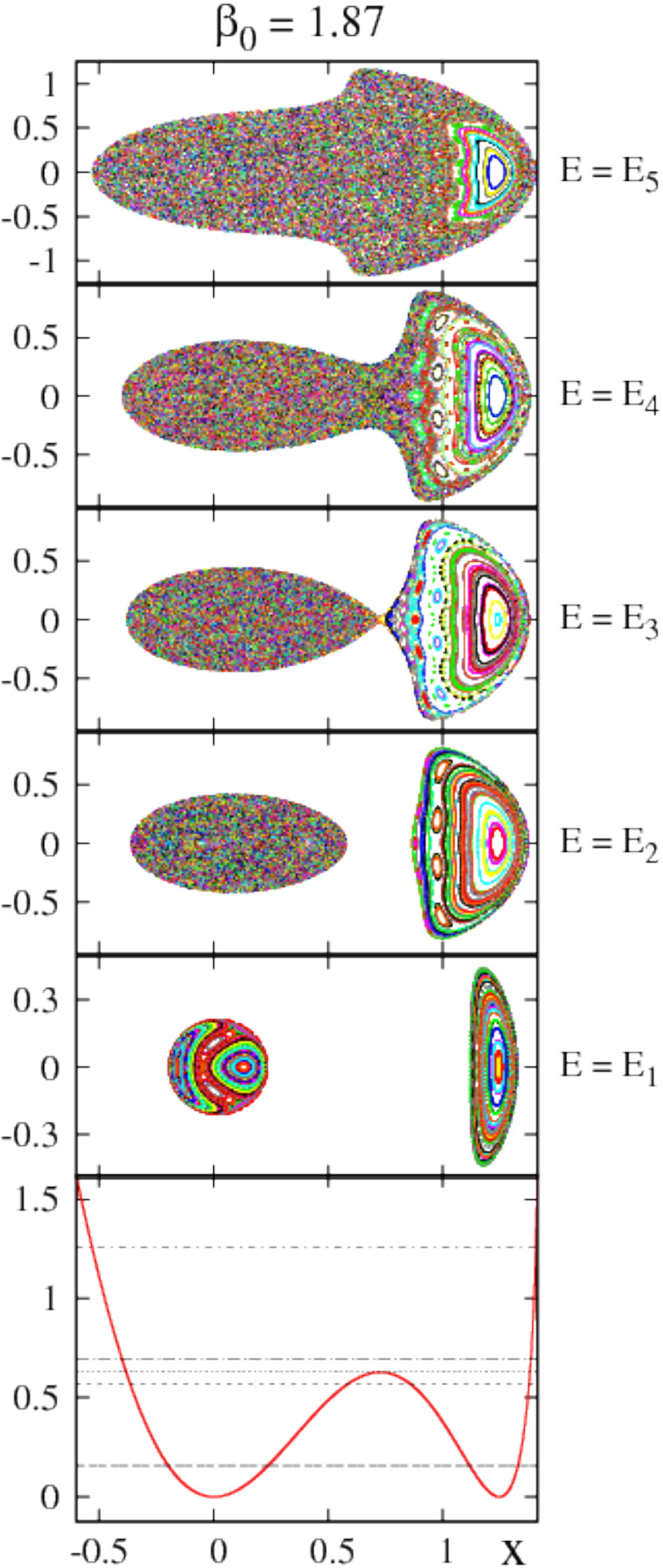,height=14.2cm}
\end{center}
\caption{
Poincar\'e sections of critical-point Hamiltonians
${\cal H}_{1}(\rho_c;\bz) = {\cal H}_{2}(\xi_c;\bz)$, Eq.~(\ref{eq:Hintcl}),
with $h_2=1$ and $\bz = 0.35, 1.5, 1.87$,
corresponding to different barrier heights
$V_{\rm bar}/h_2= {\rm 0.0018,\, 0.322,\, 1.257}$, respectively.
The value at the domain boundary is $V_{\rm lim}/h_2 = 2$.
The bottom row displays the corresponding classical potentials
$V_{1}(\rho_c;x,y=0)=V_{2}(\xi_c;x,y=0)$, Eq.~(\ref{Vclxy}).
The five energies at which the sections were calculated consecutively,
are indicated by horizontal lines.
Note the different vertical and horizontal scales for the
panels in the $\bz=0.35$ column.}
\label{fig33}
\end{figure}

Focusing the discussion to the intrinsic dynamics at
the critical point $(\rho_c,\xi_c)$,
the Poincar\'e sections for the classical Hamiltonian
${\cal H}_{1}(\rho_c)={\cal H}_{2}(\xi_c)$, Eq.~(\ref{eq:Hintcl}),
with $\bz= {\rm 0.35,\,1.5,\,1.87}$,
are displayed in the left, center and right
columns of Fig.~33, respectively.
The three cases correspond to potential barriers
$V_{\rm bar}/h_2= {\rm 0.0018,\, 0.322,\, 1.257}$,
compared to the value at the domain boundary, $V_{\rm lim}/h_2 = 2$.
The bottom row depicts the corresponding classical potentials
$V_{\rm cri}(\beta,\gamma=0)=V_{\rm cri}(x,y=0)$, Eq.~(\ref{eq:Vcri}).
Apart from an energy scale, the three cases
display similar trends, namely, a H\'enon-Heiles type of transition,
with increasing energy, from regular to chaotic motion
in the vicinity of the spherical well,
and regular dynamics in the vicinity of the deformed well.
The extremely low-barrier case displayed in the left column of Fig.~33,
is obtained for $\bz=\frac{1}{2\sqrt{2}}$, which is
the value of the equilibrium deformation for the critical-point Hamiltonian,
Eqs.~(\ref{HndQQ})-(\ref{HcrindQQ}), with $\chi=-\frac{\sqrt{7}}{2}$.
The small energy scale explains
why the simple pattern of coexisting but well-separated regular and chaotic
dynamics in the coexistence region, has escaped the attention
in all previous works which employed the Hamiltonian of Eq.~(\ref{HndQQ}).
This highlights the benefits gained by using
the intrinsic-collective resolution of the Hamiltonian, Eq.~(\ref{eq:H}),
and the ability to construct Hamiltonians accommodating a high-barrier,
in order to uncover in a transparent manner the rich dynamics in the
coexistence region of the QPT.

In spite of the overall similarity, some differences can be detected
between the classical dynamics with $\bz<1$ and $\bz>1$. In the former case,
the onset of chaos occurs at a lower energy, as demonstrated in Fig.~33.
This can be attributed to the different relative weights of the harmonic
term, $\bz^2\beta^2$, and the chaos-driving term,
$\bz\sqrt{2-\beta^2}\beta^3\cos 3\gamma$ in the Landau potential,
$V_{\rm cri}(\beta,\gamma)$, Eq.~(\ref{eq:Vcri}).
The value of $\bz$ affects also the ratio $R$ of normal-mode
frequencies of oscillations about the deformed minimum, Eq.~(\ref{R}).
As noted in Section~5.3, the number of islands in a Poincar\'e Brikhoff
chain is $2/R$, hence decreases with $R\propto(1+\bz^2)$.
Accordingly, the island chains are more visible and the resonance
structure is more pronounced for larger values of $\bz$
(see the column for $\bz=1.87$ in Fig.~33).
\begin{figure}[!t]
\vspace{-1cm}
\begin{center}
\epsfig{file=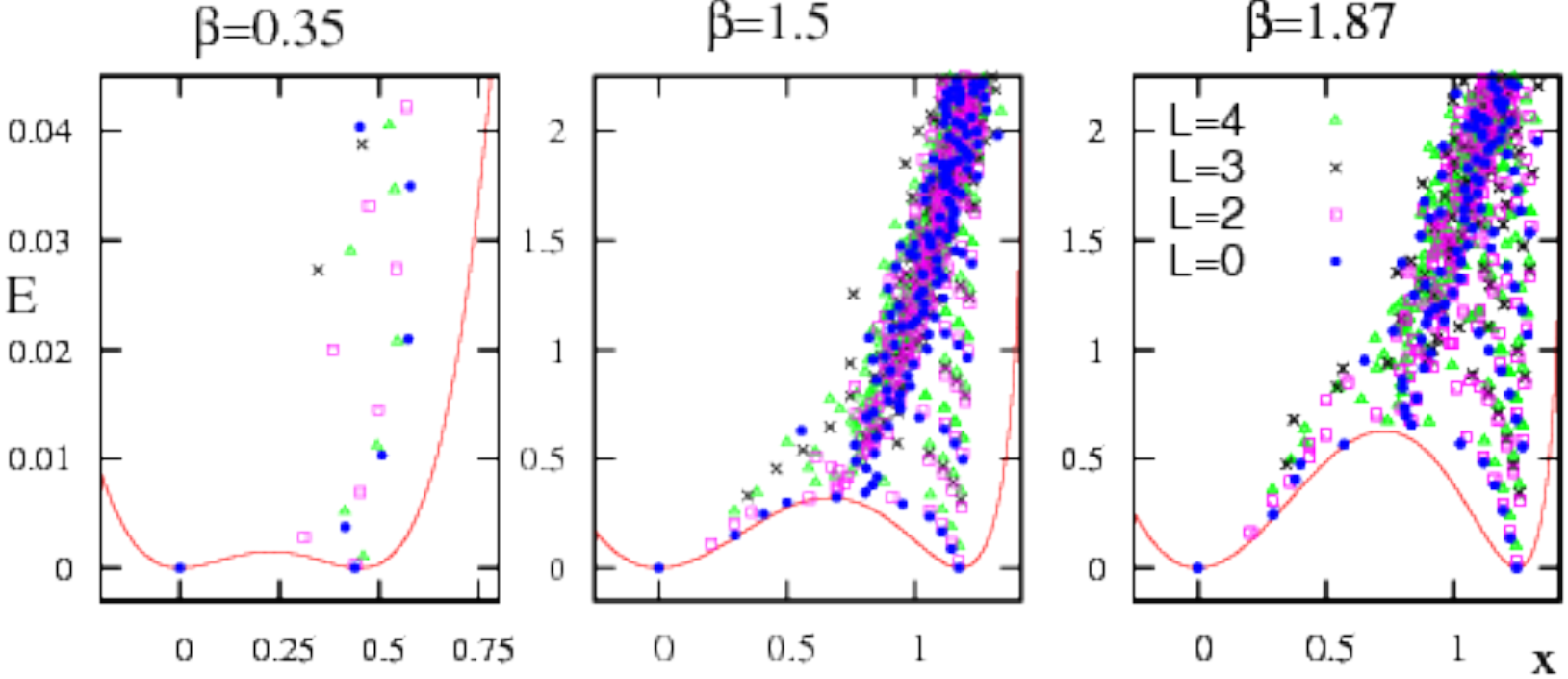,width=0.98\linewidth}
\end{center}
\caption{
Quantum Peres lattices $\{x_i,E_i\}$ for $L=0$ eigenstates $\ket{i}$
of $\hat{H}_\mathrm{cri}$~(\ref{Hcri}), with $h_2=1$, $N=50$ and
$\bz=0.35,1.5,1.87$. The lattices are overlayed on the same classical
potentials as in Fig.~33, accommodating different barrier heights.
Note the different vertical and horizontal scales for the
panel with $\bz=0.35$.}
\label{fig34}
\end{figure}

Fig.~34 presents the quantum Peres lattices calculated for
$(N=50,\,L=0,2,3,4)$
eigenstates of the critical-point intrinsic Hamiltonian, Eq.~(\ref{Hcri}),
with $\bz=0.35,\,1.5,\,1.87$,
the same values used for the classical Poincar\'e sections in Fig.~33.
In each case, one can clearly
identify regular sequences of $K=0,2,4$ bands
localized within and above the respective deformed wells,
and persisting to energies well above the barriers.
The number of such
$K$-bands is larger when the potential is deeper (larger $\bz$ values).
To the left of the barrier towards the
spherical minimum, one observes a number of low-energy
U(5)-like multiplets, Eq.~(\ref{u5mult}). This spherical
multiplet-structure is very pronounced for $\bz=1.5,\,1.87$
(high barriers) and only part of it survives for $\bz=0.35$
(extremely low barrier).

For $\bz\neq\sqrt{2}$, the intrinsic Hamiltonian,
$\hat{H}_{2}(\xi;\bz)$, Eq.~(\ref{eq:H2}),
no longer possess the SU(3) PDS property, Eq.~(\ref{solsu3}).
All eigenstates are mixed with respect to SU(3),
including member states of
the ground and gamma bands. Nevertheless, by construction,
$\hat{H}_{2}(\xi;\bz)$ still satisfies Eq.~(\ref{Hintcond0}),
and hence the states with $L=0,2,4,\ldots,2N$,
projected from the condensate, Eq.~(\ref{condgen}), with
$[\beta_{\rm eq}=\sqrt{2}\bz(1+\bz^2)^{-1/2},\gamma_{\rm eq}=0]$,
span a solvable (but SU(3)-mixed) ground band.
In general, the SU(3) mixing is stronger for larger deviations,
$\abs{\bz-\sqrt{2}}$, and the mixing is coherent for the $L$-states
in the same $K$-band.
This is illustrated in Fig.~35, which shows the SU(3) decomposition
in the solvable ground band of the critical-point intrinsic Hamiltonian
$\hat{H}_{2}(\xi_c;\bz)$, for two values of $\bz$. For $\bz=1$,
the $L=0$ bandhead state of the ground band has a high-value
for the SU(3) Shannon entropy, $S_{\rm SU3}(L=0)=0.33$, hence is less
pure compared to its counterpart with $\bz=1.5$,
for which  $S_{\rm SU3}(L=0)=0.03$.
In both cases, the ground bands exhibit SU(3) coherence
($L$-independent mixing), with SU(3) correlation coefficients
$C_{\rm SU3}(0{\rm -}6)=1$, exemplifying SU(3)-QDS.
\begin{figure}[!t]
\begin{center}
\epsfig{file=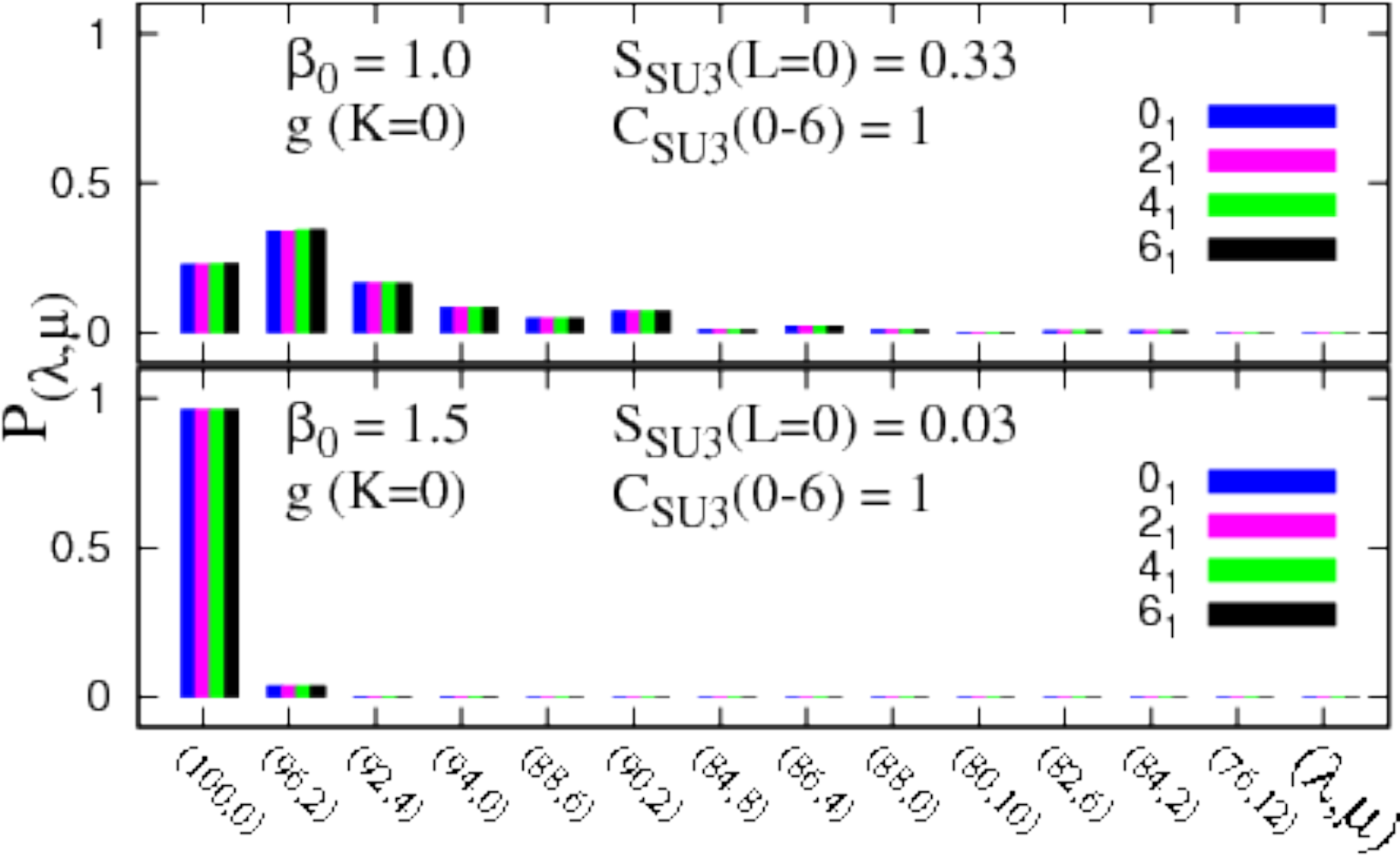,width=1.0\linewidth}
\end{center}
\caption{
SU(3) probability distribution $P_{\lambda,\mu)}^{(L)}$, Eq.~(\ref{Plammu}),
for ($N=50,\,L=0,2,4,6$) states, members of the ground band
of the intrinsic critical-point Hamiltonian
$\hat{H}_{1}(\rho_c;\bz)=\hat{H}_{2}(\xi_c;\bz)$,
Eq.~(\ref{eq:Hint}), with $\bz=1$ (top) and $\bz=1.5$ (bottom).
Each panel lists the SU(3) Shannon entropy $S_{\rm SU3}(L=0)$,
Eq.(\ref{Shannonsu3}), and the SU(3) correlator
$C_{\rm SU3}(0{\rm -}6)$, Eq.~(\ref{Pearson}),
indicating the extent of SU(3) mixing and coherence, respectively.}
\label{fig35}
\end{figure}

\section{Summary and conclusions}

We have presented a comprehensive analysis of the
dynamics evolving across a generic (high-barrier) first-order QPT,
with particular emphasis on aspects of chaos, regularity,
and symmetry. The study was conducted in the framework of the IBM,
a prototype of an algebraic model, whose phases are associated
with dynamical symmetries (DSs) and the transition between them
exemplify QPTs in an interacting many-body system.
The specific model Hamiltonian employed, describes a shape-phase transition
between spherical [U(5) DS] and deformed [SU(3) DS] quadrupole shapes,
a situation encountered in nuclei.
The resolution of the Hamiltonian into intrinsic
(vibrational) and collective (rotational) parts has allowed us to
disentangle the effects due to terms affecting the Landau potential,
from effects due to kinetic terms.
The separate treatment of the intrinsic dynamics
highlights simple features by avoiding distortions
that may arise in the presence of large rotation-vibration coupling.
The availability of IBM Hamiltonians accommodating a high barrier
in a wide range of control parameters, made it possible to uncover
a previously unrecognized pattern of competing
order and chaos, which echoes the QPT in the coexistence region.

A classical analysis of the intrinsic part of the Hamiltonian
revealed a rich mixed dynamics with distinct
features in each structural region of the QPT.
On the spherical side of the QPT,
the system is integrable at the U(5) DS limit. Near it, the phase
space portrait resembles that of a weakly perturbed anharmonic (quartic)
oscillator. In other parts of region~I, where the Landau potential has
a single spherical minimum, the phase space portrait
is similar to the H\'enon-Heiles system,
with regular dynamics at low energy and chaos at higher energy.
The non-integrability here is due to the O(5)-breaking term in
the Hamiltonian.
On the deformed side of the QPT, the system is integrable at the
SU(3) DS limit. Away from it, the integrability is lost by a different
mechanism of breaking the SU(3) symmetry. The dynamics, however, remains
robustly regular throughout region~III, where the Landau potential
supports a single deformed minimum.
The Poincar\'e  sections in this region are dominated by regular
trajectories forming a single island.
Additional chains of regular islands show up, occasionally,
due to resonances in the normal-mode oscillations.
The fact that the classical dynamics evolves differently, is
attributed to the different
topology of the Landau potential in the vicinity of the two minima.
In spite of the abrupt
structural changes taking place,
the dynamics in the phase coexistence region (region~II), exhibits a very
simple pattern where each minimum preserves, to a large extent,
its own characteristic dynamics.
The robustly ordered motion is still confined
to the deformed minimum, in marked separation from the chaotic behavior
ascribed to the spherical minimum. The coexistence of well-separated
order and chaos persists in a broad energy range, even above the barrier,
throughout region~II, and is absent outside it.
The simple pattern of mixed dynamics thus traces
the crossing of the two minima, a defining feature of a first-order QPT.
Simply divided phase spaces are known to occur in
billiard systems~\cite{Bunim01,Dietz07},
where the amount of chaoticity in the motion of a free particle is
governed by the geometry of the cavity. Here, however, they show
up in a many-body interacting system undergoing a QPT, where the onset of
chaos is governed by a variation of coupling constants
in the Hamiltonian.

The quantum manifestations of the classical inhomogeneous phase space
structure have been analyzed via Peres lattices. The latter distinguish
regular from irregular quantum states by means of ordered and disordered
meshes of points. A choice of Peres operator whose classical limit
corresponds to the deformation, allowed us to overlay
the lattices on the classical potentials and thus associate
the indicated states with a given region in phase space. The results
obtained reflect adequately the mixed nature of the classical dynamics.
The distribution of lattice points agrees with the
location of regular and chaotic domains in the classical
Poincar\'e sections. The quantum analysis has disclosed a number of
regular low-energy spherical-vibrator U(5)-like multiplets,
associated with the spherical minimum and
regular SU(3)-like rotational $K$-bands in the vicinity of the
deformed minimum. The latter bands persist to energies
well above the barrier, extend to high values of angular momenta,
and their number is larger for deeper deformed wells.
These two kinds of regular subsets of states
retain their identity amidst a complicated environment of other states,
and both are present in the coexistence region.

An important clue on the nature of the surviving regular sequences
of selected states, comes from a symmetry analysis of their wave functions.
A U(5) decomposition has shown that the above mentioned regular
U(5)-like multiplets consist of spherical type of states,
with wave functions dominated by a single $n_d$ component.
As such, they exhibit U(5) partial dynamical symmetry [U(5)-PDS],
either exactly or to a good approximation.
This enhanced U(5) purity is signaled by a low value of the
U(5) Shannon entropy, Eq.~(\ref{Shannonu5}).
In contrast, the deformed type of states exhibit a broad $n_d$-distribution.
An SU(3) decomposition of the states in the regular $K$-bands
shows a coherent ($L$-independent) SU(3) mixing, exemplifying
SU(3) quasi-dynamical symmetry [SU(3)-QDS]. This pronounced coherence
is signaled by a high value of the Pearson-based SU(3) correlation
coefficient, Eq.~(\ref{Pearson}).
The persisting regular U(5)-like [SU(3)-like] multiplets
reflect the geometry of the Landau potential, as they are associated
with the different spherical (deformed) minimum.
Accordingly, their total number $\nu_{\rm U5}$ ($\nu_{\rm SU3}$)
can be used both as a global measure of the U(5)-PDS
[SU(3)-PDS] present in the system, and as a mirror which captures
the structural evolution of the first-order QPT.
This is demonstrated in Fig.~23, showing the change
in these quantities as a function of the control parameters $(\rho,\xi)$.
The quantity $\nu_{\rm U5}$ ($\nu_{\rm SU3}$) is maximal at the U(5)-DS
[SU(3)-DS] limit and vanishes towards the anti-spinodal (spinodal) point,
where the spherical (deformed) minimum disappears.

The collective part of the Hamiltonian consists of kinetic terms
associated with O(3), O(5) and $\overline{\rm O(6)}$ rotations.
When added to the intrinsic part of the Hamiltonian they lead to
rotational splitting and mixing.
Although these kinetic terms do not affect the Landau potential,
the mixing induced by the O(5) and $\overline{\rm O(6)}$ terms,
can affect the onset of classical chaos and the regular features
of quantum states. An analysis of the classical and quantum
dynamics has shown that
in region~I, the added collective terms being O(5)-invariant,
maintain the H\'enon-Heiles type of dynamics; the onset of chaos
being largely determined by the intrinsic Hamiltonian.
The $\overline{\rm O(6)}$ rotational term, being associated with the
$\beta$ degree of freedom, was found to be significant in the coexistence
region. Its presence disrupts the regular $K$-bands built on the
deformed minimum and reduces their coherence property related to SU(3)-QDS.
The simple pattern of well-separated regular and chaotic dynamics
ascribed to each minimum, however, is not destroyed.
The O(5) rotational term, being associated with the
$\gamma$ degree of freedom, was found to be significant in region~III.
Its presence modifies the regular intrinsic dynamics associated with
the single deformed minimum and leads to significant chaoticity.
The SU(3)-QDS property here is completely dissolved at higher energies.
It is important to note that chaos can develop from purely
kinetic perturbations, without a change in the Landau potential, as
vividly demonstrated in Fig.~27. This illustrates that a criterion
for the onset of chaos cannot be based solely on the geometry of the
potential.

The study of first-order QPTs conducted in this work,
considered a finite system, whose mean-field potential involved two
asymmetric wells, one dominated by chaotic dynamics
the other by regular dynamics.
A parameter $\bz$ in the Hamiltonian governed the height of the
barrier between them. The ramifications of divided phase space
and Hilbert space structure, {\it e.g.}, simple patterns of dynamics
and intermediate symmetries (PDS and QDS),
are observed at any $\bz>0$, but are more pronounced
for higher barriers (larger $\bz$). It will be interesting to see
in future studies of first-order QPTs, whether simply divided spaces occur
also when both wells accommodate regular or chaotic motion but with distinct
characteristic features, {\it e.g.}, different phase space portraits
and dissimilar symmetry structure.
Other issues which deserve further attention are finite-size effects
and scaling behavior. Although there are initial indications that
the simple pattern of mixed dynamics characterizing the QPT, occurs
also at moderate values of $N$, a systematic study is called for.
The large-$N$ scaling behavior should be considered,
in analogy to what has been done in second-order (continuous) QPTs.
An interesting question to address is whether the global
measures of U(5)-PDS and SU(3)-QDS, $\nu_{\rm U5}$ and $\nu_{\rm SU3}$,
shown in Fig.~23, converge to a particular curve for large $N$.

Returning to the key questions posed in the Introduction, we end with
some pertinent remarks.
Based on the results obtained in the present paper, we conclude that
the interplay of order and chaos accompanying the first-order QPT
can reflect its evolution, provided the underlying phase-space is
simply divided and each minimum maintains its own characteristic dynamics.
If these conditions are met, then the resulting simple pattern of
mixed dynamics can trace the modifications in the topology of the
Landau potential inside the coexistence region.
The pattern of mixed but well-separated dynamics is particularly transparent
when considering the intrinsic dynamics, and appears to be robust.
Deviations are largely due to kinetic collective rotational terms,
which may lead to strong rotation-vibration coupling, breakdown
of adiabaticity and an onset of chaos due to purely kinetic perturbations.
The present work suggests that the remaining regularity in the system,
associated with different minima at the classical level,
and with different regular subsets of eigenstates, at the quantum level,
amidst a complicated environment, can be assigned particular intermediate
symmetries, PDS or QDS. Both the classical and quantum analysis
indicate a tendency of a system undergoing a QPT,
to retain some ``local'' regularity far away from integrable limits
and some partial- or quasi form of symmetries far way from symmetry limits.
Is this linkage between persisting regularities
and persisting symmetries a general result or
an observation valid for specific algebraic models?
What are the general conditions for a dynamical system
to have these local regions of regularities and effective symmetries
for subsets of states?
Can one incorporate the notions of quasi- and partial dynamical
symmetries in attempts~\cite{Hose83,Reic87,Wisn11}
to formulate quantum analogs of the KAM and
Poincar\'e-Birkhoff theorems? Quantum phase transitions in many-body
systems and their algebraic modeling provide a fertile ground
for addressing these deep questions. The present work is only a first
step towards accomplishing this goal.

\section*{Acknowledgments}
\label{acknow}

This work is supported by the Israel Science Foundation Grant No.~493/12. 
M.M. acknowledges the Golda Meir Fellowship Fund, the 
Czech Science Foundation (P203-13-07117S) and the partial 
support by U.S. Department of Energy Grant No. DE-FG-02-91ER-40608.

\section*{Appendix A: the IBM potential surface}
\label{Appendix}

The normal-order form of the most general IBM Hamiltonian with one- and
two-body interactions is given by
\ba
\hat{H}_{IBM} &=&
\epsilon_s\,s^{\dag}s
+ \epsilon_{d}\,d^{\dag}\cdot \tilde{d}
+ u_{0}\,(s^{\dag})^2s^2
+ u_{2}\,s^{\dag}d^{\dag}\cdot\tilde{d}s
+ v_{0}\,\left [\,
(s^{\dag})^2\tilde{d}\cdot \tilde{d} + H.c. \, \right ]
\nonumber\\
&&
+ v_{2}\,\left [\,
s^{\dag}d^{\dag}\cdot (\tilde{d} \tilde{d})^{(2)} + H.c. \, \right ]
+ \sum_{L=0,2,4}c_{L}\,(d^{\dag}d^{\dag})^{(L)}\cdot
(\tilde{d}\tilde{d})^{(L)} ~,
\label{hIBMnormal}
\ea
where $H.c.$ means Hermitian conjugate.
As noted in Section~2, a potential surface $V(\beta,\gamma)$,
Eq.~(\ref{enesurf}), is obtained by the expectation value of the Hamiltonian
in an intrinsic state $\vert\beta,\gamma ; N \rangle$,
Eq.~(\ref{condgen}), where $(\beta,\gamma)$ are quadrupole shape variables.
For the general IBM Hamiltonian (\ref{hIBMnormal})
the surface reads
\ba
V(\beta,\gamma) &=&
N(N-1)\left\{ E_0 + {\textstyle\frac{1}{2}}\beta^2
\left [ a - b{\textstyle\frac{1}{2}}\beta\sqrt{2-\beta^2}\cos 3\gamma
+ (c-a){\textstyle\frac{1}{2}}\beta^2 \right ]\right\}~.
\label{eLan2}
\ea
The coefficients $E_0,a,b,c,$ involve particular linear
combinations of the Hamiltonian's parameters~\cite{lev87}
\ba
&&a = u_2 + 2v_0 -2u_0 +\bar{\epsilon}\; , \;
b = 2\sqrt{\textstyle{\frac{2}{7}}}v_2\; , \;
c = \textstyle{\frac{1}{5}}c_0
+ \textstyle{\frac{2}{7}}c_2
+ \textstyle{\frac{18}{35}}c_4 +\bar{\epsilon}
\; , \; E_0 = u_0 + \bar{\epsilon}_s ~,\qquad
\nonumber\\
&&
\bar{\epsilon} = \bar{\epsilon}_d -\bar{\epsilon}_s \; , \;
\bar{\epsilon}_i = \epsilon_i/(N-1) ~.
\label{abc}
\ea
In analyzing properties of the IBM potential surface, it is convenient
to use a different parameterization, $\tilde{\beta}$, instead of $\beta$,
where the two variables are related by Eq.~(\ref{btlb}).
The intrinsic state in the new parameterization reads~\cite{gino80,diep80}
\bsub
\label{condgentlb}
\ba
\vert\tlb,\gamma ; N \rangle &=&
(N!)^{-1/2}[\,\Gamma^{\dagger}_{c}(\tlb,\gamma)\,]^N\vert 0\rangle ~,\\
\Gamma^{\dagger}_{c}(\tlb,\gamma) &=&
(1+\tlb^2)^{-1/2}
\left [\tlb\cos\gamma d^{\dagger}_{0} + \tlb\sin\gamma
( d^{\dagger}_{2} + d^{\dagger}_{-2})/\sqrt{2} + s^{\dagger}\right ] ~,
\label{GKcond}
\ea
\esub
and the potential surface has the form
\bsub
\ba
V(\tilde{\beta},\gamma) &=&
N(N-1)[ E_0 + f(\tilde{\beta},\gamma)] ~,\\
f(\tilde{\beta},\gamma) &=&
(1+\tilde{\beta}^2)^{-2}\tilde{\beta}^2\left
[ a - b\tilde{\beta}\cos 3\gamma + c\tilde{\beta}^2 \right ] ~.
\label{eLan}
\ea
\esub
The extremum equations,
$\partial V/\partial\tlb=\partial V/\partial\gamma=0$,
always have $\tilde{\beta}=0$ as a solution.
It is a local minimum for $a>0$ and a global minimum if in addition,
$c- b^2/4a >0$. For $b\neq 0$, a deformed extremum
$(\tilde{\beta}^{*}\neq 0,\gamma^{*})$,
has $\gamma^{*}=0$ (prolate) or $\gamma^{*}=\pi/3$ (oblate), mod($2\pi/3$).
For $\gamma=0$, $\tilde{\beta}^{*}\neq 0$ is a real solution
of the following equation
\ba
2a - 3b\tilde{\beta} + 2(2c-a)\tilde{\beta}^2 + b\tilde{\beta}^3 = 0 ~.
\label{exteq}
\ea
For $\gamma=\pi/3$, the sign of $b$ in Eq.~(\ref{exteq}) is reversed.
This equation has one real root for $D>0$; all roots real and unequal
for $D<0$; and all roots real and at least two are equal for $D=0$,
where
\ba
D = -(1+\zeta^2)^3
+ \left (\frac{3}{2}\zeta + \zeta^3 + \frac{a}{b}\right )^2
\;\;\; , \;\;\;
\zeta = \frac{2(2c-a)}{3b} ~.
\label{D}
\ea
In a local minimum, $\gamma^{*}=0\,(\pi/3)$ for $b>0\, (b<0)$
and $b\tilde{\beta}_{*}(3+\tilde{\beta}_{*}^2) - 4a > 0$.
In a global minimum, $b\tilde{\beta}_{*} - 2a > 0$.

Given an Hamiltonian $\hat{H}(\lambda)$ describing a QPT, its potential
surface coefficients, $a(\lambda)$, $b(\lambda)$ and $c(\lambda)$,
Eq.~(\ref{abc}), depend on the control parameter $\lambda$.
In case of a first-order QPT between a spherical and prolate-deformed
shape, the value of the control parameter at the critical point
$(\lambda_c)$, which defines the critical-point Hamiltonian
$\hat{H}(\lambda_c)$, is determined by the condition
\ba
b^{2}=4ac \;,\; a>0,\; b > 0\; , \;
&& f(\tlb,\gamma=0) =
c(1+\tlb^2)^{-2}\tlb^2\left ( \tlb -\bar{\beta}\,\right )^2 ~.
\label{1st}
\ea
The corresponding potential surface shown, has degenerate spherical and
deformed minima at $\tlb=0$ and $(\tlb=\bar{\beta},\gamma=0)$,
where $\bar{\beta} =2a/b$.
The value of the control parameter at the
spinodal point ($\lambda^{*}$), where the deformed minimum disappears,
is obtained by requiring $D=0$, with $D$ given in Eq.~(\ref{D}).
The value of the control parameter at the
anti-spinodal point ($\lambda^{**}$), where the spherical minimum
disappears, is obtained by requiring $a=0$.

The potential surface coefficients for the first-order
intrinsic Hamiltonian of Eq.~(\ref{eq:Hint}), are given by
\bsub
\ba
\hat{H}_1(\rho)/\bar{h}_2:\;
&&a = 2\beta_{0}^2\; , \;
b = 4\beta_{0}^2\rho \; , \;
c = 2 \;\; , \;\;
E_0 = 0 ~,\\
\hat{H}_2(\xi)/\bar{h}_2:\;
&&a = 2\beta_{0}^2[1-\xi(1+\beta_{0}^2)]\; , \;
b = 4\beta_{0}\;\; , \;\;
c=2 +\xi(1-\beta_{0}^2)\; , \; E_0 = \xi\beta_{0}^4 ~.
\qquad\qquad
\ea
\esub
The values of the control parameters at the critical $(\rho_c,\xi_c)$,
spinodal $(\rho^{*})$, and anti-spinodal $(\xi^{**})$ points, given in
Eqs.~(\ref{critical})-(\ref{antispinodal}), were obtained
by the conditions mentioned above.
The energy surface coefficients of the collective Hamiltonian,
Eq.~(\ref{eq:Hcol}), all vanish.

\section*{Appendix B: Linear correlation coefficients}

The extent to which two $n$-dimensional
vectors, $X$ and $Y$, with components
$X_m$ and $Y_m$ respectively, support a linear relation, can be measured
by the standard Pearson correlation coefficient, defined as
\begin{equation}
\pi(X,Y) = \frac{1}{n-1}\sum_{m=1}^n
\frac{(X_m - \bar{X})}{s_X}\frac{(Y_m - \bar{Y})}{s_Y} ~.
\label{piXY}
\end{equation}
Here $\bar{X}$, $\bar{Y}$ and $s_X$, $s_Y$ are the mean values and
standard deviations of the vector components, respectively.
The values of the Pearson coefficient lie in the range
$\-1\leq\pi(X,Y)\leq 1$, with
$\pi(X,Y)=1$, $\pi(X,Y)=-1$, and $\pi(X,Y)=0$ indicate a perfect
correlation, perfect anti-correlation and no linear correlation,
respectively.

In the present work, we apply the Pearson indicator
to estimate the amount of correlations between
two eigenstates of an IBM Hamiltonian,
$\ket{L_i}$ and $\ket{L'_j}$. For that purpose,
we expand both states in the SU(3) basis as in Eq.~(\ref{Li}),
and identify their SU(3) probability distributions
$P_{(\lambda,\mu)}^{(L_i)}$ and $P_{(\lambda,\mu)}^{(L'_j)}$, Eq.~(\ref{Plammu}),
with the components of the vectors $X_m$ and $Y_m$ in Eq.~(\ref{piXY}).
To ensure an equal number of components,
we assign a value zero to the component, $P_{(\lambda,\mu)}^{(L'_j)}=0$,
if the angular momentum $L'$
is not contained in a particular SU(3) irrep
$(\lambda,\mu)$ that does accommodate the angular momentum $L$.
To associate a band of states with a given state $\ket{L_i}$,
we scan the entire spectrum of states $\ket{L'_j}$,
with angular momentum $L'\neq L$, and choose the state that
maximizes the Pearson correlation coefficient
$\max_{j} \{\pi(L_i,L'_j)\}$, Eq.~(\ref{piXY}).
This identifies among the ensemble of states with angular momentum $L'$,
the most correlated state with $\ket{L_i}$, which is the favored candidate
to be its member in the same band.
This procedure, adapted from~\cite{Macek10}, was used in Eq.~(\ref{Pearson})
to identify $K=0$ bands, composed of sequences of rotational states with
$L=2,4,6,$ built on a given $\ket{L=0_i}$ bandhead state.

\end{document}